\documentclass[11pt,a4paper]{article}
\pdfoutput=1
\usepackage{jheppub}

\usepackage{amsmath}
\usepackage{amssymb}
\usepackage{graphicx}
\usepackage{bbm}
\usepackage{color}

\usepackage{indentfirst}
\usepackage{amsfonts}
\usepackage{caption}
\usepackage{subcaption}
\usepackage{hyperref} 
\usepackage{ulem} 

\usepackage{tikz}
\usetikzlibrary{arrows,positioning,shapes,mindmap}
\usetikzlibrary{decorations.pathmorphing}
\usetikzlibrary{decorations.markings}
\usepackage{comment}

\def \onehalf {\frac{1}{2}}

\def\gam{\gamma}
\def\sig{\sigma}
\def\mh{m_h}

\def\lsim{\mathrel{\raise.3ex\hbox{$<$\kern-.75em\lower1ex\hbox{$\sim$}}}}
\def\gsim{\mathrel{\raise.3ex\hbox{$>$\kern-.75em\lower1ex\hbox{$\sim$}}}}

\def\gev{\;\hbox{GeV}}
\def\tev{\;\hbox{TeV}}
\def\decay{\text{decay}}
\def\ECAL{\text{ECAL}}

\def\beq{\begin{equation}}
\def\eeq{\end{equation}}
\def\bea{\begin{eqnarray}}
\def\eea{\end{eqnarray}}

\def\bit{\begin{itemize}}
\def\eit{\end{itemize}}
\def\bec{\begin{center}}
\def\eec{\end{center}}

\title{Boosted scalar confronting 750 GeV di-photon excess
}
\author[a]{Yun~Jiang,}
\author[b]{Lingfeng~Li,}
\author[b]{Rui~Zheng}

\affiliation[a]{\,NBIA and Discovery Center, Niels Bohr Institute, University of Copenhagen, \\Blegdamsvej 17, DK-2100, Copenhagen, Denmark}
\affiliation[b]{\,Department of Physics, University of California, Davis, CA 95616, USA} 

\emailAdd{yunjiang@nbi.ku.dk}
\emailAdd{llfli@ucdavis.edu}
\emailAdd{ruizh@ucdavis.edu}
   
\abstract{
We consider the di-photon signal arises from two bunches of collimated photon jets emitting from a pair of highly boosted scalars.
Following the discussion of detecting the photon jets at the collider, we extend the two-Higgs-doublet model (2HDM) by adding a gauge singlet scalar. 
To explain the di-photon excess which is recently observed at the first 13~TeV run of the LHC, the mixing between the heavy doublet state and the newly added singlet is crucially needed. After the mixing, one can have a heavy Higgs state $Y_2$ at 750~GeV and a very singlet-like scalar $Y_1$ of sub-GeV, which would be highly boosted through the $Y_2$ decay.
Both real singlet and complex singlet extension are studied. It turns out that only the complex model can yield the 1-10~fb cross section in the di-photon final state in accompany with the decay length of the order of 1~m for the $Y_1$. This complex model parametrically predicts the width of 750 GeV resonance $\gtrsim 1$~GeV.
In addition, the pseudoscalar component of the singlet in this model is naturally stable and hence could be a dark matter candidate.

}

\begin{document}
\maketitle

\tikzstyle{every picture}+=[remember picture]
%%%%%%%%%%%%%%%%%%%%%%%%%%%%%%%%%%%%%%%%%%%%%%%%%%%%%%%%%%%%%%%%%%%%%%%%%%%%%%%%%%%%%%%%%%%%%%%%%%%%%%%%%%%%%%%%%%%%%%%%%%%%%%%%%%%%%%%%%%%%%%%
\pgfdeclaredecoration{complete sines}{initial}
{
    \state{initial}[
        width=+0pt,
        next state=sine,
        persistent precomputation={\pgfmathsetmacro\matchinglength{
            \pgfdecoratedinputsegmentlength / int(\pgfdecoratedinputsegmentlength/\pgfdecorationsegmentlength)}
            \setlength{\pgfdecorationsegmentlength}{\matchinglength pt}
        }] {}
    \state{sine}[width=\pgfdecorationsegmentlength]{
        \pgfpathsine{\pgfpoint{0.25\pgfdecorationsegmentlength}{0.5\pgfdecorationsegmentamplitude}}
        \pgfpathcosine{\pgfpoint{0.25\pgfdecorationsegmentlength}{-0.5\pgfdecorationsegmentamplitude}}
        \pgfpathsine{\pgfpoint{0.25\pgfdecorationsegmentlength}{-0.5\pgfdecorationsegmentamplitude}}
        \pgfpathcosine{\pgfpoint{0.25\pgfdecorationsegmentlength}{0.5\pgfdecorationsegmentamplitude}}
}
    \state{final}{}
}
\tikzset{
fermion/.style={thick,draw=black, line cap=round, postaction={decorate},
    decoration={markings,mark=at position 0.6 with {\arrow[black]{latex}}}},%triangle 45,stealth
photon/.style={thick, line cap=round,decorate, draw=black,
    decoration={complete sines,amplitude=4pt, segment length=6pt}},
boson/.style={thick, line cap=round,decorate, draw=black,
    decoration={complete sines,amplitude=4pt,segment length=8pt}},
gluon/.style={thick,line cap=round, decorate, draw=black,
    decoration={coil,aspect=1,amplitude=3pt, segment length=8pt}},
scalar/.style={dashed, thick,line cap=round, decorate, draw=black},
ghost/.style={dotted, thick,line cap=round, decorate, draw=black},
->-/.style={decoration={
  markings,
  mark=at position 0.6 with {\arrow{>}}},postaction={decorate}}
 }

\makeatletter
\tikzset{
    position/.style args={#1 degrees from #2}{
        at=(#2.#1), anchor=#1+180, shift=(#1:\tikz@node@distance)
    }
}
\makeatother

%%%%%%%%%%%%%%%%%%%%%%%%%%%%%%%%%%%%%%%%%%%%%%%%%%%%%%%%%%%%%
\section {Introduction}
\label{sec:Intro}
%%%%%%%%%%%%%%%%%%%%%%%%%%%%%%%%%%%%%%%%%%%%

Recently the ATLAS \cite{ATLAS:2015} and CMS \cite{CMS:2015} collaborations at the LHC presented results for the di-photon search at energy $\sqrt{s}=13\tev$. Both collaborations have found an excess over the background at $m_{\gam\gam}\sim750\gev$. The local significances of this signal   are $3.9~\sigma$ and $2.6~\sigma$ for the ATLAS and CMS collaborations, respectively. We summarize the LHC data in Table~\ref{tab:lhc} for the excess observed above expected background over an interval centered on $m_{\gamma\gamma}\approx750\gev$. Meanwhile, it is important to note that no excess is observed in any other channels, including the $t\bar t,~hh,~WW,~ZZ$ and di-jet final states. 
%In this report, we present a possible interpretation of this signal .....
\begin{table}[h]
	\caption{Summary of the LHC di-photon excess at the invariant mass of $\sim750\gev$ from the ALTAS \cite{ATLAS:2015} and CMS \cite{CMS:2015} collaborations at energy $\sqrt s=13\tev$ ($\sigma$ is the local statistical significance). }
	\label{tab:lhc}
	\centering
	\begin{tabular}{|c|c|c|}
		\hline
		&ATLAS @ $\sqrt s=13\tev$ \cite{ATLAS:2015}& CMS @ $\sqrt s=13\tev$ \cite{CMS:2015}\cr
		\hline
		Excess Events &$14$ with $3.9\sigma$ & $10$ with $2.6\sigma$ \cr
		\hline
		$\sigma(pp\to\gamma\gamma)$ &$(10\pm3)$ fb&$(6\pm3)$ fb \cr
		\hline
	\end{tabular}
\end{table}

Since the announcement of this di-photon excess, a plethora of scenarios have been studied for interpreting this excess through the direct decay of a heavy resonance (to a pair of photon) in the context of both model-independent approaches and concrete models. 
The simple and natural candidates for the 750 GeV resonance include the heavy CP-even field $H$ and the CP-odd field $A$ of the popular two-Higgs-doublet model (2HDM) or the minimal supersymmetric SM (MSSM).
However, it has been shown in~\cite{Bernon:2015qea} that having the $H$ and/or $A$ at 750 GeV, the cross section in the di-photon final state via the gluon-fusion production $\sigma(gg \to H/A) {\rm BR} (H/A \to \gamma\gamma)$ approaches at most $10^{-2}$ fb, which is three orders smaller than the level observed at the LHC 13~TeV run. This results from the fact that the branching ratio of their decay into $\gam\gam$ (mediated by the top quark loop and, for the case of the $H$, the $W^\pm$ loop)  is maximally of order of $\sim 10^{-5}$ due to the dominance of tree-level decays into $t\bar t$ and/or $W^+ W^-/ZZ$ (the latter being possibly significant away from the alignment limit for the SM-like $h$ at $\mh\simeq 125\gev$). 
The contribution of the charged Higgs to the loop-induced $H\gamma\gamma$ coupling is negligible. Therefore, the minimal 2HDM cannot explain the di-photon excess. To enhance the di-photon signal, 
one considered to extend the 2HDM by adding extra heavy colored states~\cite{Angelescu:2015uiz,Han:2015qqj,Badziak:2015zez,Altmannshofer:2015xfo,Bizot:2015qqo,Han:2016bus,Hernandez:2016rbi,Djouadi:2016eyy,Han:2016bvl,Kang:2015roj,Arbelaez:2016mhg,Bertuzzo:2016fmv,Ahriche:2016mcx} that would give additional loop contributions to the $Xgg$ and/or $X\gam\gam$ couplings of the 750 GeV resonance, which we generically denote by $X$. 
There also exists many other works related to direct di-photon production~\cite{Chiang:2016ydx,Chaudhuri:2016rwo,Delgado:2016arn,Higaki:2015jag,Ghosh:2015apa,Franceschini:2015kwy,Aloni:2015mxa,DiChiara:2015vdm,Das:2015enc,Kanemura:2015vcb,Jiang:2015oms,Tsai:2016lfg,Chakrabortty:2015hff,Kim:2015ksf,Berthier:2015vbb,Godunov:2016kqn,Kavanagh:2016pso,deBlas:2015hlv,Son:2015vfl,Kaneta:2015qpf,Buttazzo:2015txu,McDermott:2015sck,Ellis:2015oso,Dutta:2015wqh,Chao:2015ttq,Fichet:2015vvy,Falkowski:2015swt,Benbrik:2015fyz,Wang:2015kuj,Cao:2015twy,Gu:2015lxj,Cheung:2015cug,Li:2015jwd,Wang:2015omi,D'Eramo:2016mgv,Ko:2016wce,Chao:2016aer,Dolan:2016eki,Ding:2016udc,Bae:2016xni,Chakraborty:2015jvs,Dhuria:2015ufo,Zhang:2015uuo,Salvio:2015jgu,Aydemir:2016qqj,Salvio:2016hnf,Hamada:2016vwk,Cox:2015ckc,Ahmed:2015uqt,Megias:2015ory,Bardhan:2015hcr,Davoudiasl:2015cuo,Cai:2015hzc,Cao:2016udb,Abel:2016pyc,Pilaftsis:2015ycr,Molinaro:2015cwg,Bian:2015kjt,Kim:2015xyn,Ben-Dayan:2016gxw,Barrie:2016ntq,Chiang:2016eav,Higaki:2016yqk,Mambrini:2015wyu,Backovic:2015fnp,Kobakhidze:2015ldh,Han:2015cty,Han:2015dlp,Bi:2015uqd,Bauer:2015boy,Barducci:2015gtd,Dev:2015isx,Han:2015yjk,Park:2015ysf,Huang:2015svl,Ghorbani:2016jdq,Bhattacharya:2016lyg,Okada:2016rav,Cao:2016cok,Kawamura:2016idj,Ge:2016xcq,Staub:2016dxq,Redi:2016kip,Bi:2016gca,Csaki:2016kqr,Bai:2016rmn,Deppisch:2016qqd,Hewett:2016omf,Ko:2016sht,Carmona:2016jhr,Howe:2016mfq,Collins:2016pef,Frandsen:2016bke,Dillon:2016fgw,Ellis:2016yrj,Liu:2016lkj,Chakrabarty:2016hxi,Cynolter:2016jxv,McDonald:2016cdh,Lebiedowicz:2016lmn,Chala:2016mdz,Kusenko:2016vcq,Kamenshchik:2016tjz,Barrie:2016ndh,Nilles:2016bjl,Agarwal:2016gxe,Duerr:2016eme,Gopalakrishna:2016tku,Yamada:2016jgg,Takahashi:2016iph,Iwamoto:2016ral,Hamaguchi:2016umx,Dutta:2016ach,Bae:2016ewo,Chen:2016sck,Dev:2015vjd,Bardhan:2016rsb,Marzola:2015xbh,Ding:2015rxx,Ding:2016ldt,Dey:2015bur,Modak:2016ung,Bonilla:2016sgx,Heckman:2015kqk,Moretti:2015pbj,Harigaya:2015ezk,Hall:2015xds,Harigaya:2016pnu,Harigaya:2016eol}.

Other than the direct decay of a resonance $X(750)$ into di-photon,  alternative topologies that could explain the excess have been studied. 
These include the ideas that i) the di-photon arises from a three-body decay~\cite{Kim:2015ron,Alves:2015jgx,Bernon:2015abk,Huang:2015evq,Cho:2015nxy,An:2015cgp}  in association with an invisible particle 
or ii) two bunches of collimated photon jets emitting from a pair of highly boosted (pseudo)scalars~\cite{Dobrescu:2000jt,Dobrescu:2000yn,Agrawal:2015dbf,Chang:2015sdy,Bi:2015lcf,Aparicio:2016iwr,Ellwanger:2016qax,Dasgupta:2016wxw,Domingo:2016unq,Badziak:2016cfd}.
For the latter case, if there is a (pseudo)scalar $Y$ of sub-GeV scale, its decay into jets will be enormously suppressed and thereby leads to a substantial branching ratio to di-photon. Another element of achieving this scenario is to require a heavy scalar $X$ at 750~GeV which decays into two sub-GeV scalar particles. Due to the huge mass difference, the light scalar particles will be highly boosted so each photon pair could be identified as photon jet.
The CP-odd Higgs $A$ of the 2HDM could have been a good candidate for this light state. However, the presence of a very light $A$ would prevent the non-SM CP-even Higgs $H$ from being heavier than $\simeq 630 \gev$ due to the electroweak precision observables (EWPO) once the lightest CP-even Higgs $h$ is identified as the SM one observed at 125 GeV~\cite{Bernon:2014nxa}.~\footnote{This conclusion also holds for $m_A$ in the scenario where the heavier CP-Higgs $H$ being $125\gev$ is SM-like.} Therefore, we extend the 2HDM by adding a gauge singlet scalar field. Both real scalar and complex scalar models will be studied in this paper. 
For our purpose, we allow the mixing between the singlet field $S$ and the heavy Higgs $H$ and identify the heavy mass eigenstate $Y_2$ as the 750 GeV resonance, while the light mass eigenstate $Y_1$ produces photon jet.  Since no additional colored fermion is introduced, the $Y_2$ has to contain sizeable doublet component, otherwise it will be difficult to be produced via gluon fusion. On the contrary, the light state $Y_1$ tends to be singlet-like.
We find that it is possible to explain the 750 GeV di-photon excess in the complex singlet model, while the real singlet model is strongly limited by the problem of a too long decay length of the boosted light scalar. Very recently, this idea has been applied to interpret the 750 GeV di-photon excess in the context of NMSSM~\cite{Ellwanger:2016qax,Domingo:2016unq,Badziak:2016cfd}.  

This paper is organized as follows. In Sec.~\ref{sec:photonjet}, we introduce properties of photon jet and its implications on di-photon excess. In Sec.~\ref{sec:ggfusion}, we discuss di-photon signal from gluon fusion and its decay. Two attempts to explain this di-photon excess are made in Sec.~\ref{sec:realmodel} and in Sec.~\ref{sec:complexmodel} with real and complex singlet scalar extensions to 2HDM respectively. The collider prospects are discussed in Sec.~\ref{sec:prospects}. Finally, Sec.~\ref{sec:conclusion} summarizes our main results of this work.

\section{Detecting the boosted photon jets in the collider}\label{sec:photonjet}
%%%%%%%%%%%%%%%%%%%%%%%%%%%%%%%%%%%%%%%%

The photon-jet is a special object that consists of a cluster of (nearly) collinear photons which have a signature similar to that of a single photon.
This idea was initially proposed in~\cite{Ellis:2012sd} and has recently been applied to explain this di-photon excess~\cite{Agrawal:2015dbf,Chang:2015sdy,Bi:2015lcf,Aparicio:2016iwr,Ellwanger:2016qax,Dasgupta:2016wxw,Domingo:2016unq,Badziak:2016cfd}. It could be generated from the decay of a highly boosted light particle (sub-GeV). 
In this section we discuss two technical issues, angular separation and decay length, which are very crucial in detecting a photon-jet in the collider machine.

Suppose a particle $X(750)$, once produced in the hadron collider, {\it instantly} decays to two highly boosted $Y$s. For each $Y$, it decays into a pair of photons with a branching ratio close to $100\%$. The angular separation of two outgoing photons (which forms a photon-jet) will be $\theta \sim \frac{2 m_{Y}}{p_{Y}} \sim \frac{2 m_{Y}}{375}$, assuming 375 GeV is the expected momentum  for $X(750) \to Y Y$ process. If this angular separation is smaller than the resolution of the ECAL in the CMS/ATLAS, the photon jet could probably be mistagged as a single photon. Hence, the naive estimation from the parameters of the ECAL segmentation ($\Delta \phi = 0.0174$ in the CMS~\cite{Khachatryan:2015iwa} and $\Delta \phi = 0.025$ at the ATLAS~\cite{Aad:2009wy,Aad:2010sp}) places a bound on $Y$ mass: $m_{Y}\lesssim (0.0174\times375)/2=3.26$~GeV so that two photons will hit the same ECAL segment and eventually be recorded as a single photon. Of course, the actual photon jet conversion could be much more complicated and gives a stronger bound. However, this is not a big issue as $m_{Y}$ stays well below this bound in the following analysis.   

\begin{figure}[t]
\begin{center}
\includegraphics[width=0.45\textwidth]{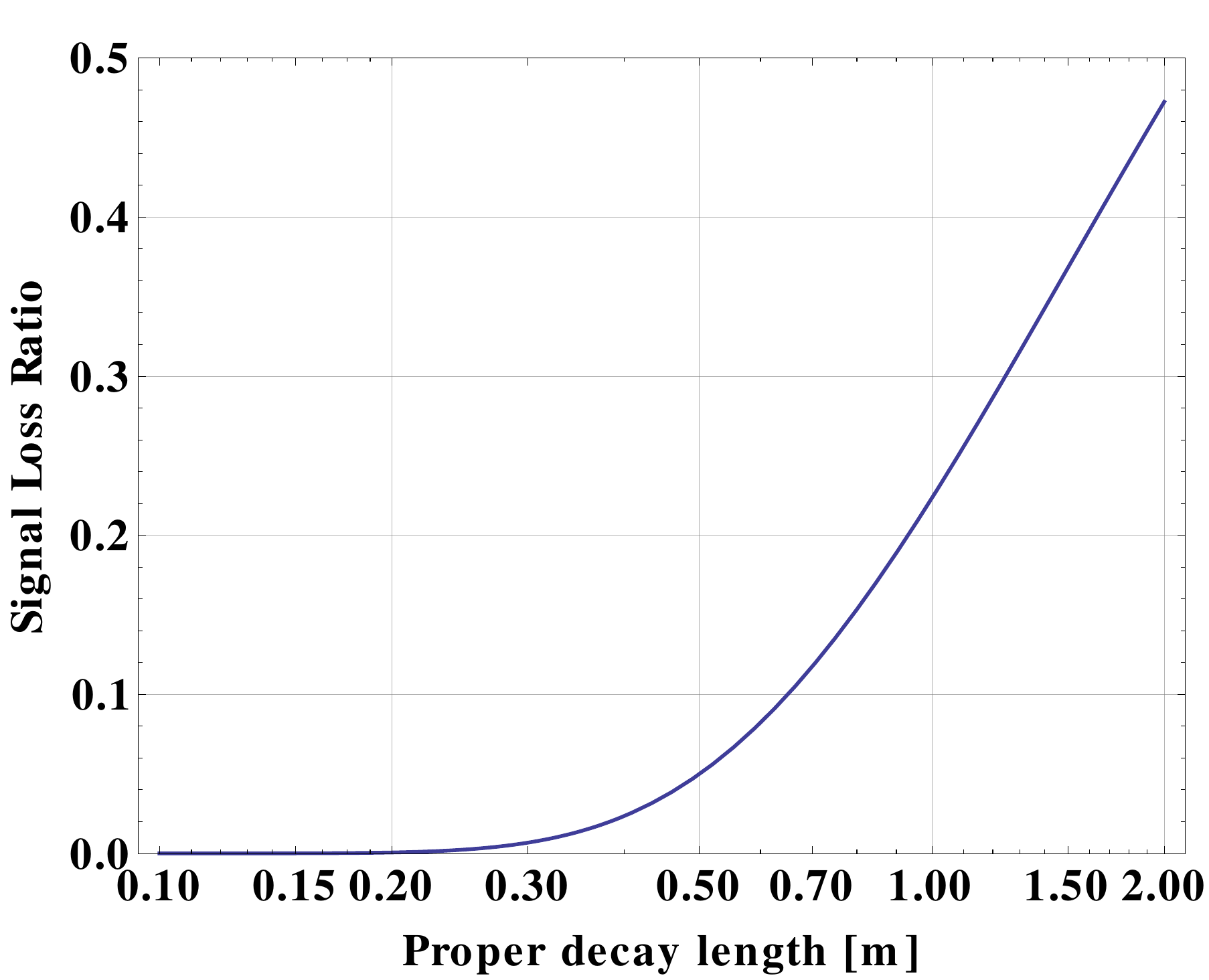}
\caption{Assuming that the typical $R_{\text{ECAL}}=1.5$ m from the beam, the signal loss ratio in percentage versus the decay length ($L_{\decay}$) of light particle $Y$. When the decay length of $Y$ exceeds 2 m, about half of the $Y$ generated would not be detected.}
\label{decaylength}
\end{center}
\vspace{1mm}
\end{figure}

The other concern about this topology in the language of photon-jet is the proper decay length of the light (pseudo)scalar $Y$. This is because a very light but highly boosted particle could have an unexpectedly long decay length. If its decay takes place after passing through the ECAL detector, it cannot be technically detected. 
Quantitatively, the proper average decay length of such a highly boosted particle is given by
\begin{equation}
	L_{\decay}=c\tau\gamma \approx \frac{375c}{\Gamma_{Y} m_{Y}}
\end{equation} 
in which $\tau=\Gamma^{-1}$ and $\gamma= \frac{E_{Y}}{m_{Y}} \approx \frac{p_{Y}}{m_{Y}}$ are employed. $\Gamma_{Y}$, the total decay width of $Y$, is generally proportional to $m_{Y}$ for the light $Y$ state.
Hence, a lighter $m_{Y}$ leads to a longer $L_{\decay}$. Once the decay length is comparable with the size of detector radius ($R_{\ECAL}$ in this case), the probability of $Y$ escaping the detection soars. The correlation of the proper decay length versus the percentage of signal loss is illustrated in Fig.~\ref{decaylength}. 

We can see that for a decay length longer than $\sim 1$~m, less than $80\%$ of the $Y_1$ decay events would happen inside (or before reaching) the ECAL and eventually be captured by the detector. In this case, a larger production rate is needed to compensate the significant event loss. However, such a long $L_{\decay}$ will cause another problem. For a considerable portion of $Y$ which pass through the ECAL, they will decay within the HCAL and their decay products will be tagged as displaced jets.
We expect that future experiments searching for resonances using the construction of displaced jets and/or photons~\cite{Aad:2013txa,Aad:2014yea,Aad:2015asa} could place an upper bound on $L_{\decay}$.

On the other hand, it was argued in~\cite{Khachatryan:2015iwa} that, if $L_{\decay}$ is close to $R_{\text{ECAL}}$, $Y$ would decay in a position near the ECAL layer, thereby, the two photons produced from a not-so-light $Y$ ($m_{Y} \gtrsim$ 3.26~GeV) can still hit the same ECAL segment simultaneously. 
Nonetheless, the probability of $Y$ decay takes place drops off exponentially along the distances away from the beam.
This infers a larger probability of  $Y$ decay occurring close to the beam rather than near the ECAL layer, giving rise to a pair of photons distinguishable when reaching the ECAL layer. 
If this were true, one would have seen a significant amount of 3-photon or 4-photon signals.

In short, $L_{\text{decay}}(Y) \lsim 1$~m and $m_{Y}\lsim 3$~GeV are viable assumptions in our scenario in order to have sufficient di-photon signal detected in the ECAL.

\vspace{2mm}

\section{Gluon fusion production and di-photon decay}\label{sec:ggfusion}
%%%%%%%%%%%%%%%%%%%%%%%%%%%%%%%%%%%%%%%%
\subsection{Doublet $X(750)$ and its production at the collider}

\begin{figure}[t]
\begin{center}
\includegraphics[width=0.45\textwidth]{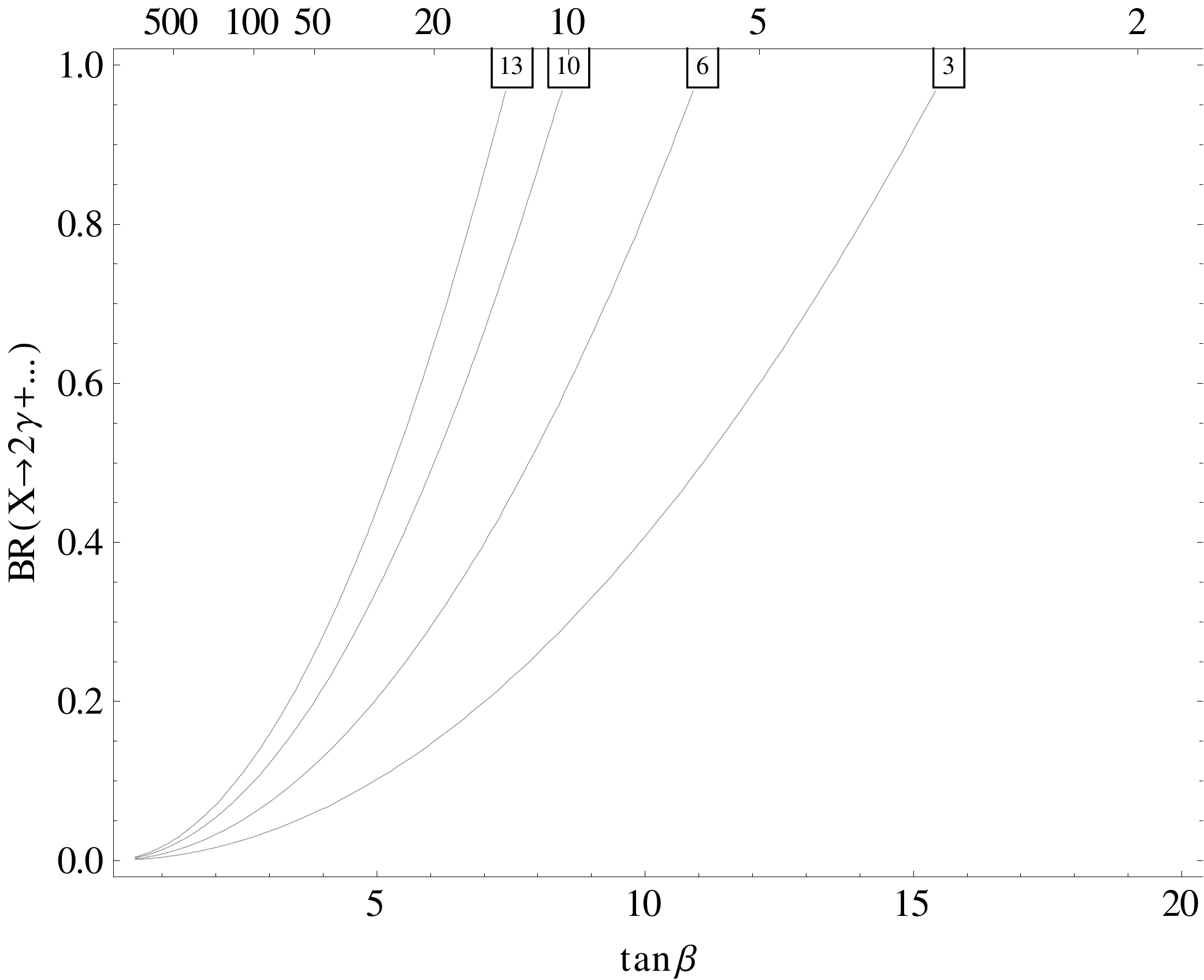}
\hspace{4mm}
\includegraphics[width=0.45\textwidth]{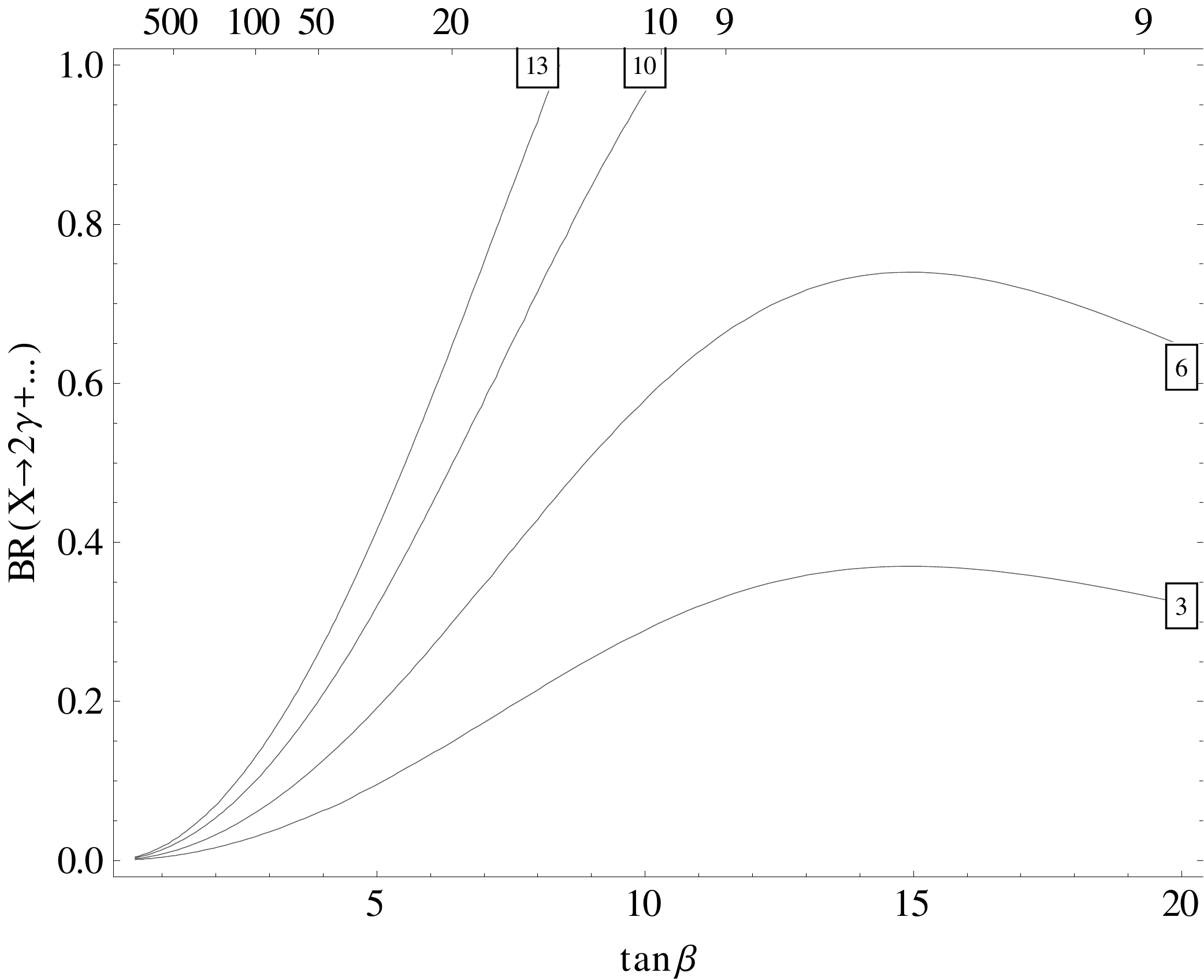}
\caption{The correlation of the predicted di-photon signal ratio (in-box numbers) to the size of the branching ratio of $X(750) \to 2 \gamma + \text{else}$. Such decay could be either a direct or cascade decay. The gluon fusion production cross section (indicated in the unit of fb on the top axis) is obtained in the assumption that $X(750)$ resonance behaves like a doublet whose Yukawa couplings given in Type~I (left)  and Type~II (right) models.}
\label{fig:ggBRplot}
\end{center}
\vspace{1mm}
\end{figure}

Let us begin with a numerical estimate to assess the possibility of realizing this scenario in the context of the 2HDM with the inclusion of a singlet scalar. 
For simplicity, we assume that the $X(750)$ resonance behaves like the heavy CP-even Higgs such as the couplings and the production modes. Thus, the $X(750)$ is dominantly produced via gluon fusion or $b-$quark associated production (we focus on $\tan\beta \lesssim 4$ and thus do not consider this production mode in the present paper). 
It is shown in~\cite{Bernon:2015qea} that the gluon-fusion production cross section decrease as $\tan\beta$ grows in Type~I, while this cross section minimizes at a modest value of $\tan\beta \sim 8$ in Type~II.
Hence, a maximal value of $\tan\beta$ that could yield the di-photon signal can be estimated by assuming the $X(750)$ decays inclusively into two photons with 100\% branching ratio. 
As shown in Fig.~\ref{fig:ggBRplot}, we evaluate the di-photon signal,
$\sigma\left(gg\to X\right) \mbox{BR}\left(X\to \gamma\gamma\right)$, as a function of $\tan\beta$ at different levels of $\text{BR}(X\to 2\gam + ... )$.

As expected, the obtainable di-photon signal diminishes as $\tan\beta$ increases, while maintaining the level of $\text{BR}(X\to 2\gam + ... )$. (This is also true for $\tan\beta \lesssim 20$ in Type~II.) This is the result of suppressing the $X(750)$ production at high $\tan\beta$ due to the fact the heavy Higgs coupling normalized to the SM value with the top quark, which dominantly mediates the gluon-fusion, is inversely proportional to $\tan\beta$. To offset this drop, the decay branching ratio $\text{BR}(X\to 2\gam + ... )$ has to be increased to acquire a larger signal rate. For instance, in order to yield a $10$~fb signal, $t_\beta \leq 8(10)$ and $t_\beta \leq 6(7)$ are required in Type~I (Type~II) in the assumption of $\text{BR}(X\to 2\gam + ... )$ at $100\%$ and more practically at $60\%$, respectively.
In particular, such an upper limit on the $\tan\beta$ would be relaxed in Type~II for a lower cross section of $6$~fb.
In both Type~I and Type~II, a relative large cross session ($\approx 700$ fb in the SM) of producing the $X(750)$ state is achieved when $t_\beta \lesssim 1$ as a result of the enhanced coupling to top-quark. Meanwhile, this will maximize the cross section in the $t\bar{t}$ and di-jet final states. Comparing with the current experiment bounds, $\sigma(pp \to X \to gg) \lesssim 10$~pb and $\sigma(pp \to X \to t\bar{t})\lesssim 700$~fb~\cite{CMS:2015neg,Aad:2015fna}, we 
find that our model stays well below the di-jet bound while the di-top bound could be marginal in the case of $t_\beta \geq 1$. Thus, we shall limit $t_\beta \geq 1$ in the following analysis.

\subsection{Singlet $Y$: scalar vs. pseudoscalar}
\label{ssec:singlet}

Concerning the light state $Y$, there are two important requirements that must be satisfied to realize this scenario. These include the presence of decay into di-photon and the substantially large total width. To accomplish the first goal, unlike the situation in the NMSSM, this state cannot be a pseudoscalar in the present study. As explained in the Introduction, the CP-odd scalar from the doublet field cannot be too light, while the pseudoscalar arising from the newly introduced singlet field does not couple to SM particles in the assumption of CP-conservation and thus has no di-photon decay mode. On the other hand, requiring $Y$ a singlet scalar is not sufficient. 
Though $Y$ can decay to di-photon via charged Higgs loop, or to $4f$ via two off-shell $H^*, A^*$ or $H^{\pm *}$, these decay channels are highly suppressed by kinematics. Tree-level decays to light quarks are absent due to its decoupling to the SM particles. 
The combination of these effects result in a extremely narrow decay width. The proper decay length for a singlet scalar is typically at the order of kilometer. Therefore, $Y$ state must be a scalar and also gains an amount of doublet composition from the mixing. The content to which the mixing is needed will be analyzed in Sec.~\ref{sec:decayY1}. Meanwhile, the singlet pseudoscalar, if present, could serve as a dark matter candidate.

In addition, the mass of $Y$ is very crucial in determining the products of $Y$ decay. When $m_Y$ is $\mathcal{O}(1)$ GeV, its main decay products are jets. 
In this case, the decay to di-photon is only mediated via loop diagrams and its branching ratio $\lsim 0.1-1\%$ (see an example of $m_{Y}=5$ GeV in Fig.~\ref{fig:doubletfraction}). In contrast, the notorious jet background is well suppressed in the decay for a sub-GeV particle. In particular, considering a $m_Y$ below the $s\bar{s}$ and $\mu^+ \mu^-$ thresholds, it will only decay to $u\bar u,d\bar{d}$ and $e^+e^-$ at tree level. Notice that $m_{Y} < \Lambda_{\text{QCD}}$, so the outgoing quark pairs would not develop into di-jet but hadronize and cascade decay into di-photon. On the other hand, $BR(Y \to e^+e^-)$ is negligible compared to quark pair production due to the small electron Yukawa coupling. Therefore, one can expect the branching ratio of the decay to di-photon for such a light $Y$ is nearly $\sim 100\%$.

Of course, the situation becomes rather subtle when $m_{Y}$ exceeds the $\mu^+ \mu^-$ threshold, $\approx 210$ MeV. Since the muon has a much larger Yukawa coupling than $u$ and $d$ quarks, there will be a considerable portion of $Y$ decaying to muon pairs. This seems to generate a heavy suppression in the branching ratio of the decay to di-photon. However, strong dynamics may have a large correction to the decay width of $Y\to \gam\gam$ when this decay is mediated via hadronic states.
Above the $\mu^+\mu^-$ threshold, once $m_{Y}$ reaches the $\pi\pi$ threshold, $\approx 280$ MeV, 
$Y$ will hadronically decay into either $\pi^+\pi^-$  or $\pi^0 \pi^0$ since $\pi$ is the lightest hadron.
By virtue of isospin conservation, only about $1/3$ of outgoing pion pairs are $\pi^0\pi^0$ pair and cascade decay to di-photon for each $\pi^0$. Whereas, the remaining $2/3$ will be $\pi^+\pi^-$ and eventually decay to $\mu^+\mu^-+4\nu$, which can be detected by the muon chamber, leaving a large missing $p_T$.

To avoid the subtlety, we will consider the case throughout the paper in which $m_{Y}$ is below the di-muon threshold. In the numerical analysis unless specified, $m_{Y}=150~\text{MeV}$ is chosen and the assumption of BR$(Y\to \gamma\gamma)=100\%$ is globally adopted.

%%%%%%%%%%%%%%%%%%%%%%%%%%%%%%%%%%%%%%%%

\section{Model I: real singlet}
\label{sec:realmodel}
%%%%%%%%%%%%%%%%%%%%%%%%%%%%%%%%%%%%%%%%

We first consider adding to the 2HDM a real scalar gauge singlet $S$.\footnote{This model has been already constructed in earlier literatures, see an example~\cite{Drozd:2014yla} in which the singlet does not acquire a vev and also is $Z_2$ odd for the sake of having a dark matter candidate.} To eliminate the substantial FCNC, we assume a $\mathbb{Z}_2$ symmetry under which $\Phi_2$ is odd. For the singlet $S$ we impose a $\mathbb{Z'}_2$ symmetry under which $S$ is the only odd field.
%\footnote{For the sake of preventing $S$ decay and thereby being a dark matter candidate, one imposes a $Z_2$ symmetry and does not allow it to acquire a vev.}
The gauge invariant and renormalizable Lagrangian for this model is 
\beq
\mathcal{L}_{\text{2HDMS}} = \mathcal{L}_{\text{2HDM}} + \mathcal{L}_{S}
\eeq
where the Lagrangian of CP-conserving but soft $Z_2$ breaking 2HDM can be found in~	\cite{Bernon:2014nxa,Drozd:2014yla,Bernon:2015qea} in which two Higgs-doublet fields are expanded as
\begin{equation}
\Phi_a = 
\begin{pmatrix}
H_a^+ \\ \frac{1}{\sqrt{2}}(v_a+\rho_a+ i\eta_a)
\end{pmatrix} \quad (a=1,2), \qquad S=v_s+\chi
\end{equation}
with the ratio of two vevs given by $\tan\beta \equiv v_2/v_1$.

While the $S$-associated part reads
\beq
\mathcal{L}_{S} = \onehalf m_s^2 S^2 +\frac{1}{4!}\lambda_s S^4+\kappa_1 \Phi^\dagger_1 \Phi_1 S^2 + \kappa_2 \Phi^\dagger_2 \Phi_2 S^2
\eeq
In contrast to~\cite{Drozd:2014yla}, in the present study we allow the $S$ to acquire a vev so that it will mix with the Higgs doublets.
Three mass-squared parameters can be replaced by three vevs through the corresponding minimization conditions such that 
\begin{align}
m_{11}^2 &= m^2_{12} \frac{v_2}{v_1}-(\onehalf \lambda_1 v_1^2 +\onehalf \lambda_{345}v_2^2 +\kappa_1 v_s^2) \label{eq: minh1}\\
m_{22}^2 &= m^2_{12} \frac{v_1}{v_2}-(\onehalf \lambda_2 v_2^2 +\onehalf \lambda_{345}v_1^2 +\kappa_2 v_s^2) \label{eq: minh2}\\
m_s^2 &= -\frac{1}{6} \lambda_s v_s^2 -\kappa_1 v_1^2 -\kappa_2 v_2^2 \label{eq: mins}
\end{align}

%%%%%%%%%%%%%%%%%%%%%%%%
\subsection{Mass eigenstates and spectrum}
%%%%%%%%%%%%%%%%%%%%%%%%

For the CP-even neutral states $(\rho_1, \rho_2, \chi)$ in the $Z_2$ basis, the mass matrix can be written as
\beq
\mathcal{M}^2 = \begin{pmatrix}
	m^2_{12} \frac{v_2}{v_1}+\lambda_1 v_1^2 &  -m^2_{12} +\lambda_{345}v_1 v_2 &  2 \kappa_1 v_1 v_s\\
	-m^2_{12} +\lambda_{345}v_1 v_2 & m^2_{12} \frac{v_1}{v_2} +\lambda_2 v_2^2  &2 \kappa_2 v_2 v_s \\
	2 \kappa_1 v_1 v_s & 2 \kappa_2 v_2 v_s & m^2_\chi
\end{pmatrix}\label{m square of real singlet}
\eeq
where $m^2_\chi=\frac{1}{3} \lambda_s v_s^2$ given Eq.~(\ref{eq: mins}) has been used. In the presence of non-negligible off-diagonal elements $(v_s \neq 0)$ in the above mass matrix, $(\rho_1, \rho_2, \chi)$ are apparently not the mass eigenstates.
We first rotate two doublet components (upper $2 \times 2$ block) into the basis $(\hat{h}, H)$ via an angle $\alpha$. 

\beq
\label{2hdmHiggsstate}
\begin{pmatrix}
	\hat{h} \\ H
\end{pmatrix}
=
\begin{pmatrix}
	\begin{array}{ccc}
		-s_\alpha & c_\alpha \\
		c_\alpha & s_\alpha  \\
	\end{array}
\end{pmatrix}
\begin{pmatrix}
	\rho_1 \\  \rho_2
\end{pmatrix}
\eeq
In fact, they are mass eigenstates in the pure 2HDM but no longer true in the model we consider here due to the doublet-singlet mixture induced by $v_s \neq 0$.
This can be seen explicitly from the full $3 \times 3$ mass matrix under the unitary rotation.
\begin{equation}\label{eq:massmatrix}
\hat{\mathcal{M}}^2=
\begin{pmatrix}
-s_\alpha & c_\alpha & 0 \\
c_\alpha & s_\alpha &0 \\
0 & 0 & 1
\end{pmatrix}
\mathcal{M}^2
\begin{pmatrix}
-s_\alpha & c_\alpha & 0 \\
c_\alpha & s_\alpha &0 \\
0 & 0 & 1
\end{pmatrix} \\
= \begin{pmatrix}
m^2_{\hat{h}} &  0 & \Delta\\
0 & m_H^2 &  D\\
\Delta & D & m^2_\chi
\end{pmatrix}
\end{equation}
where 
\begin{align}
D &= 2v v_s (\kappa_1 c_\alpha c_\beta + \kappa_2 s_\alpha s_\beta) \label{eq:Ddef}\\
\Delta &=2 v v_s (-\kappa_1 s_\alpha c_\beta + \kappa_2 c_\alpha s_\beta) \label{eq:deltadef} 
\end{align}
Clearly, both off-diagonal elements, $\Delta$ and $D$, are not vanishing due to the presence of non-zero $v_s$.

To fit the LHC Higgs data, we expect the SM Higgs $h$ with $m_h=125$~GeV to be nearly pure doublet. This demands the mixing parameter $\Delta$ very small. For simplicity we will take $\Delta=0$ in the following discussion, which then gives us 
\begin{equation}
\label{smallmixing}
\kappa_1= \kappa_2t_\beta/t_\alpha
\end{equation}
Whereas, we allow an arbitrary mixing between $H$ and $\chi$ for our purpose.  
Applying diagonalization between them, we find three resulting mass eigenstates which are formed by:
\beq
\label{eq:eigenstatecomp}
\begin{pmatrix}
	h \\ Y_2\\ Y_1
\end{pmatrix}
=
\begin{pmatrix}
	\begin{array}{ccc}
		1 & 0 & 0  \\
		0 & c_\theta & s_\theta \\
		0 & -s_\theta & c_\theta  \\
	\end{array}
\end{pmatrix}
\begin{pmatrix}
	\hat{h} \\ H \\ \chi
\end{pmatrix}
\eeq
where the states $(\hat{h}, H)$ defined in Eq.~(\ref{2hdmHiggsstate}) are expressed in the $Z_2$ basis.
The mixing angle between $H$ and $\chi$ is given by
\begin{equation}
\label{mixingangle}
s_{2\theta}=\frac{2D}{\sqrt{4D^2 + \left( m^2_\chi -m_H^2\right)^2}}, \quad c_{2\theta} =\frac{ m^2_\chi-m_H^2}{\sqrt{4D^2 + \left( m^2_\chi -m_H^2\right)^2}}
\end{equation}
Alternatively, one can parameterize the mixing angle Eq.~(\ref{mixingangle}) 
as, 
\begin{equation}\label{eq:mixinganglev2}
c_\theta=\sqrt{\frac{m_{Y_2}^2-m^2_\chi}{m_{Y_2}^2-m_{Y_1}^2}},\qquad s_\theta=\mbox{Sign}(\kappa_1 v c_\beta c_\alpha+\kappa_2 v s_\beta s_\alpha)\sqrt{\frac{m^2_\chi -m_{Y_1}^2}{m_{Y_2}^2-m_{Y_1}^2}}
\end{equation}
As discussed in the Introduction, to realize the idea of photon jet we consider the scenario where $Y_1$ is a sub-GeV singlet-like state and $Y_2$ a heavy doublet-like resonance at 750 GeV. 
To this end, we examine the lower $2 \times 2$ block in the mass matrix, Eq.~(\ref{eq:massmatrix}). Since both $m_\chi$ and $D$ are proportional to the singlet VEV $v_s$, the mixing $s_\theta$ is small for most of the parameter space if ${Y_1}$ is required to be very light. (The exception occurs in the region where the potential stability is violated.) This suppression in $s_\theta$ is a fatal weakness for the phenomenology of the di-photon excess in this model as will be shown in Sec.~\ref{subsec:realdiphoton}.

%%%%%%%%%%%%%%%%%%%%%%%%

Finally, we present the masses for three scalar mass eigenstates
\bea
m_{h}^2 & = & m_{\hat{h}}^2 \label{eq:hmass}%+ \mathcal{O}(\delta^2) \label{mheq}
\\
m_{Y_{1,2}}^2 &=& \frac{1}{2}\left[m_H^2+m^2_\chi \pm \sqrt{4D^2 + \left( m^2_\chi -m_H^2\right)^2}  \right] \label{mYeq}
\eea
where the expression of $m_{\hat{h}}^2$ and $m_H^2$ for two CP-even Higgs states in the 2HDM can be found in~\cite{Bernon:2015qea}. $m_{Y_2}$ and $m_{Y_1}$ in Eq.~(\ref{mYeq}) take $+$ and $-$ signs, respectively.

%%%%%%%%%%%%%%%%%%%%%%%%
\subsection{Higgs couplings and decays}
%%%%%%%%%%%%%%%%%%%%%%%

Though the $\chi$ component does not directly couple to any SM fermions, both $Y_1$ and $Y_2$ that arise from the $H-\chi$ mixing couple to SM particles as well as the charged Higgs. Some relevant couplings for the three CP-even mass eigenstates, normalized to SM values, are listed in Table~\ref{table:scalarcoupling}, where the couplings of $h, H$ to the charge Higgs are defined in the 2HDM context as
\bea
g_{hH^\pm H^\mp} &=& -\frac{1}{v} \left\lbrace [m_h^2 + 2(m_{H^\pm}^2-\bar{m}^2)]s_{\beta-\alpha} + 2 \cot2\beta(m_h^2-\bar{m}^2)c_{\beta-\alpha} \right\rbrace \\
g_{HH^\pm H^\mp} &=& -\frac{1}{v} \left\lbrace [m_H^2 + 2(m_{H^\pm}^2-\bar{m}^2)]c_{\beta-\alpha} - 2 \cot2\beta(m_H^2-\bar{m}^2)s_{\beta-\alpha} \right\rbrace \\
g_{\chi H^\pm H^\mp} &=& 2 v_s (\kappa_1 s_\beta^2 + \kappa_2 c_\beta^2) \label{eq:chiHH}
\eea
with $\bar{m}^2 = \frac{2m_{12}^2}{s_{2\beta}}$.

\begin{table}[t]
	\caption{The couplings of scalars normalized to the SM values except for those to the charged Higgs. 
		}
	\vspace*{-5mm}
	\begin{center}
		\begin{tabular}{|c|c|c|c|}
			\hline
			Scalar states& SM gauge bosons & SM fermions & charged Higgs \cr
			\hline
			$h$  &  $s_{\beta-\alpha}$ &  $C^h_f$ & $g_{hH^\pm H^\mp}$ \cr
			$Y_1$  &  $-c_{\beta-\alpha} s_\theta$  & $-C^H_f s_\theta$  & $-s_\theta g_{HH^\pm H^\mp} + c_\theta g_{\chi H^\pm H^\mp} $ \cr
			$Y_2$  &  $c_{\beta-\alpha} c_\theta$  & $C^H_f c_\theta$  & $c_\theta g_{HH^\pm H^\mp} + s_\theta g_{\chi H^\pm H^\mp} $ \cr
			\hline
		\end{tabular}
	\end{center}
	\label{table:scalarcoupling}
\end{table}%

It is noticeable that the coupling of the Higgs $h$ to gauge bosons is proportional to $\sin(\beta-\alpha)$ while that for the other two states $Y_1$ and $Y_2$, after mixing, display the $\cos(\beta-\alpha)$ dependence. 
This implies that most relations in the 2HDM, in particular the alignment limit, still holds in this extended model. To fit the 125~GeV Higgs data, for simplicity we take the alignment limit ($c_{\beta-\alpha} \to 0$), in which the $h$ has SM-like couplings while the $Y_1$ and $Y_2$ decouple with gauge bosons. This reduces  Eq.~(\ref{smallmixing}) to 
\begin{equation}\label{eq:k1k2v2}
\kappa_1 \sim - \kappa_2 \tan^2\beta
\end{equation}
Taken this relation, the coupling $hY_1Y_1$ vanishes at small mixing limit. Hence, the decay of SM Higgs $h$ to new light scalars such as $h\to Y_1Y_1\to 4\gamma$, which is already constrained by the Higgs invisible decay search~\cite{ATLAS:2012soa}, is switched off automatically.

As for Yukawa couplings, they display an overall dependence on $c_\theta$ and $s_\theta$ for the $Y_2$ and $Y_1$ states, respectively. This means the mixing between them is crucial in determining their decays to the SM fermions.
Other than the couplings listed in Table~\ref{table:scalarcoupling}, the $Y_2 Y_1 Y_1$ coupling is most relevant to our discussion, which reads in the alignment limit 
\begin{align}\label{eq:gy2y1y1}
g_{Y_2 Y_1 Y_1} &=3c_{\theta } s_{\theta }^2 g_{HHH}+2 \kappa _2 v c_{\theta } s_{\theta }^2 s_\beta c_{\beta }\left(t^2_\beta+1\right)  -\kappa _2 v s_\beta  c_{\beta } c_{\theta }^3 \left(t_{\beta }^2+1\right)\nonumber\\
&+\kappa _2 v_s  s_{\theta }^3
\left(c_{\beta }^2-s_\beta^2 t_\beta^2\right)+\frac{1}{2} c_{\theta }^2
s_{\theta } v_s \left(-4 \kappa _2 c_{\beta }^2+\lambda_s +4 \kappa _2 s_\beta^2 t_\beta^2\right)
\end{align}
In addition, $Y_2 \to hh$ should have been another important decay mode considering the fact that in the 2HDM context the heavy Higgs $H$ generally has a sizable decay branching ratio into a pair of SM-like Higgs $hh$. However, we examine that this coupling $g_{Y_2 hh}$ is vanishing at the exact alignment limit~\cite{Bernon:2015qea}. Therefore, we do not consider this decay mode in the present analysis. 
It is also necessary to notice that the couplings of the singlet $\chi$ to massless Goldstone bosons are proportional to:
$v_s (\kappa_1 c_\beta^2 + \kappa_2 s_\beta^2)$. As a result, nearly singlet-like $Y_1$ does not couple to the longitudinal modes of $W^\pm/Z$ in the limit $\Delta\to0$.

\subsection{The decay of $Y_1$}
\label{sec:decayY1}
%%%%%%%%%%%%%%%%%%%%%%%%%%%%%%%%%%%%%%%%

\begin{figure} [t]
\centering
\includegraphics[height=0.3\textheight,width=0.48\textwidth]{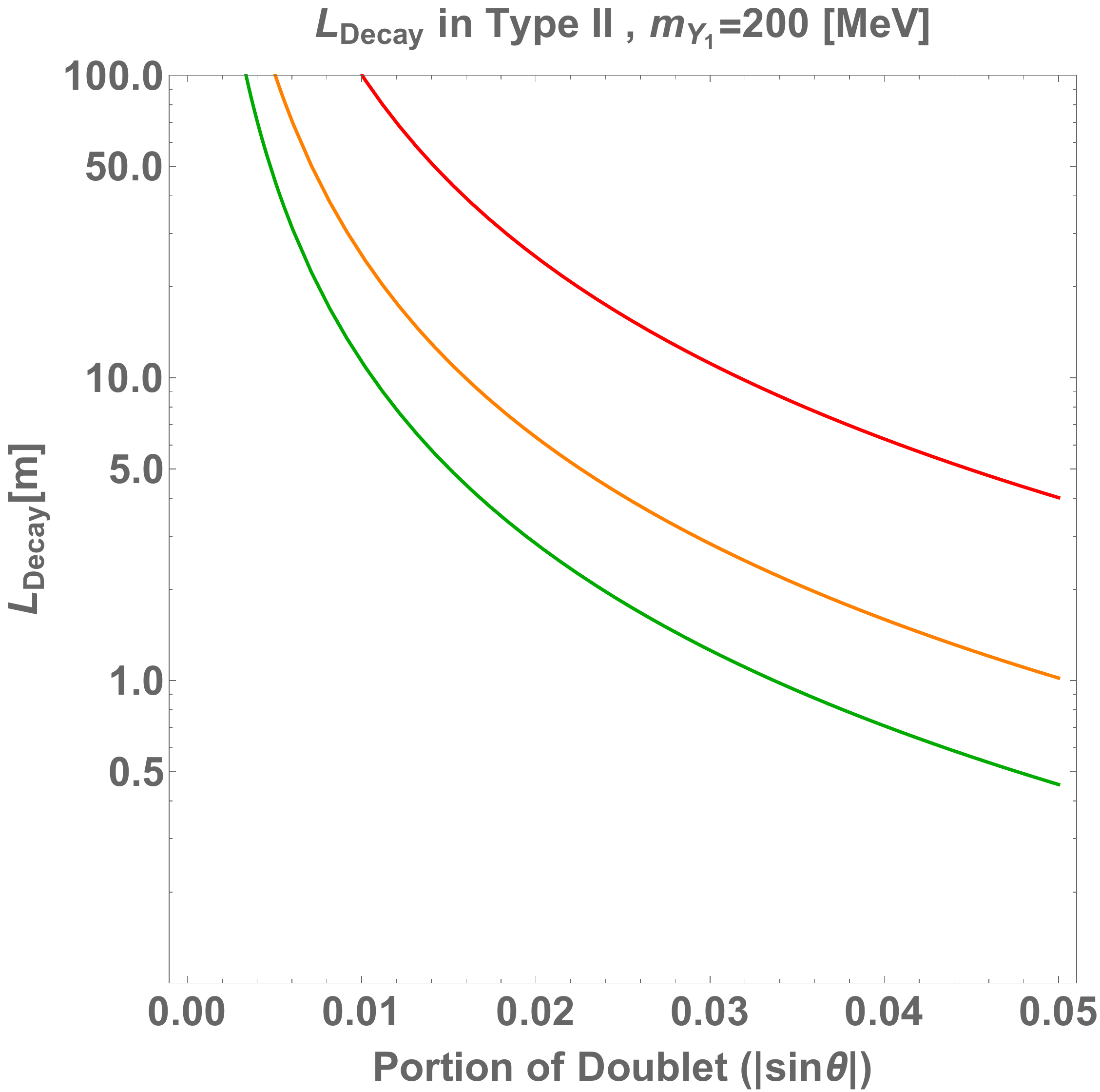}
\caption{Decay length of $Y_1$ 
as a function of the doublet fraction in $Y_1$ for $m_{Y_1}=200$~MeV 
by choosing various values of $\tan\beta$. Red, orange and green curves correspond to $\tan\beta=2, 4, 6$, respectively.
}
\label{fig:doubletfraction}
\end{figure}

We now turn to study $Y_1$ decay. Recall that $Y_1$ is a mixture of CP-even heavy Higgs $H$ and real singlet $\chi$, so it has Yukawa couplings as a doublet (the fraction of which is described by $s_\theta$) and also couplings with scalars present in the Higgs sector. As a result, it can decay to SM light quarks. The presence of these tree level decays greatly increases the total width and in turns shorten the decay length $L_{\decay}$ within the range of $R_{ECAL}$ scale. The influence of mixing $s_\theta$ on the decay length of $Y_1$ is illustrated in Fig.~\ref{fig:doubletfraction} for two choices of $m_{Y_1}$. 
In order to maximally enhance the decay width of $Y_1$, we consider the Yukawa patterns obeying the Type~II model in the following numerical analysis. 

First, as expected, BR$(Y_1\to \gamma\gamma)\lsim 0.1\%$ for $m_{Y_1}=5$ GeV. This is the result of the presence of tree-level decay to light quarks. 
To better understand the $Y_1$ decay, it is useful to analyze the explicit form of $Y_1$ coupling to fermions, $g_{Y_1 ff}$ (c.f.~Table~\ref{table:scalarcoupling}). This coupling is proportional to $s_\theta$, which depends monotonically on $D=-2v v_s \kappa_2(1+ \tan^2\beta)$. Thus, as $s_\theta$, or essentially $\kappa_2$, increases, the total decay width of $Y_1$ grows up. This can be easily understood from the fact that  $Y_1$ acquires more doublet component in the large mixing. The remaining factors $C^H_d\sim t_\beta$,  $C^H_u\sim t_\beta^{-1}$ are specified at the exact alignment limit in the Type~II model. In this case $Y_1 \to d\bar d$ decay is enhanced for $t_\beta>1$. Of course, $Y_1 \to u\bar u$ is simultaneously suppressed but not so efficiently as an offset since $m_d/m_u \simeq 2$. Therefore, it could be expected that the total decay width increases as $t_\beta$ becomes large, which in turn leads to a shorter decay length. All these behaviors are clearly reflected in Fig.~\ref{fig:doubletfraction}.

%%%%%%%%%%%%%%%%%% Need Graphics Below %%%%%%%%%%%%%%%%

%%%%%%%%%%%%%%%%%%%%%%%%%%%%%%%%%%%%%%%%

\subsection{Di-photon cross section}\label{subsec:realdiphoton}
%%%%%%%%%%%%%%%%%%%%%%%%%%%%%%%%%%%%%%%%

According to the preceding discussion, there are five independent parameters in the numerical analysis, which include $m_H$, $v_s$, $\lambda_s$, $\kappa_2$ and $t_\beta$.  However, since $m_{Y_2} = 750 $~GeV$\gg m_{Y_1}$, the mixing effect on the mass of the heavy eigenstate would be very small, allowing us to take $m_H=750$~GeV as a good approximation. While $v_s$ can be determined by using Eq.~(\ref{mYeq}) once the value of $m_{Y_1}$ is chosen. In the end, we are left with three free parameters $\lambda_s$, $\kappa_2$ and $t_\beta$.

To compute the di-photon cross section of our interest, we examine the decay modes of $Y_2$.  As discussed in Sec.~\ref{sec:decayY1}, we know that a moderate mixing (small $s_\theta$) is required to render the decay length of $Y_1$ sufficiently short (also see Fig.~\ref{fig:doubletfraction}). 
In this case, $|c_\theta|\approx\pm 1$ and the leading contribution to $g_{Y_2Y_1Y_1}$ in Eq.~(\ref{eq:gy2y1y1}) reads
\begin{equation}\label{eq:gy2y1y1realv2}
	g_{Y_2Y_1Y_1} = -\text{sign}(s_\theta)vc_\beta s_\beta\kappa_2(1+t_\beta^2)
\end{equation}
It implies that this coupling, or equivalently the decay $Y_2 \to Y_1 Y_1$ tends to be important as either $|\kappa_2|$ or $\tan\beta$ becomes large. However, one cannot achieve both two large because of the constraint from the stability, as you will see shortly.  
In the alignment limit, $Y_2$ decays mainly into $Y_1 Y_1$ and $t\bar t$, the latter one is sensitive to $t_\beta$ as in the 2HDM. In particular, when $t_\beta \lesssim 6$, $Y_2 \to t\bar{t}$ is a predominant decay channel. 

\begin{figure}[t]
\begin{center}
\includegraphics[width=0.5\textwidth]{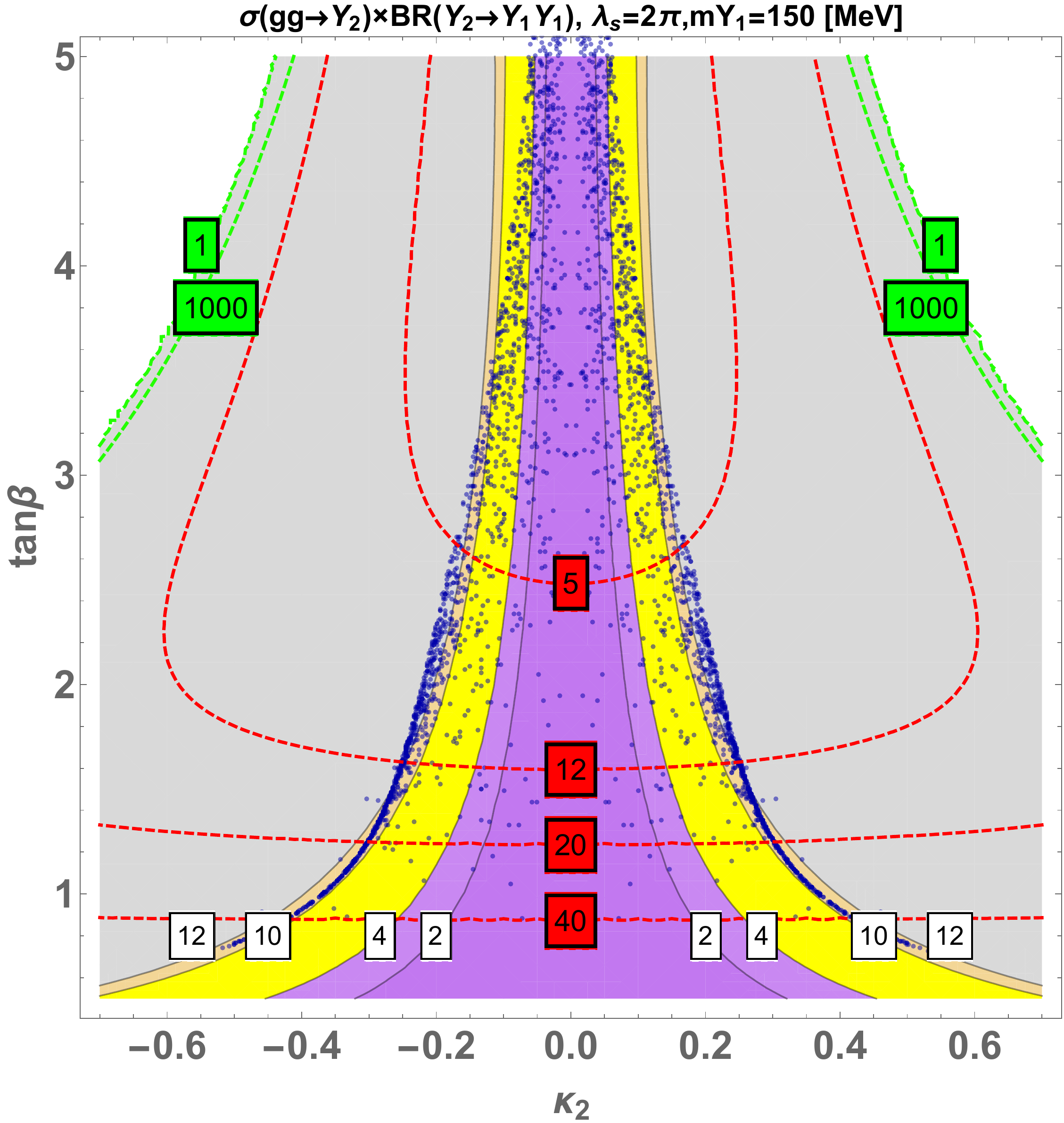}
\caption{The di-photon cross section $\sig(gg\to Y_2) \text{BR}(Y_2 \to Y_1 Y_1)$ (in the unit of fb) produced by the 750 GeV resonance $Y_2$ under the assumption that  $\text{BR}(Y_1 \to \gam\gam)=100\%$, see the contours with white boxes. Only the region covered by blue scattering points is allowed by the potential stability condition for $\lambda_s=2\pi$. The dashed green and red lines show the contours of the proper decay length of $Y_1$ (in the unit of meter) and the total decay width of $Y_2$ (in the unit of GeV), respectively. 
}
\label{fig:realscalar_di-photon}
\end{center}
\vspace{1mm}
\end{figure}

The viability of this scenario is illustrated in Fig.~\ref{fig:realscalar_di-photon}. There, we vary two free parameters $t_\beta$ and $\kappa_2$, and draw the contours of the production cross session $\sigma(gg \to Y_2) \text{BR}(Y_2 \to Y_1 Y_1)$, assuming the $\text{BR}(Y_1 \to \gam\gam)=100\%$ for each $Y_1$ as argued in Sec.~\ref{ssec:singlet}. The decay length of $Y_1$, $L_{\decay}$ and the total width of $Y_2$ are also presented. In this figure, $m_{Y_1}=150$ MeV is adopted as a typical value and $\lambda_s=2\pi$ is chosen to make our model compatible with the unitarity conditions~\cite{Drozd:2014yla}. 

Clearly, there exists parameter space which could yield the 750 GeV di-photon signal, comparable to those observed at the LHC. 
The yellow shaded strip indicates the $\sigma(gg \to Y_2) \text{BR}(Y_2 \to Y_1 Y_1)$ within 4-10 fb.
However, it is not necessarily acceptable because the allowed value of $|\kappa_2|$ for a fixed $\tan\beta$ is constrained by the vacuum stability, which has been shown to play the most important role in eliminating the parameter space~\cite{Drozd:2014yla}. 
In essence, the upper bound on $|\kappa_2|$ is determined by $t_\beta, \lambda_s$ as well as quartic couplings $\lambda_{1-5}$ in the Higgs sector. A simple derivation to obtain this bound can be found in the Appendix.
The allowed region is sketched by blue scattering points in the figure and this band displays a tendency of compression as $\tan\beta$ increases. 
Though it is still possible to find a value of $\kappa_2$ for $\tan\beta \lesssim 4$ producing the desired signal rate which can fit the data, 
the decay length of $Y_1$ for such value of $\kappa_2$ within the stability bound is incredibly long due to the insufficient mixing. 
Therefore, we conclude that this model containing a real singlet is difficult to simultaneously yield the di-photon signal comparable to the observed data and also achieve a reasonable proper decay length.

%%%%%%%%%%%%%%%%%%%%%%%%%%%%%%%%%%%%%%%%
\section{Model II: complex singlet embedding a pseudoscalar dark matter}
\label{sec:complexmodel}
%%%%%%%%%%%%%%%%%%%%%%%%%%%%%%%%%%%%%%%%

To remedy the problem of too long decay length present in the real singlet model, we are now introducing a {\it complex} scalar gauge singlet field $\mathbb{S}$ to the 2HDM in this section. The $\mathbb{S}$-associated part is  given by
\beq
\label{complexL}
    \mathcal{L}_{\mathbb{S}} = \onehalf m_0^2 \mathbb{S}^2 +\frac{1}{4!}\lambda_s \mathbb{S}^4+\kappa_1 \Phi^\dagger_1 \Phi_1 \ \mathbb{S}^\dagger \mathbb{S} + \kappa_2 \Phi^\dagger_2 \Phi_2 \ \mathbb{S}^\dagger \mathbb{S}+\omega_1 \Phi^\dagger_1 \Phi_1 (\mathbb{S}^\dagger+\mathbb{S})+\omega_2 \Phi^\dagger_2 \Phi_2 (\mathbb{S}^\dagger+\mathbb{S})
\eeq
 Here $\omega_1$ and $\omega_2$ are dimensionful parameters.~\footnote{We confine our analysis to the CP conserving model where the CP phases of these two interaction terms are taken zero, resulting in real $\omega_1$ and $\omega_2$.} Unlike the real singlet model discussed in Sec.~\ref{sec:realmodel}, we do not impose the $Z'_2$ on the complex filed $\mathbb{S}$ so that the linear terms of $\mathbb{S}$ are present in the above Lagrangian. The singlet scalar can be expanded as
\begin{equation}
\mathbb{S}=\chi_S+i\chi_A
\end{equation}
We stress that the complex singlet field $\mathbb{S}$ cannot acquire a VEV: $\langle \mathbb{S} \rangle =0 $,  otherwise the CP-odd mode $\chi_A$ would be a massless Goldstone boson. 
Consequently, the minimization conditions in the scalar potential takes the same form as Eqs.~(\ref{eq: minh1}) and (\ref{eq: minh2}) with the elimination of the $v_s$ term, while the one with respect to $v_s$ ({\it i.e.} Eq.~(\ref{eq: mins})) is absent. 

After EWSB, this complex singlet model has an additional CP-odd state $\chi_A$, that could be a candidate for dark matter.
However, among the three CP-even states there are many similarities between Model~I and Model~II considered in this paper even if the $\mathbb{S}$ in Model~II does not acquire a vev. It is useful to comment that the single $\chi_S$ interactions with two doublets appearing in Eq.~(\ref{complexL}) are also present in Model~I when the real field $S$ gets vev. This implies that the coupling $g_{\chi_S H^\pm H^\mp}$ can be obtained from Eq.~(\ref{eq:chiHH}) with a simple substitution of $\kappa_a v_s$ by $\omega_a$. This replacement is also true in the $(\rho_1,\rho_2,\chi_S)$ mixing. 
To avoid redundancy, we shall transmit the results which have been derived in the real singlet model to the complex singlet model. 
A particularly important exception is the $Y_2 Y_1 Y_1$ coupling. In the present model, this coupling depends on both $\omega_a$ and $\kappa_a$ $(a=1,2)$. As a result. such an interplay provides us the possibility of simultaneously ensuring the potential stability which is crucially determined by $\kappa_a$ and achieving a desired doublet-singlet compound for the light scalar through an essential increment on the $\omega_2$. In the end, the difficulty of too-long decay length can be overcome.

\subsection{Spectrum and couplings}

Without the $Z'_2$ protection, $\mathbb{S}$ can singly couple to the doublets. This results in the doublet-singlet mixing which occurs only when the real singlet gets vev as discussed in the previous section. 
As already argued, the mass matrix in the $(\rho_1, \rho_2, \chi_S)$ basis can be obtained from Eq.~(\ref{m square of real singlet}) by replacing $k_a v_s$ with $\omega_a$
and $m_\chi^2=m_0^2+\kappa_1 v_1^2+\kappa_2 v_2^2$ in this model. By virtue of this similarity, one can simply follow the procedures described in Sec.~\ref{sec:realmodel} to derive the mixing parameters $\Delta$, $D$, the composition of the resulting mass eigenstates $h, Y_1, Y_2$ and their mass spectrum. The corresponding results are analogous to Eqs.~(\ref{eq:Ddef}), (\ref{eq:deltadef}), Eq.~(\ref{eq:eigenstatecomp}) and Eqs.~(\ref{eq:hmass}), (\ref{mYeq}) with the only substitution of $k_a v_s \to \omega_a$. This implies that in this scenario the mixing between CP-even fields as described by $s_\theta$ depends on $\omega_2$ as opposed to $\kappa_2$.

In accordance with the LHC data for 125~GeV Higgs, we also employ the alignment limit and require the zero singlet fraction for the $h$ as discussed in Sec.~\ref{sec:realmodel}. In this context, $D=-2 v\omega_2 t_\beta$ and $\Delta=0$ because $\omega_1=-\omega_2 t_\beta^2$. 
In contrast to the CP-even fields, the mixing in the CP-odd sector is absent in this scenario and hence $\chi_A$ itself is a mass eigenstate with mass $m_{\chi_A}=m_\chi$. Once it is light enough,  both the SM Higgs $h$ and the 750~GeV resonance state $Y_2$ in the model could decay into a pair of them. The $h$ to $\chi_A\chi_A$ decay, if kinematically allowed, can be switched off by the vanishing relevant coupling $g_{h \chi_A\chi_A}$. This actually gives rise to Eq.~(\ref{eq:k1k2v2}), $\kappa_1=-\kappa_2 t_\beta^2$, under which the couplings given in Table~\ref{table:scalarcoupling} are maintained except the substitution of $g_{\chi H^+ H^-}$ by
\begin{equation}
g_{\chi_S H^+ H^-} =2\omega_1 s_\beta^2+2 \omega_2 c_\beta^2=2\omega_2(1-t_\beta^2)
\end{equation}
In addition, similar to the real singlet case discussed in Sec.~\ref{sec:realmodel}, the decay channel $h\to Y_1 Y_1$ is also closed by Eq.~(\ref{eq:k1k2v2}) in this scenario. 
%%%%%%%%%%%%%%%%%%%%%%%%%%%%%%%%%%%%%%%%
\subsection{$Y_2$ decay}
\label{sec:compY2decay}

\begin{figure} [t]
	\centering
	\includegraphics[height=0.3\textheight,width=0.32\textwidth]{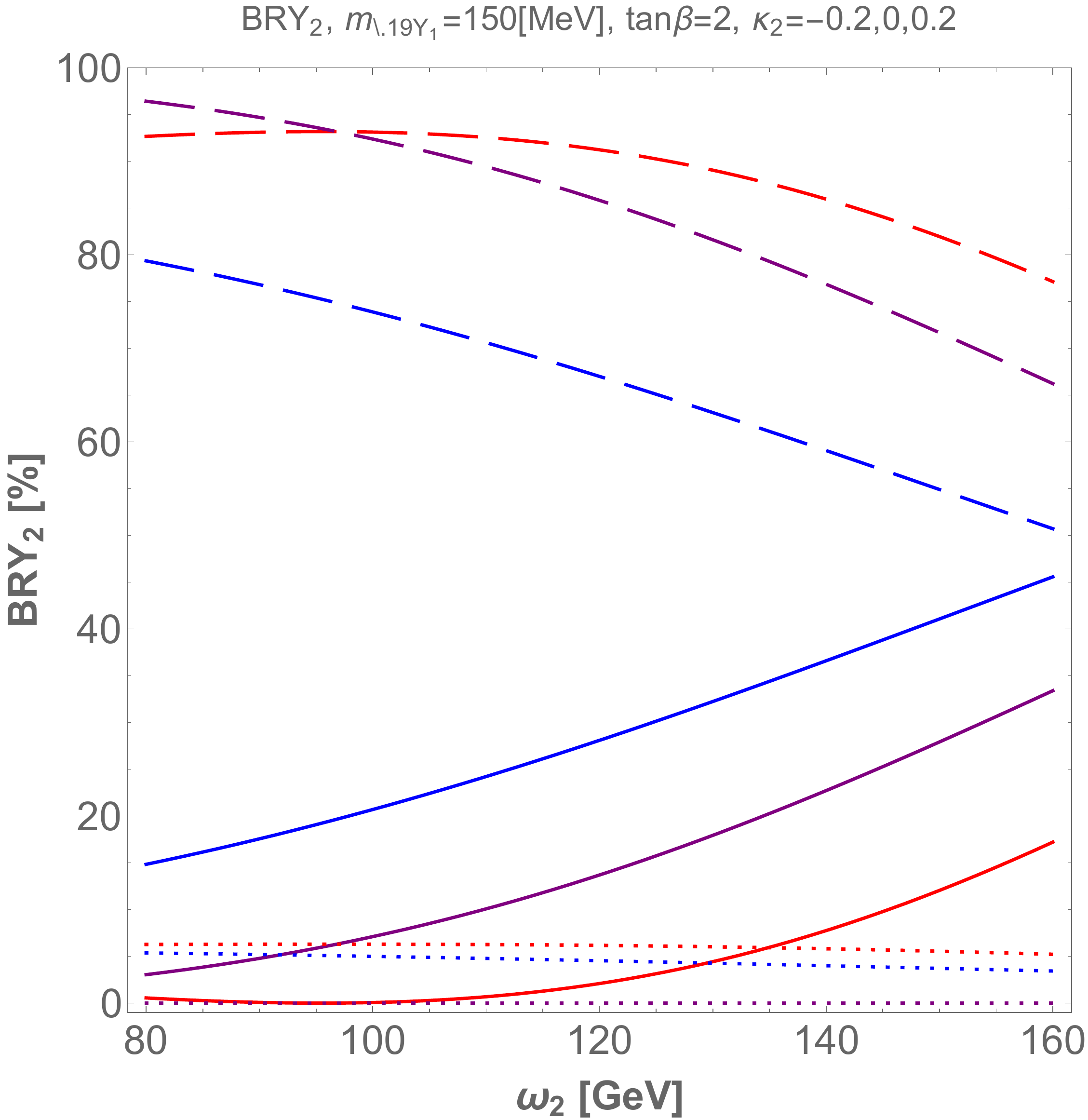}
	\includegraphics[height=0.3\textheight,width=0.32\textwidth]{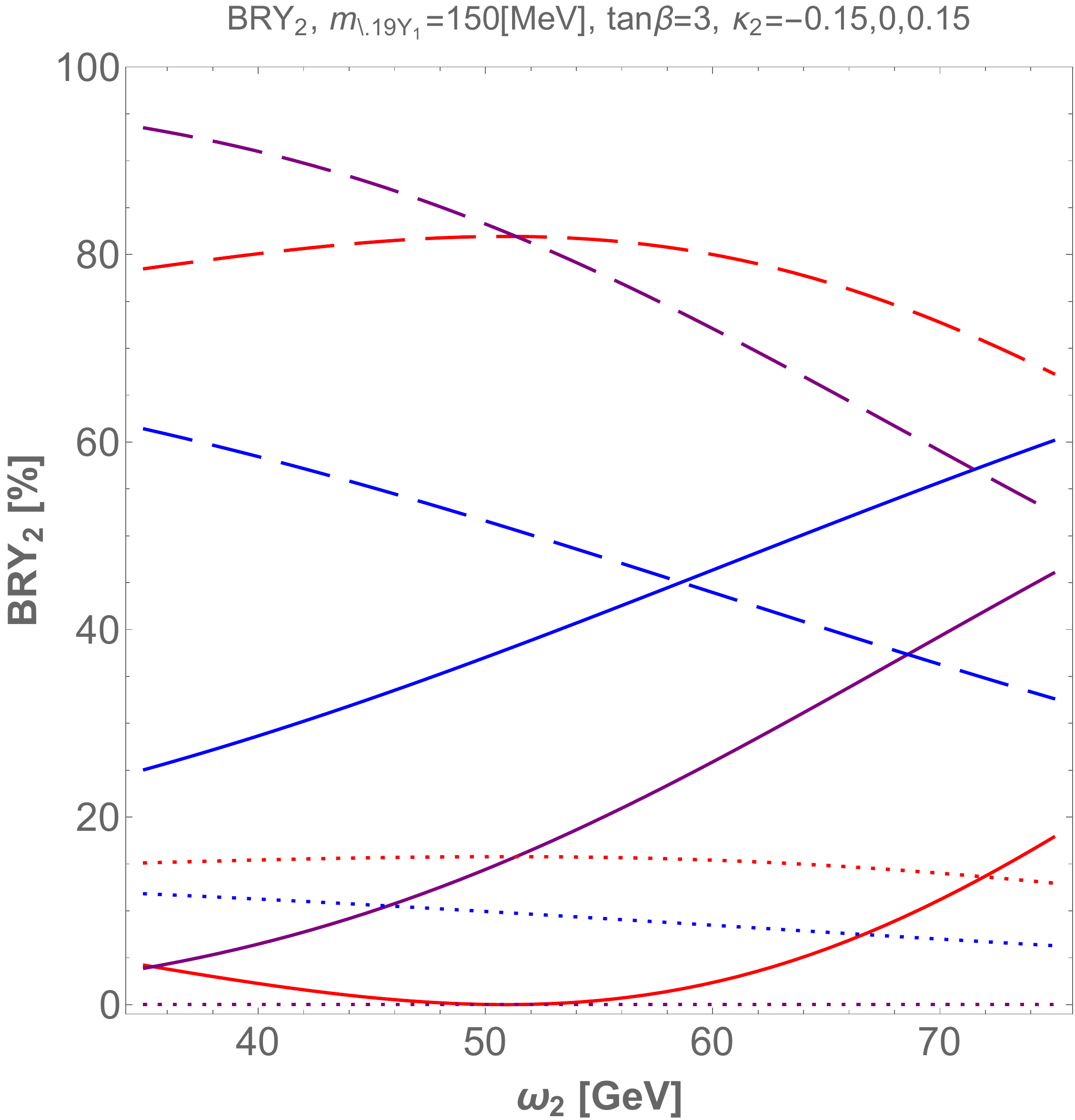}
	\includegraphics[height=0.3\textheight,width=0.32\textwidth]{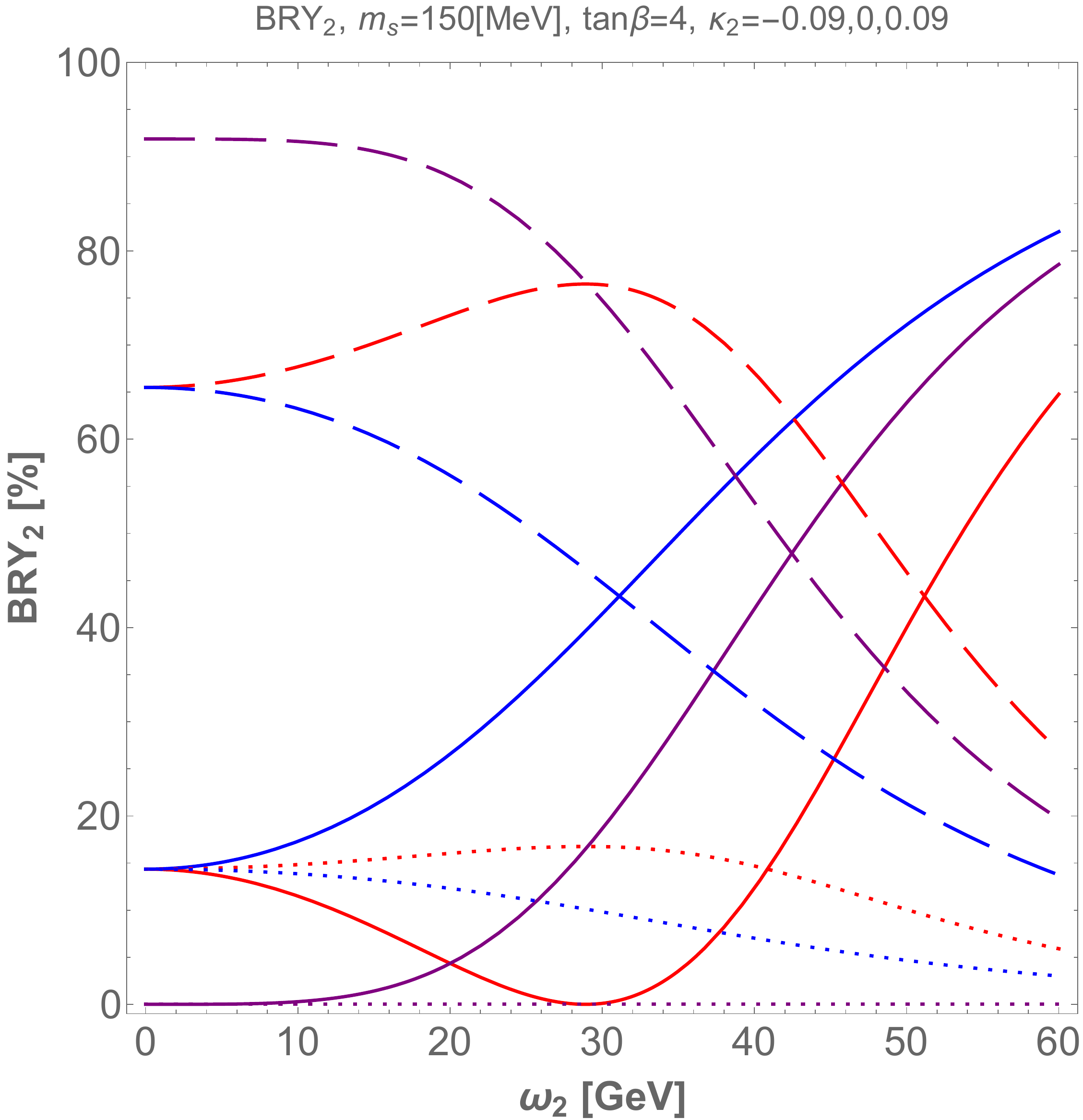}
	\caption{The $\omega_2$ dependence of BR($Y_2\to Y_1 Y_1$) (solid), BR($Y_2\to t\bar t$) (dashed)  and BR($Y_2\to \chi_A\chi_A$) (dotted) are indicated. 
In each graph, the branching ratio curves are drawn corresponding to $\kappa_2=0$ (purple) and the minimal/maximal (blue/red) value of $\kappa_2$ (which are given on top of each graph) such that the stability condition is obeyed for each $\tan\beta$.
	}
	\label{fig:Y2Br}
\end{figure}

In addition to the main decay channels (to $Y_1 Y_1$ and $t \bar t$), $Y_2$ in this model can invisibly decays into $\chi_A\chi_A$. Of the three relevant couplings,   $g_{Y_2 t t}$ is given in Table~\ref{table:scalarcoupling}, while the other two can be easily obtained at the exact alignment limit $c_{\beta-\alpha}=0$.
\begin{align}
g_{Y_2 Y_1 Y_1} 
%&=3 c_\theta^2 s_\theta g_{HHH}+c_\alpha c_\beta s_\theta(s_\theta^2-2c_\theta^2)v \kappa_1+s_\alpha s_\beta s_\theta(s_\theta^2-2c_\theta^2)v \kappa_2\nonumber\\
%& +c_\alpha^2 c_\theta(c_\theta^2-2s_\theta^2)\omega_1+s_\alpha^2 c_\theta(c_\theta^2-2s_\theta^2)\omega_2\nonumber\\
&=3 c_\theta s_\theta^2 g_{HHH}-\kappa_2 t_\beta (c_\theta^2-2s_\theta^2)c_\theta v+s_\theta(s_\theta^2-2c_\theta^2)(1-t_\beta^2)\omega_2\label{eq:gy211}\\
g_{Y_2 \chi_A\chi_A} &= (\kappa_1-\kappa_2)c_\beta s_\beta c_\theta v=-\kappa_2 t_\beta c_\theta v
\end{align} 
It should be noted that the heavy Higgs $H$ in the 2HDM context generally has a sizable decay branching ratio into a pair of SM-like Higgs $h$. However, we examine that this coupling $g_{Y_2 hh}$ is vanishing at the exact alignment limit~\cite{Bernon:2015qea}. Therefore, we do not consider this decay mode in the present analysis. 
For illustration, we display the branching ratios of $Y_2$ decay in Fig.~\ref{fig:Y2Br} by taking $\tan\beta=2,3,4$ (from left to right) in the range that could yield an observed di-photon cross session as we will see. In each graph, branching ratio curves are drawn in different colors corresponding to $\kappa_2=0$ and the maximal value of $|\kappa_2|$ such that the stability condition is obeyed for each $\tan\beta$. As seen from Fig.~\ref{fig:Y2Br}, in the small mixing case when $\omega_2$ is small, $t\bar t$ channel dominates the decay of $Y_2$ for small $\tan\beta$ as long as $\kappa_2$ 
stays within the stability bound, whereas $BR(Y_2\to Y_1Y_1)$ and $BR(Y_2\to \chi_A \chi_A)$ are not substantial. 
Reversely, increasing $\omega_2$ will invoke a larger mixing, which leads to the reduction of the doublet fraction in the $Y_2$, while $Y_1$ gains more doublet component. This results in a quick grow on the coupling $g_{Y_2 Y_1 Y_1}$, but little change on the coupling $g_{Y_2 t t}$. Consequently, BR$(Y_2 \to t\bar t)$ drops drastically while BR($Y_2\to Y_1Y_1$) becomes important as the mixing increases.
Thus, large $\omega_2$ is favored in order to accomplish a sufficiently large BR($Y_2\to Y_1 Y_1$) for our purpose.
In the plots we also observe that BR($Y_2\to Y_1Y_1$) vanishes at a certain value of $\omega_2$ when $\kappa_2<0$.
This is the consequence of the interplay between $\kappa_2$ and $\omega_2$ terms in the coupling $g_{Y_2 Y_1 Y_1}$ given in Eq.~(\ref{eq:gy211}).
Besides, the particularly presented decay $Y_2$ into $\chi_A \chi_A$ is always sub-dominant. This can be attributed to the smallness of $\kappa_2$ as demanded by the stability.

\subsection{Phenomenology of 750~GeV state}

\begin{figure}[t]
\begin{center}
\includegraphics[height=0.3\textheight,width=0.48\textwidth]{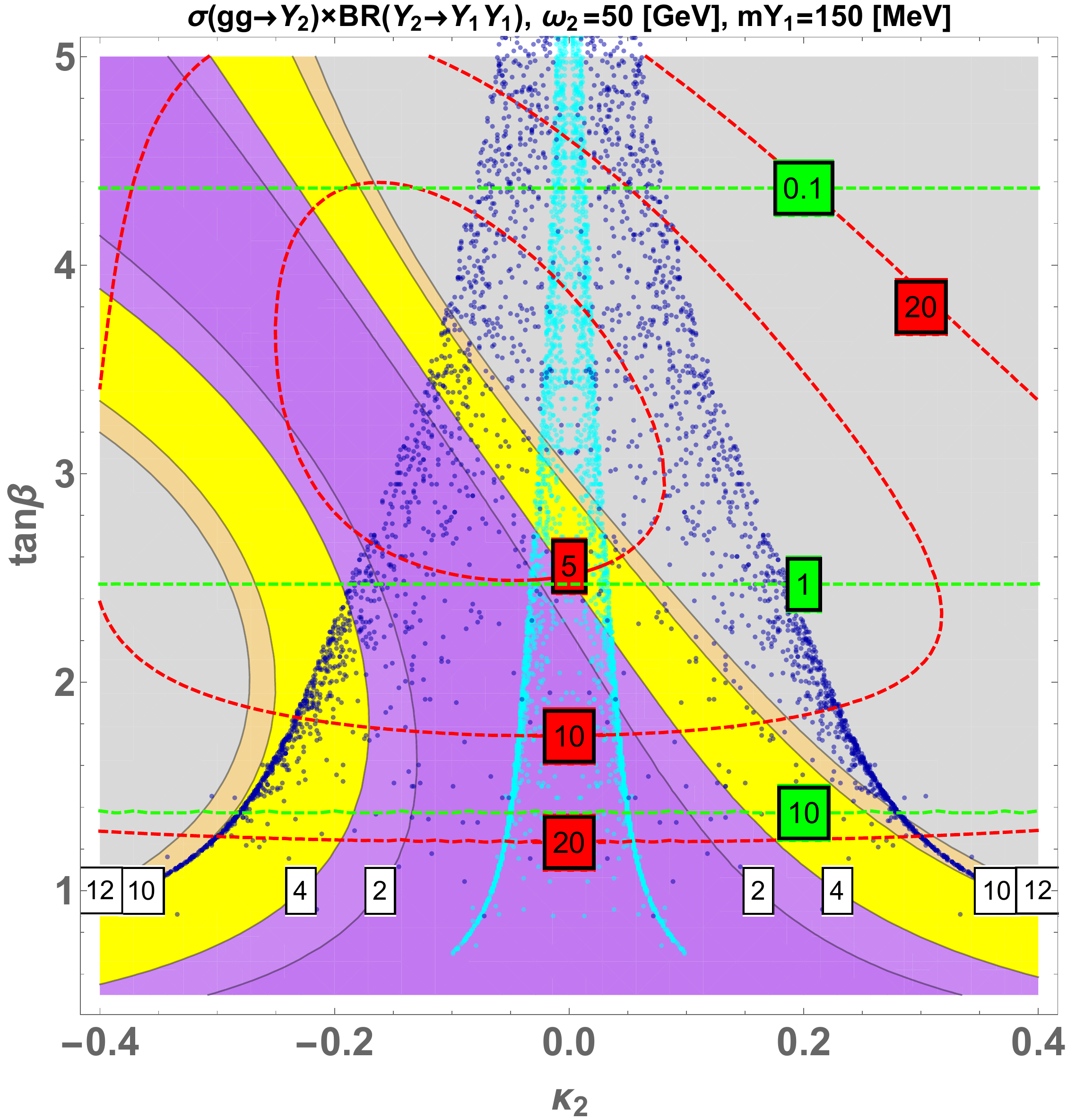}
\hspace{1mm}
\includegraphics[height=0.3\textheight,width=0.48\textwidth]{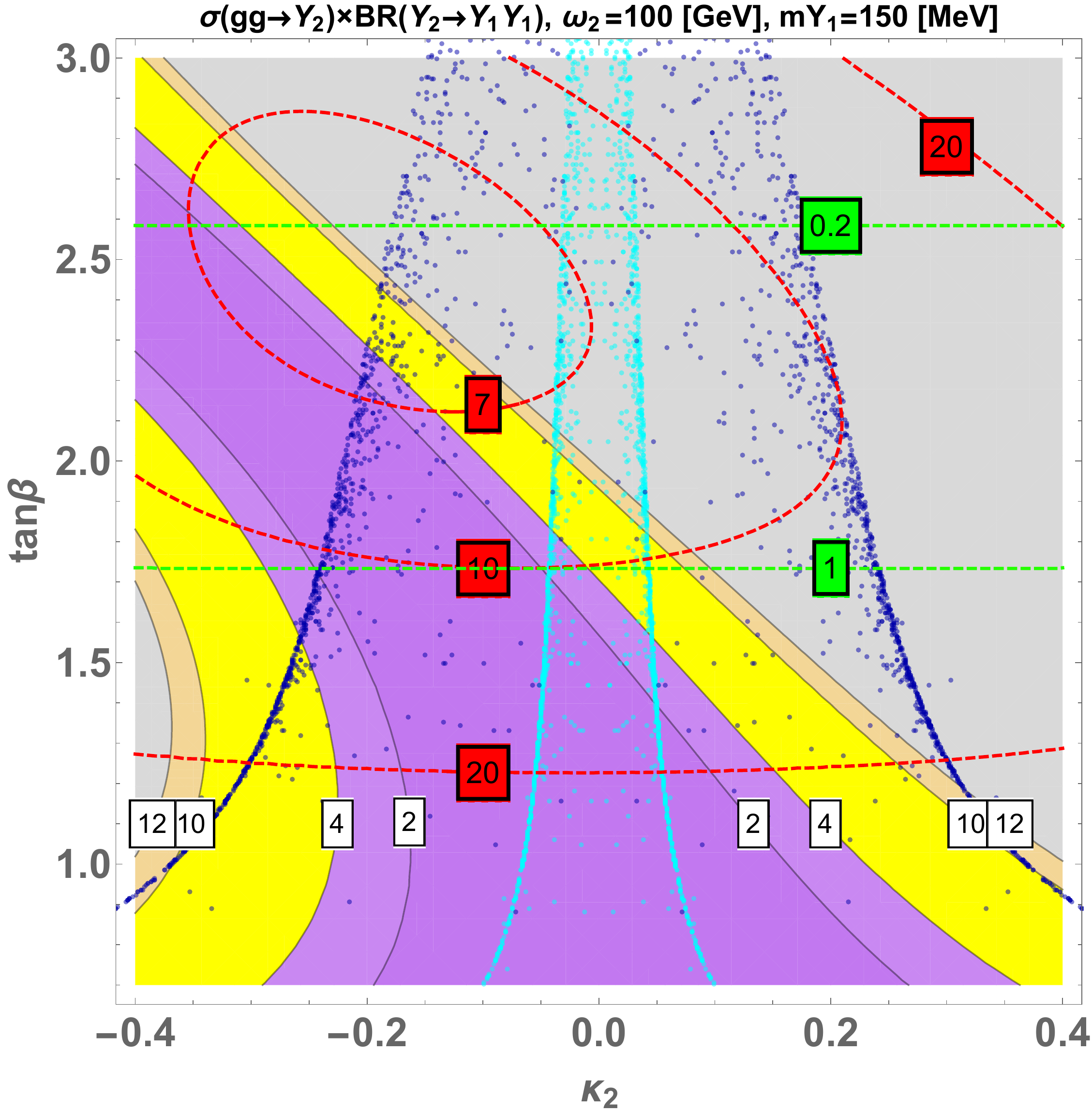}
\includegraphics[height=0.3\textheight,width=0.48\textwidth]{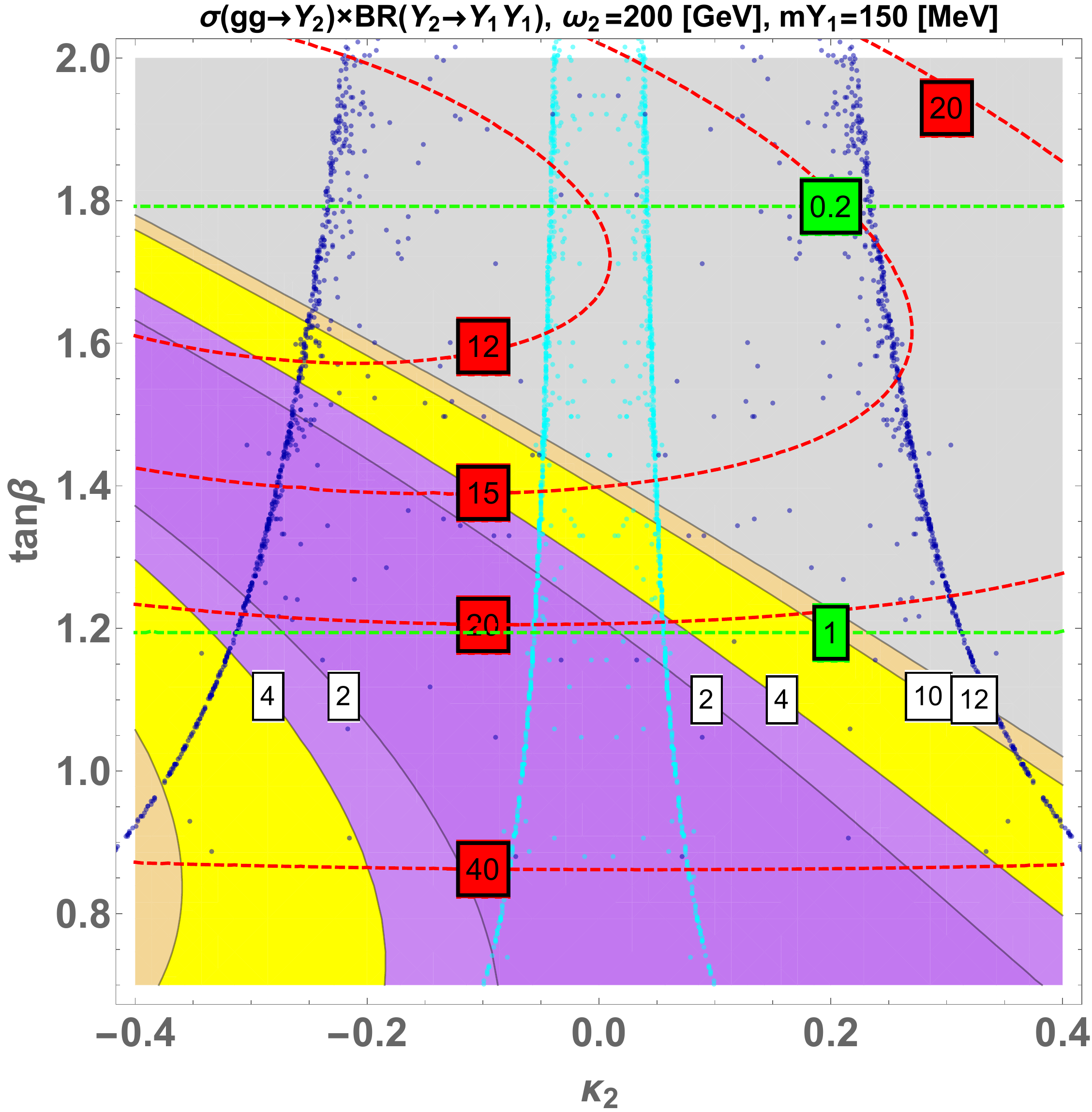}
\hspace{1mm}
\includegraphics[height=0.3\textheight,width=0.48\textwidth]{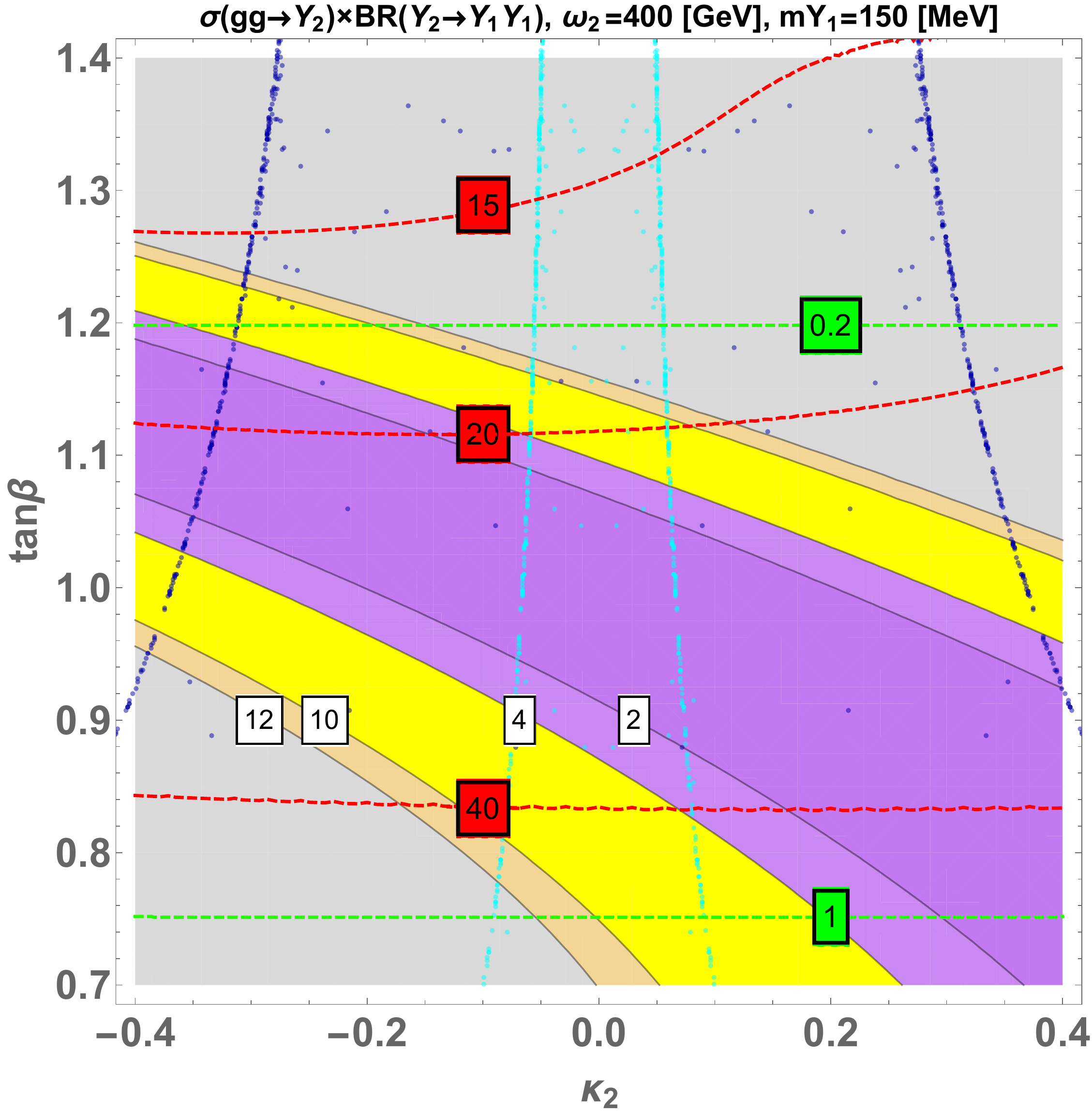}
\caption{The di-photon cross section $\sig(gg\to Y_2) \text{BR}(Y_2 \to Y_1 Y_1)$ produced by the 750 GeV resonance $Y_2$ under the assumption that $\text{BR}(Y_1 \to \gam\gam)=100\%$, see the contours with white box. $\omega_2$ is chosen different values specified on the top of each graph. Only the region which is covered by blue (cyan) scattering points for $\lambda_s=2\pi$ $(0.2)$ is allowed by the potential stability condition. The dashed lines with green and red box label the contours of the proper decay length of the light singlet $Y_1$ and the total decay width of $Y_2$, respectively. }
\label{fig:di-photon_o2}
\end{center}
\vspace{1mm}
\end{figure}

We begin our discussion for 750~GeV state by presenting in Fig.~\ref{fig:di-photon_o2} the contours of  cross section $\sigma(gg\to Y_2\to Y_1Y_1)$. For illustration, we take $m_{Y_1}=150$~MeV and $\lambda_s=2\pi$~\footnote{Here we adopt the somewhat more conservative value of  $\lambda_S=2\pi$ so that our model would remain valid to at least a moderately higher scale before additional new physics would need to be included to obtain a theory valid at all energy scales. The impact of $\lambda_S$ value on the stability was discussed in Ref.~\cite{Drozd:2014yla} in detail.}. 
The four graphs are produced by choosing $\omega_2=50, 100, 200, 400$~GeV in sequence. The yellow shaded band corresponds to $\sigma(gg\to Y_2\to Y_1Y_1)$ within 4-10 fb that could fit the ALTAS+CMS data.
In the figure, the decay length of $Y_1$ and the total decay width of $Y_2$ are also shown in green and red dashed lines, respectively. The contour numbers are uniformly indicated in the corresponding colored boxes. 
In addition, we examine the stability condition of this model and mark the allowed region which is covered by blue scattering points. The trapezoid shape indicates that a smaller value of $\tan\beta$ is able to accommodate a larger $|\kappa_2|$.
This can be understood from the fact that $\kappa_1=-\kappa_2 \tan\beta$ is employed and the stability condition essentially places upper bounds on $|\kappa_1|$ and $|\kappa_2|$ as well as their ratio~\cite{Drozd:2014yla}.

The presence of blue scattering points in the yellow shaded band tells us that 
this complex singlet model could easily yield the observed cross section $\sigma(gg\to Y_2\to Y_1 Y_1)$, while obeying the stability condition. Recall that BR$(Y_1\to \gamma\gamma)=100\%$ is assumed in the estimate. Next, we examine the decay length of $Y_1$.  
This figure shows the decay length of $Y_1$ (green dashed line) has no dependence on $\kappa_2$ but is very sensitive to $\tan\beta$ and $\omega_2$. 
Particularly, the decay length of $Y_1$ increases as $\tan\beta$ goes down. Both phenomena are analogous to the real singlet model presented in Sec.~\ref{sec:decayY1}.
Apparently, the requirement that the decay length  $L_{\decay}\lesssim 1$~m has significant impact on eliminating the small $\tan\beta$ region.
Another important measurement that characterizes the potential 750~GeV di-photon resonance is its total width (red dashed lines), which varies from a few to tens of GeV seen from Fig.~\ref{fig:di-photon_o2}. Thus, this could be used as a critical signature in examining this scenario or determining the model parameter if confirmed.
Two interesting observations regarding the cross section are placed in order. First, the cross section contours display an asymmetry with respect to $\kappa_2=0$. This is actually a result of Eq.~(\ref{eq:gy211}).  
Second, the magnitude of cross section becomes less sensitive to $\kappa_2$ as $\omega_2$ increases, as seen from the fact that the contours keep tilting to the left. 

\begin{figure}[t]
\begin{center}
\includegraphics[height=0.3\textheight,width=0.48\textwidth]{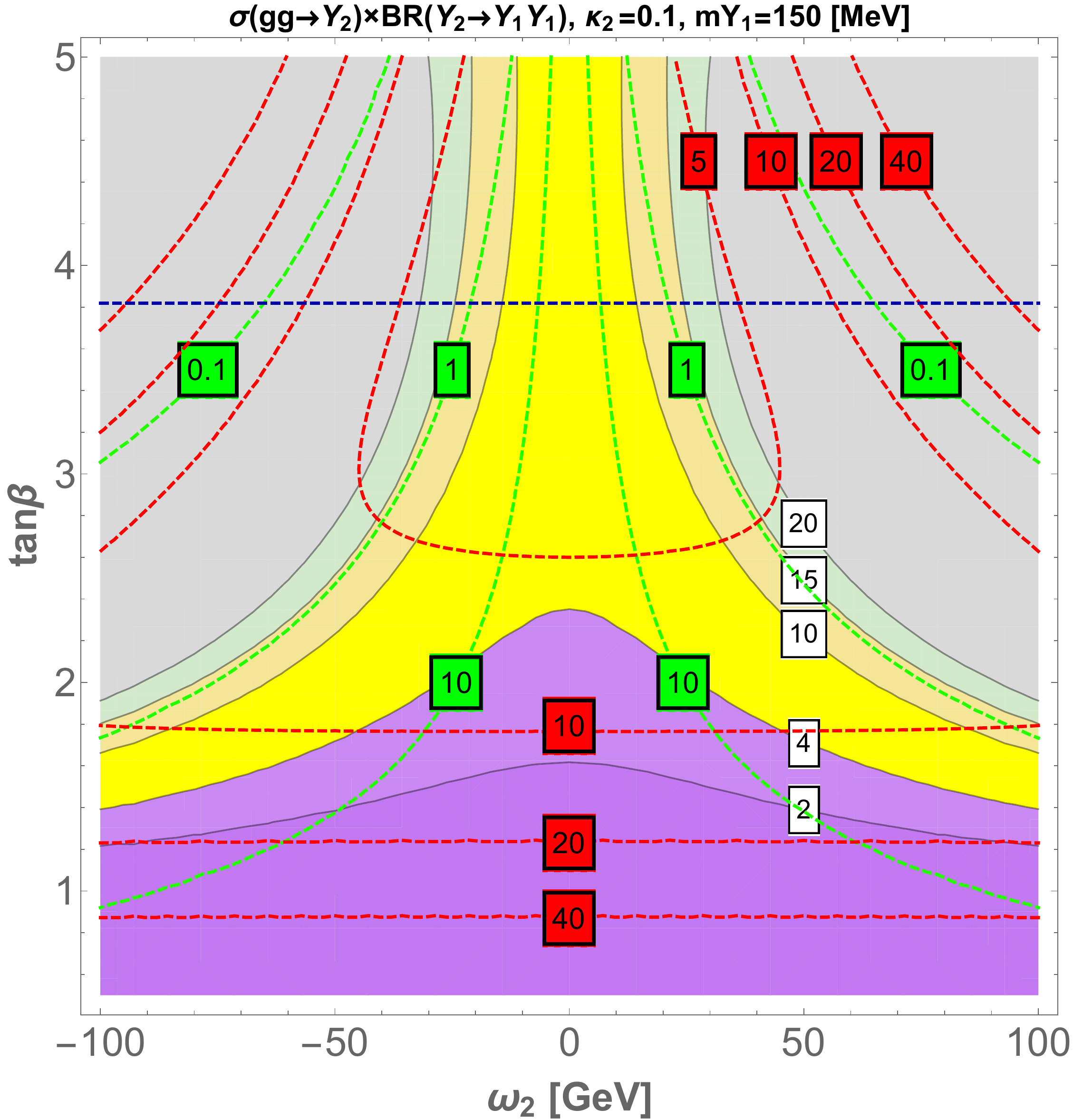}
\includegraphics[height=0.3\textheight,width=0.48\textwidth]{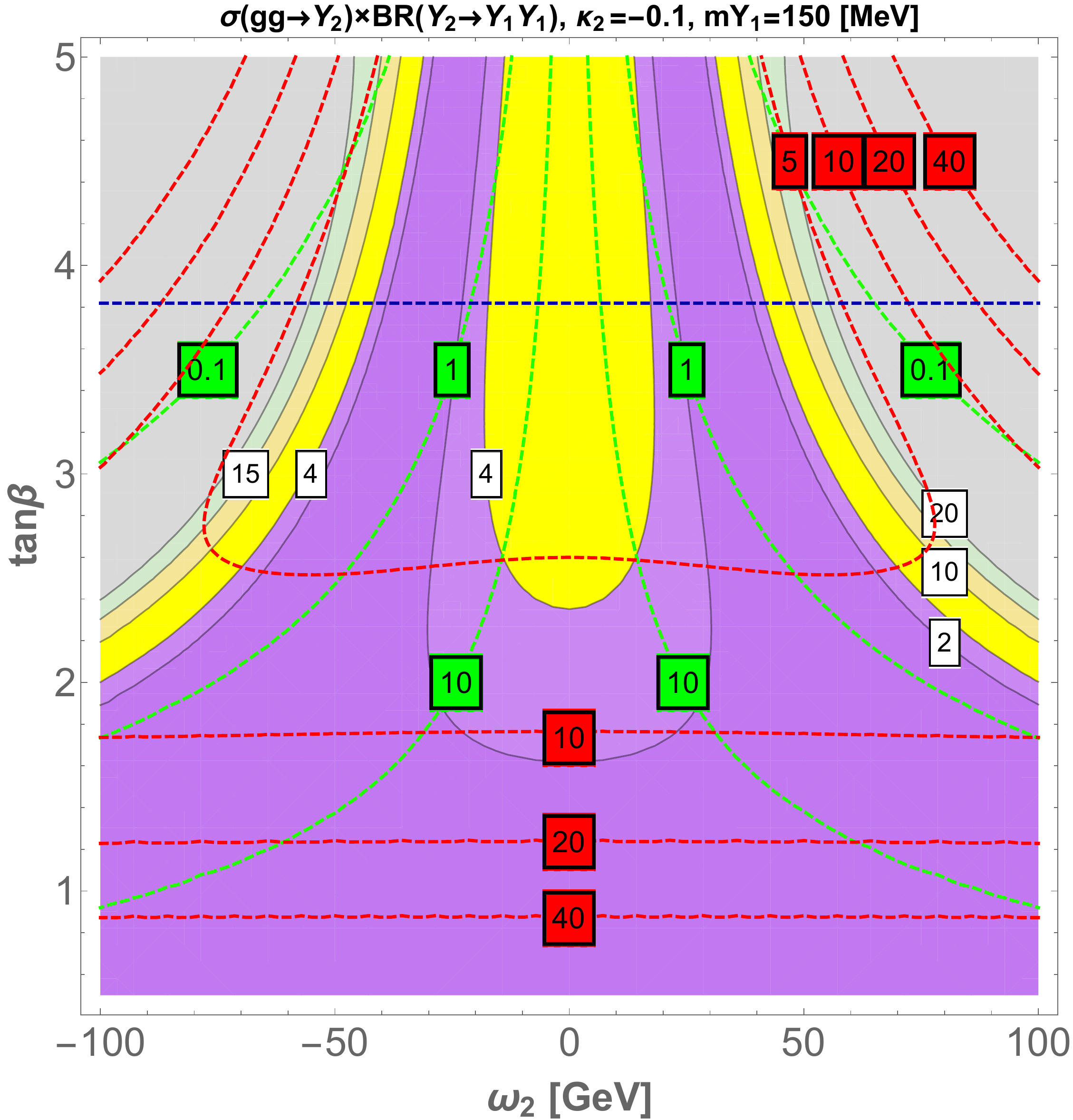}\\
\includegraphics[height=0.3\textheight,width=0.48\textwidth]{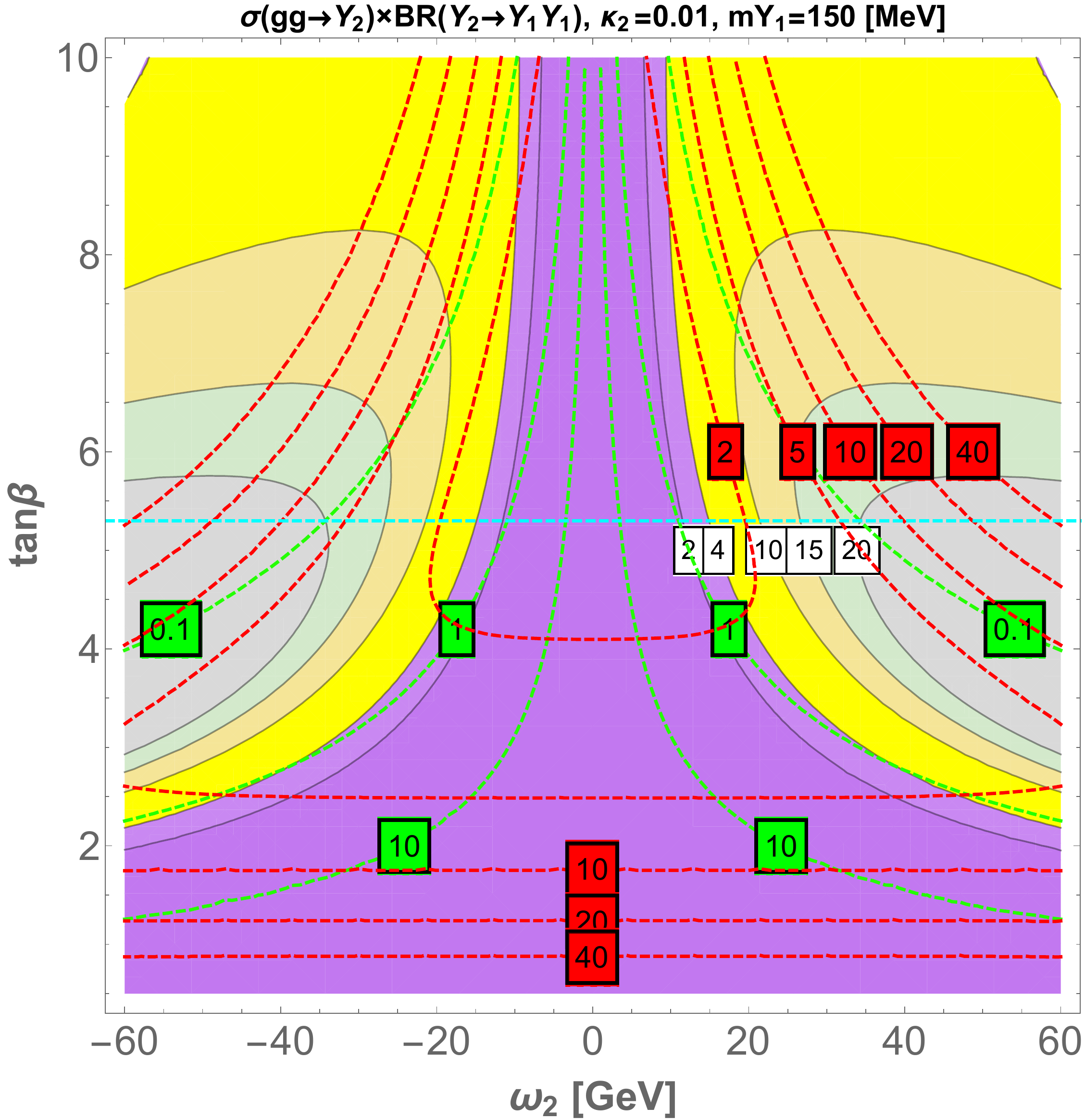}
\caption{The di-photon cross section $\sig(gg\to Y_2) \text{BR}(Y_2 \to Y_1 Y_1)$ in the $\omega_2 - \tan\beta$ plane with $\kappa_2= 0.1, -0.1, 0.01$, under the assumption that $\text{BR}(Y_1 \to \gam\gam)=100\%$. The legend is the same as in Fig.~\ref{fig:di-photon_o2} except the blue/cyan horizontal dashed line which gives the approximate upper bound on $\tan\beta$ coming from the stability. }
\label{fig:di-photon_k2}
\end{center}
\vspace{1mm}
\end{figure}

Alternatively, the result can be projected onto $\kappa_2 - \omega_2$ plane. We show two examples of $\kappa_2=\pm 0.1$ in Fig.~\ref{fig:di-photon_k2}.  
One can easily gain the additional information from this figure regarding the $\omega_2$ dependence. 
As $|\omega_2|$ goes large, the decay width of $Y_1$ increases because $Y_1$ is composed of more doublet fraction. This eventually results in a shorter decay length. 
On the other hand, for $\tan\beta \leq 2$, the total width of $Y_2$ is marginally sensitive to $\omega_2$. This is because $t\bar t$ channel dominates the decay of $Y_2$ in the small $\tan\beta$ region as seen in Fig.~\ref{fig:Y2Br}. 
Finally, it is important to mention that the minimal width of $Y_2$ is $\gtrsim 1~\text{GeV}$ in this scenario. This is in contrast to many models where the width of 750 GeV state is of sub-GeV scale.

%%%%%%%%%%%%%%%%%%%%%%%%%%%%%%%%%%%%%%%%%

%%%%%%%%%%%%%%%%%%%%%%%%%%%%%%%%%%%%%%%%%
%%%%%%%%%%%%%%%%%%%%%%%%%%%%%%%%%%%%%%%%%%%%
%

\section {Discovery prospects at the collider}
\label{sec:prospects}
%%%%%%%%%%%%%%%%%%%%%%%%%%%%%%%%%%%%%%%%%%%%
\subsection{Searching for $Y_2 (750)$ in other channels}
Aside from into the $Y_1Y_1$, the $Y_2$ state of 750~GeV largely decays into $f\bar{f}$ final state. This has been shown in Sec.~\ref{sec:compY2decay}. The cross section of producing a $Y_2$ decaying to the $t\bar t$ or $b\bar b$ final states could be as large as hundreds of fb. Hence, they could be possible channels to search for $Y_2$ at the future run of the LHC. Compared to the tree-level $f\bar f$ decays, the branching ratio of the loop-induced decay $Y_2\to gg$ is negligible. This channel also suffers from a large QCD background at the LHC. Therefore, it is less likely to be a promising discovery channel.

\subsection{Pseudo-scalar Higgs and charged Higgs}

\begin{figure}[t]
\centering
\includegraphics[width=0.45\textwidth]{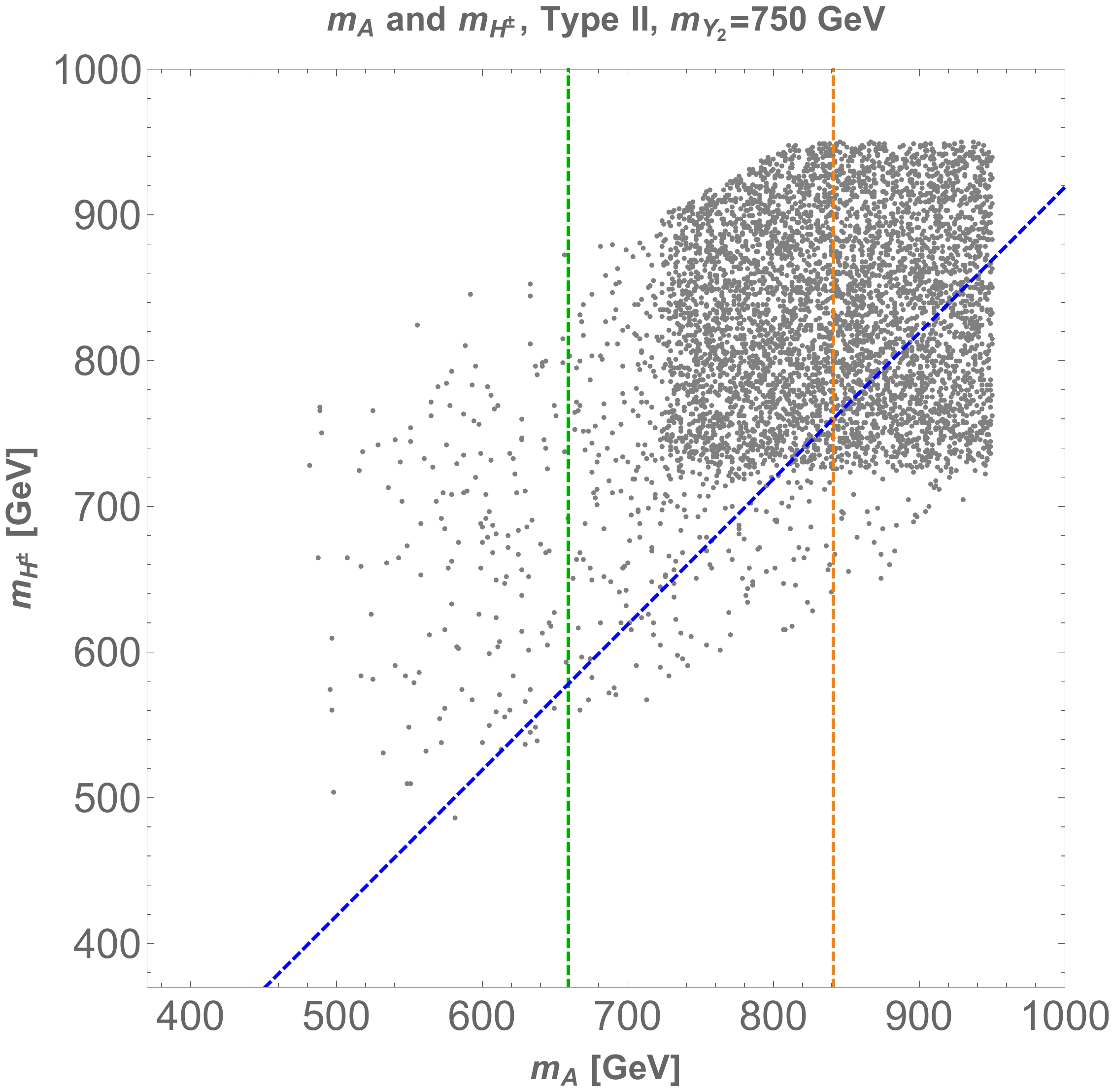}
\caption{The allowed range on $m_A$ and $m_{H^\pm}$. 
The gray points are compatible with both theoretical constraints and EWPO within the $\pm 3\sigma$ range. The green dashed line here indicates the mass relation $m_{A}=750-m_Z$. Thus, $Y_2$ cannot decay to $AZ$ on-shell on the right hand side of this line. Similarly, the orange and blue dashed line represent $m_{A}=750+m_Z$ and $m_{A}=m_{H^{\pm}}+m_W$
(The result is generically the same as in both Type~I and Type~II models.) 
}
\label{fig:mAmHpmscan}
\end{figure}

The range of the rest of two scalars, the pseudo-scalar Higgs mass $m_A$ and charged Higgs mass $m_{H^{\pm}}$, is also interesting. In principle, $m_A$ can be above or below either of two CP-even Higgs bosons, and even $m_A < m_h/2$ is possible and consistent with the data~\cite{Bernon:2014nxa}. However, once the heavy Higgs mass ($m_{Y_2}$ in our model) is fixed, the allowed range of $m_A$ is limited. 
The interrelation between $m_A$ and $m_{H^{\pm}}$ for the case of $m_{Y_2}=750$~GeV is illustrated in Fig.~\ref{fig:mAmHpmscan}. 
There, we observe that $m_{H^{\pm}},m_A$ are bounded in the $400-950$~GeV. As expected, this result is identical to what is displayed in the 2HDM~\cite{Bernon:2015qea}. This is not surprising because the introduced complex singlet does not generate the mixing in the CP-odd states.

Due to kinematical suppression, $Y_2 \to AA/H^{\pm}H^{\mp}$ is always forbidden. The absence of these two decay modes is crucially important in the success of explaining the di-photon excess. Otherwise, they will eat a large amount of branching ratio of the $Y_2$ decay, so that the $Y_2 \to Y_1 Y_1$ decay will be heavily suppressed. While $Y_2 \to ZA$ and/or $Y_2 \to W^{\pm}H^{\mp}$ decay is kinematically possible for $m_{H^{\pm}},m_A \lesssim 670$~GeV, their contributions to the $Y_2$ decay are so small that could be neglected. 

As regards for the $A$, $AZh$ coupling is vanishing in the exact alignment limit and thus the exotic decay $A \to Zh$ is not present. 
However, $A\to Y_2 Z$ (right to the orange line) and $A\to W^\pm H^\mp$ (above the blue line) could be important in addition to the fermionic decay being the potential discovery channels for the $A$ including the $t\bar{t}$, $b\bar{b}$ and $\tau^- \tau^+$ final states. In general, the cross section of producing the $A$ via gluon-fusion is proportional to $\tan\beta$ and varies from few pb and 10 fb at the 13 TeV, depending on the exact mass.  
Since the decay into $t\bar{t}$ dominates for moderate $\tan\beta$ we consider here, the cross section $gg\to A\to b\bar{b}$ can at most reach $\sim 10$~fb, an order which hardly enables them to compete with the large QCD background.

%\newpage

%%%%%%%%%%%%%%%%%%%%%%%%%%%%%%%%%%%%%%%%

%%%%%%%%%%%%%%%%%%%%%%%%%%%%%%%%%%%%%%%%
\section{Conclusions}\label{sec:conclusion}
%%%%%%%%%%%%%%%%%%%%%%%%%%%%%%%%%%%%%%%%

First, we find that a pure doublet state at 750 GeV generally has a very limited branching ratio for the loop-induced di-photon decay when the tree-level decay to $t\bar{t}$ is present. As a result, 
the process of $gg\to H/A \to \gamma\gamma$ with $CP$-even or $CP$-odd Higgs $H/A$ being identified as the 750 GeV resonance in the minimal version of 2HDM cannot reproduce the di-photon signal that is comparable to the observed level.

Other than through direct decay, we alternatively consider the di-photon signal that arises from two bunches of collimated photon jets emitting from a pair of highly boosted particles in this work. In particular, we studied the impact of $m_{Y_1}$ and its proper decay length on collider phenomenology. It turns out that $m_{Y_1}$ should be $\lesssim 210$~MeV to efficiently produce a photon jet as narrow as it must be. However, such a light boosted state cannot be the $A$ in the 2HDM, once the $H$ is identified as the 750 GeV resonance. Therefore, we extend the 2HDM by adding a gauge singlet scalar field.
Though the singlet scalar field generically mixes with two doublet fields, it is possible to accomplish a SM-like Higgs $h$ with another two mass eigenstates. They include one heavy doublet-like $Y_2$ and one light singlet-like $Y_1$ with mass of sub-GeV. 
The $Y_2(750)$ can decay to a pair of $Y_1$, each of which can further decay into two photon jets.  On the other hand, the presence of this mixing, although rather small, is the key of controlling the proper decay length of $Y_1$.

Two specific models containing an extra real or complex singlet scalar are studied. In both models, the SM-like $h$ decay to $Y_1Y_1$ or invisible final states are swiched off. 
For the real singlet model, 
the mixing between heavy Higgs and the singlet state is strongly constrained so that it is difficult to simultaneously yield the di-photon signal comparable to the observed data and also achieve a reasonable proper decay length. In contrast, in the absence of the $Z_2$ symmetry the complex model has linear singlet terms, which are irrelevant to the Higgs invisible decay but play an essential role in generating the mixing in the scalar sector. As a result, this model is easy to yield the 1-10~fb cross section in the di-photon final state with a decay length of $\mathcal{O}(1)$~m for the $Y_1$, meanwhile parametrically predicts the width of 750 GeV resonance $\gtrsim 1$~GeV. In addition, the pseudoscalar component of the singlet is naturally stable and hence could be a dark matter candidate.

Finally, we have discussed the discovery prospects of other scalar states such as the $CP-$odd $A$ and the charged Higgs $H^\pm$ at the future LHC run.

%%%%%%%%%%%%%%%%%%%%%%%%%%%%%%%%%%%%%%%%
\section*{Acknowledgements} 

We thank John Conway and Yong Yang for useful discussions regarding the details of collider tracker to verify the possibility of this scenario. 
We are also grateful to John~F.~Gunion and Zhen~Liu for their particular attention as well as helpful comments on the manuscript. 
YJ is supported by the Villum Foundation.

%%%%%%%%%%%%%%%%%%%%%%%%%%%%%%%%%%%%%%%%

\section*{Appendix: Stability constraint}
%%%%%%%%%%%%%%%%%%%%%%%%%%%%%%%%%%%%%%%%
%%%%%%%%%%% %%%%%%%%%%%%%%%%%%%

To derive the stability bound we use the inequalities of $\kappa_{1,2}$ in \cite{Drozd:2014yla}. Given in Eq.~(\ref{eq:k1k2v2}), $\kappa_1$ and $\kappa_2$ are of the opposite signs: $\kappa_1=-t_\beta^2 \kappa_2$. Employing this relation, the upper limit of $|\kappa_2|$ reads
\begin{equation}
	|\kappa_2|\leqslant \text{Min}(A_1,A_2)
\end{equation}
where $A_1=\sqrt{\lambda_1 \lambda_s /12}/ t_\beta^2$, $A_2=\sqrt{\lambda_2\lambda_s/12}$. When $\lambda_3 < 0$, one should ensure

\begin{eqnarray}\label{eq:k2constraint}
	\left\lbrace \begin{array}{ll}
		|\kappa_2|\leqslant \sqrt{\frac{A_1^2A_2^2-A_3^2}{A_1^2+A_2^2-2A_3}},& \text{if} \quad A_3 < A_1A_2\\
		|\kappa_2|=0,& \text{if} \quad A_3 \geqslant A_1A_2
	\end{array}\right.
\end{eqnarray}
where $A_3=-\lambda_3 \lambda_s/(12 t_\beta^2)$. 
Additionally, if $\lambda_3+\lambda_4-|\lambda_5| < 0$ is satisfied, one more constraint is required which can be obtained by replacing $A_3$ by $A_4$ with 
\beq
A_4=-(\lambda_3+\lambda_4-|\lambda_5|)\lambda_s/(12 t_\beta^2).
\eeq
%%%%%%%%%%%%%%%%%%%%%%%%%%%%%%%%%%%%%%%%%%%%%%%%%%%%%%%%%%%%
%-----------------------------------------------------------------------------------------------------------------------------
%\bibliographystyle{elsarticle-num}
%\bibliography{references}

\begin{thebibliography}{10}
	\bibitem{ATLAS:2015}
	ATLAS Collaboration, G.~Aad {\it et~al.}, {\it  ``Search for high-mass diphoton
		resonances in $pp$ collisions at $\sqrt{s}=13$ TeV with the ATLAS detector''},
	ATLAS-CONF-2015-081.
	
	\bibitem{CMS:2015}
	CMS Collaboration, S.~Chatrchyan {\it et~al.}, {\it ``Search for High-Mass
		Diphoton Resonances in pp Collisions at $\sqrt s=13$ TeV with the CMS
		Detector''},  CMS PAS EXO-15-004.
	
	
	
	
	
	
	%%%%% 2HDM %%%%%%
	
	%\cite{Bernon:2015qea}
	\bibitem{Bernon:2015qea}
	J.~Bernon, J.~F.~Gunion, H.~E.~Haber, Y.~Jiang and S.~Kraml,
	{\it ``Scrutinizing the alignment limit in two-Higgs-doublet models: m$_h$=125 GeV''},
	Phys.\ Rev.\ D {\bf 92} (2015) 7,  075004
	%doi:10.1103/PhysRevD.92.075004
	[arXiv:1507.00933 [hep-ph]].
	
	
	%%%%% 2HDM for 750 %%%%%%
	
	\bibitem{Angelescu:2015uiz} 
	A.~Angelescu, A.~Djouadi and G.~Moreau,
	{\it``Scenarii for interpretations of the LHC diphoton excess: two Higgs doublets and vector-like quarks and leptons,''}
	Phys.\ Lett.\ B {\bf 756} (2016) 126,
	%doi:10.1016/j.physletb.2016.02.064
	[arXiv:1512.04921 [hep-ph]].
	%%CITATION = doi:10.1016/j.physletb.2016.02.064;%%
	%156 citations counted in INSPIRE as of 14 Mar 2016
	
	\bibitem{Han:2015qqj} 
	X.~F.~Han and L.~Wang,
	{\it``Implication of the 750 GeV diphoton resonance on two-Higgs-doublet model and its extensions with Higgs field,''}
	[arXiv:1512.06587 [hep-ph]].
	%%CITATION = [arXiv:1512.06587;%%
	%90 citations counted in INSPIRE as of 14 Mar 2016
	
	\bibitem{Badziak:2015zez} 
	M.~Badziak,
	{\it``Interpreting the 750 GeV diphoton excess in minimal extensions of Two-Higgs-Doublet models,''}
	[arXiv:1512.07497 [hep-ph]].
	%%CITATION = [arXiv:1512.07497;%%
	%69 citations counted in INSPIRE as of 14 Mar 2016
	
	
	\bibitem{Altmannshofer:2015xfo} 
	W.~Altmannshofer, J.~Galloway, S.~Gori, A.~L.~Kagan, A.~Martin and J.~Zupan,
	{\it``On the 750 GeV di-photon excess,''}
	[arXiv:1512.07616 [hep-ph]].
	%%CITATION = [arXiv:1512.07616;%%
	%86 citations counted in INSPIRE as of 14 Mar 2016
	
	\bibitem{Bizot:2015qqo} 
	N.~Bizot, S.~Davidson, M.~Frigerio and J.-L.~Kneur,
	{\it``Two Higgs doublets to explain the excesses $pp\rightarrow \gamma\gamma(750\ {\rm GeV})$ and $h \to \tau^\pm \mu^\mp$,''}
	[arXiv:1512.08508 [hep-ph]].
	%%CITATION = [arXiv:1512.08508;%%
	%59 citations counted in INSPIRE as of 14 Mar 2016
	
	%\cite{Han:2016bus}
	\bibitem{Han:2016bus} 
	X.~F.~Han, L.~Wang, L.~Wu, J.~M.~Yang and M.~Zhang,
	{\it``Explaining 750 GeV diphoton excess from top/bottom partner cascade decay in two-Higgs-doublet model extension,''}
	[arXiv:1601.00534 [hep-ph]].
	%%CITATION = [arXiv:1601.00534;%%
	%29 citations counted in INSPIRE as of 14 Mar 2016
	
	
	\bibitem{Hernandez:2016rbi} 
	A.~E.~C.~Hernández, I.~d.~M.~Varzielas and E.~Schumacher,
	{\it``The $750\,\text{GeV}$ diphoton resonance in the light of a 2HDM with $S_3$ flavour symmetry,''}
	[arXiv:1601.00661 [hep-ph]].
	%%CITATION = [arXiv:1601.00661;%%
	%31 citations counted in INSPIRE as of 14 Mar 2016
	
	\bibitem{Djouadi:2016eyy} 
	A.~Djouadi, J.~Ellis, R.~Godbole and J.~Quevillon,
	{\it``Future Collider Signatures of the Possible 750 GeV State,''}
	[arXiv:1601.03696 [hep-ph]].
	%%CITATION = [arXiv:1601.03696;%%
	%28 citations counted in INSPIRE as of 14 Mar 2016
	
	%\cite{Han:2016bvl}
	\bibitem{Han:2016bvl} 
	X.~F.~Han, L.~Wang and J.~M.~Yang,
	{\it``An extension of two-Higgs-doublet model and the excesses of 750 GeV diphoton, muon g-2 and $h\to\mu\tau$,''}
	[arXiv:1601.04954 [hep-ph]].
	%%CITATION = [arXiv:1601.04954;%%
	%19 citations counted in INSPIRE as of 14 Mar 2016
	
	
	\bibitem{Kang:2015roj}
	S.~K.~Kang and J.~Song,
	{\it``Top-phobic heavy Higgs boson as the 750 GeV diphoton resonance,''}
	[arXiv:1512.08963 [hep-ph]].
	%%CITATION = [arXiv:1512.08963;%%
	%18 citations counted in INSPIRE as of 13 Jan 2016
	
	
	%	\bibitem{Han:2016bus}
	%	X.~F.~Han, L.~Wang, L.~Wu, J.~M.~Yang and M.~Zhang,
	%	{\it``Explaining 750 GeV diphoton excess from top/bottom partner cascade decay in two-Higgs-doublet model extension,''}
	%	[arXiv:1601.00534 [hep-ph]].
	%	%%CITATION = [arXiv:1601.00534;%%
	%	%6 citations counted in INSPIRE as of 13 Jan 2016
	
	%\cite{Arbelaez:2016mhg}
	\bibitem{Arbelaez:2016mhg}
	C.~Arbeláez, A.~E.~C.~Hernández, S.~Kovalenko and I.~Schmidt,
	{\it``Linking radiative seesaw-type mechanism of fermion masses and non-trivial quark mixing with the 750 GeV diphoton excess,''}
	[arXiv:1602.03607 [hep-ph]].
	%%CITATION = [arXiv:1602.03607;%%
	%10 citations counted in INSPIRE as of 24 Mar 2016
	
	%	%\cite{Han:2016bvl}
	%	\bibitem{Han:2016bvl}
	%	X.~F.~Han, L.~Wang and J.~M.~Yang,
	%	{\it``An extension of two-Higgs-doublet model and the excesses of 750 GeV diphoton, muon g-2 and $h\to\mu\tau$,''}
	%	[arXiv:1601.04954 [hep-ph]].
	%	%%CITATION = [arXiv:1601.04954;%%
	%	%22 citations counted in INSPIRE as of 23 Mar 2016
	
	
	%\cite{Bertuzzo:2016fmv}
	\bibitem{Bertuzzo:2016fmv}
	E.~Bertuzzo, P.~A.~N.~Machado and M.~Taoso,
	{\it``Di-Photon excess in the 2HDM: hasting towards the instability and the non-perturbative regime,''}
	[arXiv:1601.07508 [hep-ph]].
	%%CITATION = [arXiv:1601.07508;%%
	%12 citations counted in INSPIRE as of 24 Mar 2016
	
	%\cite{Ahriche:2016mcx}
	\bibitem{Ahriche:2016mcx}
	A.~Ahriche, G.~Faisel, S.~Nasri and J.~Tandean,
	{\it``Addressing the LHC 750 GeV diphoton excess without new colored states,''}
	[arXiv:1603.01606 [hep-ph]].
	%%CITATION = [arXiv:1603.01606;%%
	%1 citations counted in INSPIRE as of 24 Mar 2016
	
	%%%%%%%%%% triplet Higgs %%%%%%%%%%%
	%\cite{Chiang:2016ydx}
	\bibitem{Chiang:2016ydx}
	C.~W.~Chiang and A.~L.~Kuo,
	{\it``Can the 750-GeV diphoton resonance be the singlet Higgs boson of custodial Higgs triplet model?,''}
	[arXiv:1601.06394 [hep-ph]].
	%%CITATION = [arXiv:1601.06394;%%
	%15 citations counted in INSPIRE as of 24 Mar 2016
	
	%\cite{Chaudhuri:2016rwo}
	\bibitem{Chaudhuri:2016rwo}
	A.~Chaudhuri and B.~Mukhopadhyaya,
	{\it``A CP-violating phase in a two Higgs triplet scenario : some phenomenological implications,''}
	[arXiv:1602.07846 [hep-ph]].
	%%CITATION = [arXiv:1602.07846;%%
	
	%%%% other Higgs %%%%%%
	%\cite{Delgado:2016arn}
	\bibitem{Delgado:2016arn}
	A.~Delgado, M.~Garcia-Pepin, M.~Quiros, J.~Santiago and R.~Vega-Morales,
	{\it``Diphoton and Diboson Probes of Fermiophobic Higgs Bosons at the LHC,''}
	[arXiv:1603.00962 [hep-ph]].
	%%CITATION = [arXiv:1603.00962;%%
	
	%%%%%%%%%% Related 750 Works %%%%%%%
	
	%\cite{Higaki:2015jag}
	\bibitem{Higaki:2015jag}
	T.~Higaki, K.~S.~Jeong, N.~Kitajima and F.~Takahashi,
	{\it``The QCD Axion from Aligned Axions and Diphoton Excess,''}
	[arXiv:1512.05295 [hep-ph]].
	%%CITATION = [arXiv:1512.05295;%%
	%94 citations counted in INSPIRE as of 13 Jan 2016
	
	%\cite{Ghosh:2015apa}
	\bibitem{Ghosh:2015apa}
	S.~Ghosh, A.~Kundu and S.~Ray,
	{\it``On the potential of a singlet scalar enhanced Standard Model,''}
	[arXiv:1512.05786 [hep-ph]].
	%%CITATION = [arXiv:1512.05786;%%
	%7 citations counted in INSPIRE as of 13 Jan 2016
	
	%\cite{Franceschini:2015kwy}
	\bibitem{Franceschini:2015kwy} 
	R.~Franceschini {\it et al.},
	{\it``What is the gamma gamma resonance at 750 GeV?,''}
	[arXiv:1512.04933 [hep-ph]].
	%%CITATION = [arXiv:1512.04933;%%
	%78 citations counted in INSPIRE as of 25 Dec 2015
	
	%\cite{Aloni:2015mxa}
	\bibitem{Aloni:2015mxa} 
	D.~Aloni, K.~Blum, A.~Dery, A.~Efrati and Y.~Nir,
	{\it``On a possible large width 750 GeV diphoton resonance at ATLAS and CMS,''}
	[arXiv:1512.05778 [hep-ph]].
	%%CITATION = [arXiv:1512.05778;%%
	%41 citations counted in INSPIRE as of 25 Dec 2015
	
	%\cite{DiChiara:2015vdm}
	\bibitem{DiChiara:2015vdm} 
	S.~Di Chiara, L.~Marzola and M.~Raidal,
	{\it``First interpretation of the 750 GeV di-photon resonance at the LHC,''}
	[arXiv:1512.04939 [hep-ph]].
	%%CITATION = [arXiv:1512.04939;%%
	%67 citations counted in INSPIRE as of 25 Dec 2015
	%\cite{Das:2015enc}
	
	\bibitem{Das:2015enc}
	K.~Das and S.~K.~Rai,
	{\it``750 GeV diphoton excess in a U(1) hidden symmetry model,''}
	Phys.\ Rev.\ D {\bf 93} (2016) no.9,  095007
	%doi:10.1103/PhysRevD.93.095007
	[arXiv:1512.07789 [hep-ph]].
	%%CITATION = doi:10.1103/PhysRevD.93.095007;%%
	%86 citations counted in INSPIRE as of 27 May 2016
	
	%\cite{Kanemura:2015vcb}	
	\bibitem{Kanemura:2015vcb}
	S.~Kanemura, N.~Machida, S.~Odori and T.~Shindou,
	{\it``Diphoton excess at 750 GeV in an extended scalar sector,''}
	[arXiv:1512.09053 [hep-ph]].
	%%CITATION = [arXiv:1512.09053;%%
	%12 citations counted in INSPIRE as of 13 Jan 2016  
	
	%\cite{Jiang:2015oms}
	\bibitem{Jiang:2015oms}
	Y.~Jiang, Y.~Y.~Li and T.~Liu,
	{\it``750 GeV Resonance in the Gauged $U(1)'$-Extended MSSM,''}
	[arXiv:1512.09127 [hep-ph]].
	%%CITATION = [arXiv:1512.09127;%%
	%9 citations counted in INSPIRE as of 13 Jan 2016
	
	%\cite{Tsai:2016lfg}
	\bibitem{Tsai:2016lfg}
	Y.~Tsai, L.~T.~Wang and Y.~Zhao,
	{\it``Faking The Diphoton Excess by Displaced Dark Photon Decays,''}
	[arXiv:1603.00024 [hep-ph]].
	%%CITATION = [arXiv:1603.00024;%%
	%6 citations counted in INSPIRE as of 24 Mar 2016
	
	%%%%%%%%%%%%% EFT fit  %%%%%%%%%%%%%%
	
	%\cite{Chakrabortty:2015hff}
	\bibitem{Chakrabortty:2015hff} 
	J.~Chakrabortty, A.~Choudhury, P.~Ghosh, S.~Mondal and T.~Srivastava,
	{\it``Di-photon resonance around 750 GeV: shedding light on the theory underneath,''}
	[arXiv:1512.05767 [hep-ph]].
	%%CITATION = [arXiv:1512.05767;%%
	%42 citations counted in INSPIRE as of 25 Dec 2015
	
	%\cite{Kim:2015ksf}
	\bibitem{Kim:2015ksf} 
	J.~S.~Kim, K.~Rolbiecki and R.~R.~de Austri,
	{\it``Model-independent combination of diphoton constraints at 750 GeV,''}
	[arXiv:1512.06797 [hep-ph]].
	%%CITATION = [arXiv:1512.06797;%%
	%13 citations counted in INSPIRE as of 25 Dec 2015
	
	%\cite{Berthier:2015vbb}
	\bibitem{Berthier:2015vbb} 
	L.~Berthier, J.~M.~Cline, W.~Shepherd and M.~Trott,
	{\it``Effective interpretations of a diphoton excess,''}
	[arXiv:1512.06799 [hep-ph]].
	%%CITATION = [arXiv:1512.06799;%%
	%13 citations counted in INSPIRE as of 25 Dec 2015
	
	%\cite{Godunov:2016kqn}
	\bibitem{Godunov:2016kqn}
	S.~I.~Godunov, A.~N.~Rozanov, M.~I.~Vysotsky and E.~V.~Zhemchugov,
	{\it``New Physics at 1 TeV?,''}
	[arXiv:1602.02380 [hep-ph]].
	%%CITATION = [arXiv:1602.02380;%%
	%7 citations counted in INSPIRE as of 24 Mar 2016
	
	%%% data ?????????????? %%%%
	
	%\cite{Kavanagh:2016pso}
	\bibitem{Kavanagh:2016pso}
	B.~J.~Kavanagh,
	{\it``Re-examining the significance of the 750 GeV diphoton excess at ATLAS,''}
	[arXiv:1601.07330 [hep-ph]].
	%%CITATION = [arXiv:1601.07330;%%
	%9 citations counted in INSPIRE as of 24 Mar 2016
	
	%%%%%%%%%%%%% New Vector %%%%%%%%%%%%%%
	
	%\cite{deBlas:2015hlv}
	\bibitem{deBlas:2015hlv} 
	J.~de Blas, J.~Santiago and R.~Vega-Morales,
	{\it``New vector bosons and the diphoton excess,''}
	[arXiv:1512.07229 [hep-ph]].
	%%CITATION = [arXiv:1512.07229;%%
	%7 citations counted in INSPIRE as of 25 Dec 2015
	
	%\cite{Son:2015vfl}
	\bibitem{Son:2015vfl} 
	M.~Son and A.~Urbano,
	{\it``A new scalar resonance at 750 GeV: Towards a proof of concept in favor of strongly interacting theories,''}
	[arXiv:1512.08307 [hep-ph]].
	%%CITATION = [arXiv:1512.08307;%%
	
	%\cite{Kaneta:2015qpf}
	\bibitem{Kaneta:2015qpf}
	K.~Kaneta, S.~Kang and H.~S.~Lee,
	{\it``Diphoton excess at the LHC Run 2 and its implications for a new heavy gauge boson,''}
	[arXiv:1512.09129 [hep-ph]].
	%%CITATION = [arXiv:1512.09129;%%
	%14 citations counted in INSPIRE as of 13 Jan 2016     
	
	%%%% SM+ singlet context %%%%%%%%%%%%%%%%%%%%%%%%%%%%%%%%%%%%%%
	
	%\cite{Buttazzo:2015txu}
	\bibitem{Buttazzo:2015txu} 
	D.~Buttazzo, A.~Greljo and D.~Marzocca,
	{\it``Knocking on New Physics' door with a Scalar Resonance,''}
	[arXiv:1512.04929 [hep-ph]].
	%%CITATION = [arXiv:1512.04929;%%
	%65 citations counted in INSPIRE as of 25 Dec 2015
	
	%\cite{McDermott:2015sck}
	\bibitem{McDermott:2015sck} 
	S.~D.~McDermott, P.~Meade and H.~Ramani,
	{\it``Singlet Scalar Resonances and the Diphoton Excess,''}
	[arXiv:1512.05326 [hep-ph]].
	%%CITATION = [arXiv:1512.05326;%%
	%63 citations counted in INSPIRE as of 25 Dec 2015
	
	%\cite{Ellis:2015oso}
	\bibitem{Ellis:2015oso} 
	J.~Ellis, S.~A.~R.~Ellis, J.~Quevillon, V.~Sanz and T.~You,
	{\it``On the Interpretation of a Possible $\sim 750$ GeV Particle Decaying into $\gamma \gamma$,''}
	[arXiv:1512.05327 [hep-ph]].
	%%CITATION = [arXiv:1512.05327;%%
	%62 citations counted in INSPIRE as of 25 Dec 2015
	
	%\cite{Dutta:2015wqh}
	\bibitem{Dutta:2015wqh} 
	B.~Dutta, Y.~Gao, T.~Ghosh, I.~Gogoladze and T.~Li,
	{\it``Interpretation of the diphoton excess at CMS and ATLAS,''}
	[arXiv:1512.05439 [hep-ph]].
	%%CITATION = [arXiv:1512.05439;%%
	%41 citations counted in INSPIRE as of 25 Dec 2015
	
	%\cite{Chao:2015ttq}
	\bibitem{Chao:2015ttq} 
	W.~Chao, R.~Huo and J.~H.~Yu,
	{\it``The Minimal Scalar-Stealth Top Interpretation of the Diphoton Excess,''}
	[arXiv:1512.05738 [hep-ph]].
	%%CITATION = [arXiv:1512.05738;%%
	%44 citations counted in INSPIRE as of 25 Dec 2015
	
	%\cite{Fichet:2015vvy}
	\bibitem{Fichet:2015vvy} 
	S.~Fichet, G.~von Gersdorff and C.~Royon,
	{\it``Scattering Light by Light at 750 GeV at the LHC,''}
	[arXiv:1512.05751 [hep-ph]].
	%%CITATION = [arXiv:1512.05751;%%
	%45 citations counted in INSPIRE as of 25 Dec 2015
	
	%\cite{Falkowski:2015swt}
	\bibitem{Falkowski:2015swt} 
	A.~Falkowski, O.~Slone and T.~Volansky,
	{\it``Phenomenology of a 750 GeV Singlet,''}
	[arXiv:1512.05777 [hep-ph]].
	%%CITATION = [arXiv:1512.05777;%%
	%49 citations counted in INSPIRE as of 25 Dec 2015
	
	%\cite{Benbrik:2015fyz}
	\bibitem{Benbrik:2015fyz} 
	R.~Benbrik, C.~H.~Chen and T.~Nomura,
	{\it``Higgs singlet as a diphoton resonance in a vector-like quark model,''}
	[arXiv:1512.06028 [hep-ph]].
	%%CITATION = [arXiv:1512.06028;%%
	%27 citations counted in INSPIRE as of 25 Dec 2015
	
	%\cite{Wang:2015kuj}
	\bibitem{Wang:2015kuj} 
	F.~Wang, L.~Wu, J.~M.~Yang and M.~Zhang,
	{\it``750 GeV Diphoton Resonance, 125 GeV Higgs and Muon g-2 Anomaly in Deflected Anomaly Mediation SUSY Breaking Scenario,''}
	arXiv:1512.06715 [hep-ph].
	%%CITATION = ARXIV:1512.06715;%%
	%102 citations counted in INSPIRE as of 09 May 2016
	
	%\cite{Cao:2015twy}
	\bibitem{Cao:2015twy} 
	J.~Cao, C.~Han, L.~Shang, W.~Su, J.~M.~Yang and Y.~Zhang,
	{\it``Interpreting the 750 GeV diphoton excess by the singlet extension of the Manohar-Wise Model,''}
	[arXiv:1512.06728 [hep-ph]].
	%%CITATION = [arXiv:1512.06728;%%
	%13 citations counted in INSPIRE as of 25 Dec 2015
	
	%	%\cite{Altmannshofer:2015xfo}
	%	\bibitem{Altmannshofer:2015xfo} 
	%	W.~Altmannshofer, J.~Galloway, S.~Gori, A.~L.~Kagan, A.~Martin and J.~Zupan,
	%	{\it``On the 750 GeV di-photon excess,''}
	%	[arXiv:1512.07616 [hep-ph]].
	%	%%CITATION = [arXiv:1512.07616;%%
	%	%4 citations counted in INSPIRE as of 25 Dec 2015
	
	%\cite{Gu:2015lxj}
	\bibitem{Gu:2015lxj} 
	J.~Gu and Z.~Liu,
	{\it``Running after Diphoton,''}
	[arXiv:1512.07624 [hep-ph]].
	%%CITATION = [arXiv:1512.07624;%%
	%2 citations counted in INSPIRE as of 25 Dec 2015
	
	%\cite{Cheung:2015cug}
	\bibitem{Cheung:2015cug} 
	K.~Cheung, P.~Ko, J.~S.~Lee, J.~Park and P.~Y.~Tseng,
	{\it``A Higgcision study on the 750 GeV Di-photon Resonance and 125 GeV SM Higgs boson with the Higgs-Singlet Mixing,''}
	[arXiv:1512.07853 [hep-ph]].
	%%CITATION = [arXiv:1512.07853;%%
	
	%\cite{Li:2015jwd}
	\bibitem{Li:2015jwd} 
	G.~Li, Y.~n.~Mao, Y.~L.~Tang, C.~Zhang, Y.~Zhou and S.~h.~Zhu,
	{\it``A Loop-philic Pseudoscalar,''}
	arXiv:1512.08255 [hep-ph]].
	%%CITATION = [arXiv:1512.08255;%%
	
	%\cite{Wang:2015omi}
	\bibitem{Wang:2015omi} 
	F.~Wang, W.~Wang, L.~Wu, J.~M.~Yang and M.~Zhang,
	{\it``Interpreting 750 GeV diphoton resonance as degenerate Higgs bosons in NMSSM with vector-like particles,''}
	arXiv:1512.08434 [hep-ph].
	%%CITATION = ARXIV:1512.08434;%%
	%74 citations counted in INSPIRE as of 09 May 2016
	
	%\cite{D'Eramo:2016mgv}
	\bibitem{D'Eramo:2016mgv}
	F.~D'Eramo, J.~de Vries and P.~Panci,
	{\it``A 750 GeV Portal: LHC Phenomenology and Dark Matter Candidates,''}
	[arXiv:1601.01571 [hep-ph]].
	%%CITATION = [arXiv:1601.01571;%%
	%3 citations counted in INSPIRE as of 13 Jan 2016
	
	
	%\cite{Ko:2016wce}
	\bibitem{Ko:2016wce}
	P.~Ko and T.~Nomura,
	{\it``Dark sector shining through 750 GeV dark Higgs boson at the LHC,''}
	[arXiv:1601.02490 [hep-ph]].
	%%CITATION = [arXiv:1601.02490;%%
	
	%\cite{Chao:2016aer}
	\bibitem{Chao:2016aer}
	W.~Chao,
	{\it``The Diphoton Excess Inspired Electroweak Baryogenesis,''}
	[arXiv:1601.04678 [hep-ph]].
	%%CITATION = [arXiv:1601.04678;%%
	%23 citations counted in INSPIRE as of 23 Mar 2016
	
	%\cite{Dolan:2016eki}
	\bibitem{Dolan:2016eki}
	M.~J.~Dolan, J.~L.~Hewett, M.~Krämer and T.~G.~Rizzo,
	{\it``Simplified Models for Higgs Physics: Singlet Scalar and Vector-like Quark Phenomenology,''}
	[arXiv:1601.07208 [hep-ph]].
	%%CITATION = [arXiv:1601.07208;%%
	%8 citations counted in INSPIRE as of 24 Mar 2016
	
	%\cite{Ding:2016udc}
	\bibitem{Ding:2016udc}
	R.~Ding, Y.~Fan, L.~Huang, C.~Li, T.~Li, S.~Raza and B.~Zhu,
	{\it``Systematic Study of Diphoton Resonance at 750 GeV from Sgoldstino,''}
	[arXiv:1602.00977 [hep-ph]].
	%%CITATION = [arXiv:1602.00977;%%
	%13 citations counted in INSPIRE as of 24 Mar 2016
	
	%\cite{Bae:2016xni}
	\bibitem{Bae:2016xni} 
	K.~J.~Bae, M.~Endo, K.~Hamaguchi and T.~Moroi,
	{\it``Diphoton Excess and Running Couplings,''}
	[arXiv:1602.03653 [hep-ph]].
	%%CITATION = [arXiv:16t02.03653;%%
	%11 citations counted in INSPIRE as of 24 Mar 2016
	
	
	
	
	
	%%% link to high-scale %%%%%%%%%%%%%%%%%%%%%%%%%%%%%%%%%%%%%%%
	
	%\cite{Chakraborty:2015jvs}
	\bibitem{Chakraborty:2015jvs} 
	I.~Chakraborty and A.~Kundu,
	{\it``Diphoton excess at 750 GeV: Singlet scalars confront naturalness,''}
	[arXiv:1512.06508 [hep-ph]].
	%%CITATION = [arXiv:1512.06508;%%
	%11 citations counted in INSPIRE as of 25 Dec 2015
	
	%\cite{Dhuria:2015ufo}
	\bibitem{Dhuria:2015ufo} 
	M.~Dhuria and G.~Goswami,
	{\it``Perturbativity, vacuum stability and inflation in the light of 750 GeV diphoton excess,''}
	[arXiv:1512.06782 [hep-ph]].
	%%CITATION = [arXiv:1512.06782;%%
	%9 citations counted in INSPIRE as of 25 Dec 2015
	
	%\cite{Zhang:2015uuo}
	\bibitem{Zhang:2015uuo} 
	J.~Zhang and S.~Zhou,
	{\it``Electroweak Vacuum Stability and Diphoton Excess at 750 GeV,''}
	[arXiv:1512.07889 [hep-ph]].
	%%CITATION = [arXiv:1512.07889;%%
	
	%\cite{Salvio:2015jgu}
	\bibitem{Salvio:2015jgu} 
	A.~Salvio and A.~Mazumdar,
	{\it``Higgs Stability and the 750 GeV Diphoton Excess,''}
	[arXiv:1512.08184 [hep-ph]].
	%%CITATION = [arXiv:1512.08184;%%
	
	%\cite{Aydemir:2016qqj}
	\bibitem{Aydemir:2016qqj}
	U.~Aydemir and T.~Mandal,
	{\it``Interpretation of the 750 GeV diphoton excess with colored scalars in $\mathbf{SO(10)}$ grand unification,''}
	[arXiv:1601.06761 [hep-ph]].
	%%CITATION = [arXiv:1601.06761;%%
	%19 citations counted in INSPIRE as of 24 Mar 2016
	
	%\cite{Salvio:2016hnf}
	\bibitem{Salvio:2016hnf}
	A.~Salvio, F.~Staub, A.~Strumia and A.~Urbano,
	{\it``On the maximal diphoton width,''}
	[arXiv:1602.01460 [hep-ph]].
	%%CITATION = [arXiv:1602.01460;%%
	%14 citations counted in INSPIRE as of 24 Mar 2016
	
	%\cite{Hamada:2016vwk}
	\bibitem{Hamada:2016vwk}
	Y.~Hamada, H.~Kawai, K.~Kawana and K.~Tsumura,
	{\it``Models of LHC Diphoton Excesses Valid up to the Planck scale,''}
	[arXiv:1602.04170 [hep-ph]].
	%%CITATION = [arXiv:1602.04170;%%
	%12 citations counted in INSPIRE as of 24 Mar 2016
	
	%%%%%%%%%%%%% Extra dimension %%%%%%%%%%%%%%
	
	%\cite{Cox:2015ckc}
	\bibitem{Cox:2015ckc} 
	P.~Cox, A.~D.~Medina, T.~S.~Ray and A.~Spray,
	{\it``Diphoton Excess at 750 GeV from a Radion in the Bulk-Higgs Scenario,''}
	[arXiv:1512.05618 [hep-ph]].
	%%CITATION = [arXiv:1512.05618;%%
	%45 citations counted in INSPIRE as of 25 Dec 2015
	
	%\cite{Ahmed:2015uqt}
	\bibitem{Ahmed:2015uqt} 
	A.~Ahmed, B.~M.~Dillon, B.~Grzadkowski, J.~F.~Gunion and Y.~Jiang,
	{\it``Higgs-radion interpretation of 750 GeV di-photon excess at the LHC,''}
	[arXiv:1512.05771 [hep-ph]].
	%%CITATION = [arXiv:1512.05771;%%
	%42 citations counted in INSPIRE as of 25 Dec 2015
	
	%\cite{Megias:2015ory}
	\bibitem{Megias:2015ory} 
	E.~Megias, O.~Pujolas and M.~Quiros,
	{\it``On dilatons and the LHC diphoton excess,''}
	[arXiv:1512.06106 [hep-ph]].
	%%CITATION = [arXiv:1512.06106;%%
	%27 citations counted in INSPIRE as of 25 Dec 2015
	
	%\cite{Bardhan:2015hcr}
	\bibitem{Bardhan:2015hcr} 
	D.~Bardhan, D.~Bhatia, A.~Chakraborty, U.~Maitra, S.~Raychaudhuri and T.~Samui,
	{\it``Radion Candidate for the LHC Diphoton Resonance,''}
	[arXiv:1512.06674 [hep-ph]].
	%%CITATION = [arXiv:1512.06674;%%
	%12 citations counted in INSPIRE as of 25 Dec 2015
	
	%\cite{Davoudiasl:2015cuo}
	\bibitem{Davoudiasl:2015cuo} 
	H.~Davoudiasl and C.~Zhang,
	{\it``A 750 GeV Messenger of Dark Conformal Symmetry Breaking,''}
	[arXiv:1512.07672 [hep-ph]].
	%%CITATION = [arXiv:1512.07672;%%
	
	%\cite{Cai:2015hzc}
	\bibitem{Cai:2015hzc} 
	C.~Cai, Z.~H.~Yu and H.~H.~Zhang,
	{\it``The 750 GeV diphoton resonance as a singlet scalar in an extra dimensional model,''}
	[arXiv:1512.08440 [hep-ph]].
	%%CITATION = [arXiv:1512.08440;%%
	
	%\cite{Cao:2016udb}
	\bibitem{Cao:2016udb}
	J.~Cao, L.~Shang, W.~Su, Y.~Zhang and J.~Zhu,
	{\it``Interpreting the 750 GeV diphoton excess in the Minimal Dilaton Model,''}
	[arXiv:1601.02570 [hep-ph]].
	%%CITATION = [arXiv:1601.02570;%%
	
	%\cite{Abel:2016pyc}
	\bibitem{Abel:2016pyc}
	S.~Abel and V.~V.~Khoze,
	{\it``Photo-production of a 750 GeV di-photon resonance mediated by Kaluza-Klein leptons in the loop,''}
	[arXiv:1601.07167 [hep-ph]].
	%%CITATION = [arXiv:1601.07167;%%
	%21 citations counted in INSPIRE as of 24 Mar 2016
	
	%%%%%%%%%%%%% Axion %%%%%%%%%%%%%%
	
	%\cite{Pilaftsis:2015ycr}
	\bibitem{Pilaftsis:2015ycr} 
	A.~Pilaftsis,
	{\it``Diphoton Signatures from Heavy Axion Decays at LHC,''}
	[arXiv:1512.04931 [hep-ph]].
	%%CITATION = [arXiv:1512.04931;%%
	%66 citations counted in INSPIRE as of 25 Dec 2015
	
	%\cite{Molinaro:2015cwg}
	\bibitem{Molinaro:2015cwg} 
	E.~Molinaro, F.~Sannino and N.~Vignaroli,
	{\it``Minimal Composite Dynamics versus Axion Origin of the Diphoton excess,''}
	[arXiv:1512.05334 [hep-ph]].
	%%CITATION = [arXiv:1512.05334;%%
	%59 citations counted in INSPIRE as of 25 Dec 2015
	
	%\cite{Bian:2015kjt}
	\bibitem{Bian:2015kjt} 
	L.~Bian, N.~Chen, D.~Liu and J.~Shu,
	{\it``A hidden confining world on the 750 GeV diphoton excess,''}
	arXiv:1512.05759 [hep-ph].
	%%CITATION = ARXIV:1512.05759;%%
	%45 citations counted in INSPIRE as of 25 Dec 2015
	
	%\cite{Kim:2015xyn}
	\bibitem{Kim:2015xyn} 
	J.~E.~Kim,
	{\it``Is an axizilla possible for di-photon resonance?,''}
	[arXiv:1512.08467 [hep-ph]].
	%%CITATION = [arXiv:1512.08467;%%
	
	%\cite{Ben-Dayan:2016gxw}
	\bibitem{Ben-Dayan:2016gxw}
	I.~Ben-Dayan and R.~Brustein,
	{\it``Hypercharge Axion and the Diphoton $750$ GeV Resonance,''}
	[arXiv:1601.07564 [hep-ph]].
	%%CITATION = [arXiv:1601.07564;%%
	%13 citations counted in INSPIRE as of 24 Mar 2016
	
	%\cite{Barrie:2016ntq}
	\bibitem{Barrie:2016ntq}
	N.~D.~Barrie, A.~Kobakhidze, M.~Talia and L.~Wu,
	{\it``750 GeV Composite Axion as the LHC Diphoton Resonance,''}
	Phys.\ Lett.\ B {\bf 755} (2016) 343
	%doi:10.1016/j.physletb.2016.02.010
	[arXiv:1602.00475 [hep-ph]].
	%%CITATION = doi:10.1016/j.physletb.2016.02.010;%%
	%12 citations counted in INSPIRE as of 24 Mar 2016
	
	%\cite{Chiang:2016eav}
	\bibitem{Chiang:2016eav}
	C.~W.~Chiang, H.~Fukuda, M.~Ibe and T.~T.~Yanagida,
	{\it``750 GeV diphoton resonance in a visible heavy QCD axion model,''}
	[arXiv:1602.07909 [hep-ph]].
	%%CITATION = [arXiv:1602.07909;%%
	%5 citations counted in INSPIRE as of 24 Mar 2016
	
	%\cite{Higaki:2016yqk}
	\bibitem{Higaki:2016yqk}
	T.~Higaki, K.~S.~Jeong, N.~Kitajima and F.~Takahashi,
	{\it``Quality of the Peccei-Quinn symmetry in the Aligned QCD Axion and Cosmological Implications,''}
	[arXiv:1603.02090 [hep-ph]].
	%%CITATION = [arXiv:1603.02090;%%
	
	%%%%%%%% dark matter %%%%%%%%%
	%%%%%%%%%%%%%%%%%
	
	%\cite{Mambrini:2015wyu}
	\bibitem{Mambrini:2015wyu} 
	Y.~Mambrini, G.~Arcadi and A.~Djouadi,
	{\it``The LHC diphoton resonance and dark matter,''}
	[arXiv:1512.04913 [hep-ph]].
	%%CITATION = [arXiv:1512.04913;%%
	%69 citations counted in INSPIRE as of 25 Dec 2015
	
	%\cite{Backovic:2015fnp}
	\bibitem{Backovic:2015fnp} 
	M.~Backovic, A.~Mariotti and D.~Redigolo,
	{\it``Di-photon excess illuminates Dark Matter,''}
	[arXiv:1512.04917 [hep-ph]].
	%%CITATION = [arXiv:1512.04917;%%
	%65 citations counted in INSPIRE as of 25 Dec 2015
	
	%\cite{Kobakhidze:2015ldh}
	\bibitem{Kobakhidze:2015ldh} 
	A.~Kobakhidze, F.~Wang, L.~Wu, J.~M.~Yang and M.~Zhang,
	{\it``750 GeV diphoton resonance in a top and bottom seesaw model,''}
	Phys.\ Lett.\ B {\bf 757}, 92 (2016)
	%doi:10.1016/j.physletb.2016.03.067
	[arXiv:1512.05585 [hep-ph]].
	%%CITATION = doi:10.1016/j.physletb.2016.03.067;%%
	%156 citations counted in INSPIRE as of 09 May 2016
	
	
	%\cite{Han:2015cty}
	\bibitem{Han:2015cty} 
	C.~Han, H.~M.~Lee, M.~Park and V.~Sanz,
	{\it``The diphoton resonance as a gravity mediator of dark matter,''}
	[arXiv:1512.06376 [hep-ph]].
	%%CITATION = [arXiv:1512.06376;%%
	%14 citations counted in INSPIRE as of 25 Dec 2015
	
	%\cite{Han:2015dlp}
	\bibitem{Han:2015dlp} 
	H.~Han, S.~Wang and S.~Zheng,
	{\it``Scalar Explanation of Diphoton Excess at LHC,''}
	[arXiv:1512.06562 [hep-ph]].
	%%CITATION = [arXiv:1512.06562;%%
	%11 citations counted in INSPIRE as of 25 Dec 2015
	
	%\cite{Bi:2015uqd}
	\bibitem{Bi:2015uqd} 
	X.~J.~Bi, Q.~F.~Xiang, P.~F.~Yin and Z.~H.~Yu,
	{\it``The 750 GeV diphoton excess at the LHC and dark matter constraints,''}
	[arXiv:1512.06787 [hep-ph]].
	%%CITATION = [arXiv:1512.06787;%%
	%12 citations counted in INSPIRE as of 25 Dec 2015
	
	%\cite{Bauer:2015boy}
	\bibitem{Bauer:2015boy} 
	M.~Bauer and M.~Neubert,
	{\it``Flavor Anomalies, the Diphoton Excess and a Dark Matter Candidate,''}
	[arXiv:1512.06828 [hep-ph]].
	%%CITATION = [arXiv:1512.06828;%%
	%16 citations counted in INSPIRE as of 25 Dec 2015
	
	%\cite{Barducci:2015gtd}
	\bibitem{Barducci:2015gtd} 
	D.~Barducci, A.~Goudelis, S.~Kulkarni and D.~Sengupta,
	{\it``One jet to rule them all: monojet constraints and invisible decays of a 750 GeV diphoton resonance,''}
	[arXiv:1512.06842 [hep-ph]].
	%%CITATION = [arXiv:1512.06842;%%
	%14 citations counted in INSPIRE as of 25 Dec 2015
	
	%\cite{Dev:2015isx}
	\bibitem{Dev:2015isx} 
	P.~S.~B.~Dev and D.~Teresi,
	{\it``Asymmetric Dark Matter in the Sun and the Diphoton Excess at the LHC,''}
	[arXiv:1512.07243 [hep-ph]].
	%%CITATION = [arXiv:1512.07243;%%
	%9 citations counted in INSPIRE as of 25 Dec 2015
	
	%\cite{Han:2015yjk}
	\bibitem{Han:2015yjk} 
	H.~Han, S.~Wang and S.~Zheng,
	{\it``Dark Matter Theories in the Light of Diphoton Excess,''}
	[arXiv:1512.07992 [hep-ph]].
	%%CITATION = [arXiv:1512.07992;%%
	
	%\cite{Park:2015ysf}
	\bibitem{Park:2015ysf} 
	J.~C.~Park and S.~C.~Park,
	{\it``Indirect signature of dark matter with the diphoton resonance at 750 GeV,''}
	[arXiv:1512.08117 [hep-ph]].
	%%CITATION = [arXiv:1512.08117;%%
	
	%\cite{Huang:2015svl}
	\bibitem{Huang:2015svl}
	X.~J.~Huang, W.~H.~Zhang and Y.~F.~Zhou,
	{\it``A 750 GeV dark matter messenger at the Galactic Center,''}
	[arXiv:1512.08992 [hep-ph]].
	%%CITATION = [arXiv:1512.08992;%%
	%11 citations counted in INSPIRE as of 13 Jan 2016
	
	%\cite{Ghorbani:2016jdq}
	\bibitem{Ghorbani:2016jdq}
	K.~Ghorbani and H.~Ghorbani,
	{\it``The 750 GeV Diphoton Excess from a Pseudoscalar in Fermionic Dark Matter Scenario,''}
	[arXiv:1601.00602 [hep-ph]].
	%%CITATION = [arXiv:1601.00602;%%
	%8 citations counted in INSPIRE as of 13 Jan 2016   
	
	%\cite{Bhattacharya:2016lyg}
	\bibitem{Bhattacharya:2016lyg}
	S.~Bhattacharya, S.~Patra, N.~Sahoo and N.~Sahu,
	{\it``750 GeV Di-photon excess at CERN LHC from a dark sector assisted scalar decay,''}
	[arXiv:1601.01569 [hep-ph]].
	%%CITATION = [arXiv:1601.01569;%%
	%4 citations counted in INSPIRE as of 13 Jan 2016
	
	%\cite{Okada:2016rav}
	\bibitem{Okada:2016rav}
	H.~Okada and K.~Yagyu,
	{\it``Renormalizable Model for Neutrino Mass, Dark Matter, Muon $g-2$ and 750 GeV Diphoton Excess,''}
	[arXiv:1601.05038 [hep-ph]].
	%%CITATION = [arXiv:1601.05038;%%
	%20 citations counted in INSPIRE as of 24 Mar 2016
	
	%\cite{Cao:2016cok}
	\bibitem{Cao:2016cok}
	Q.~H.~Cao, Y.~Q.~Gong, X.~Wang, B.~Yan and L.~L.~Yang,
	{\it``One Bump or Two Peaks? The 750 GeV Diphoton Excess and Dark Matter with a Complex Mediator,''}
	[arXiv:1601.06374 [hep-ph]].
	%%CITATION = [arXiv:1601.06374;%%
	%15 citations counted in INSPIRE as of 24 Mar 2016
	
	%\cite{Kawamura:2016idj}
	\bibitem{Kawamura:2016idj}
	J.~Kawamura and Y.~Omura,
	{\it``Diphoton excess at 750 GeV and LHC constraints in models with vector-like particles,''}
	[arXiv:1601.07396 [hep-ph]].
	%%CITATION = [arXiv:1601.07396;%%
	%15 citations counted in INSPIRE as of 24 Mar 2016
	
	%\cite{Ge:2016xcq}
	\bibitem{Ge:2016xcq}
	S.~F.~Ge, H.~J.~He, J.~Ren and Z.~Z.~Xianyu,
	{\it``Realizing Dark Matter and Higgs Inflation in Light of LHC Diphoton Excess,''}
	[arXiv:1602.01801 [hep-ph]].
	%%CITATION = [arXiv:1602.01801;%%
	%13 citations counted in INSPIRE as of 24 Mar 2016


%\cite{Staub:2016dxq}
\bibitem{Staub:2016dxq}
F.~Staub {\it et al.},
{\it``Precision tools and models to narrow in on the 750 GeV diphoton resonance,''}
arXiv:1602.05581 [hep-ph].
%%CITATION = ARXIV:1602.05581;%%
	
	%\cite{Redi:2016kip}
	\bibitem{Redi:2016kip}
	M.~Redi, A.~Strumia, A.~Tesi and E.~Vigiani,
	{\it``Di-photon resonance and Dark Matter as heavy pions,''}
	[arXiv:1602.07297 [hep-ph]].
	%%CITATION = [arXiv:1602.07297;%%
	%6 citations counted in INSPIRE as of 24 Mar 2016
	
	%\cite{Bi:2016gca}
	\bibitem{Bi:2016gca}
	X.~J.~Bi, Z.~Kang, P.~Ko, J.~Li and T.~Li,
	{\it``ADMonium: Asymmetric Dark Matter Bound State,''}
	[arXiv:1602.08816 [hep-ph]].
	%%CITATION = [arXiv:1602.08816;%%
	%1 citations counted in INSPIRE as of 24 Mar 2016
	
	%\cite{Chen:2016sck}
	\bibitem{Chen:2016sck}
	C.~Y.~Chen, M.~Lefebvre, M.~Pospelov and Y.~M.~Zhong,
	{\it``Diphoton Excess through Dark Mediators,''}
	arXiv:1603.01256 [hep-ph]].
	%%CITATION = [arXiv:1603.01256;%%
	%2 citations counted in INSPIRE as of 24 Mar 2016
	
	
	
	%\cite{Csaki:2016kqr}
	\bibitem{Csaki:2016kqr}
	C.~Csaki and L.~Randall,
	{\it``A Diphoton Resonance from Bulk RS,''}
	arXiv:1603.07303 [hep-ph].
	%%CITATION = ARXIV:1603.07303;%%
	%4 citations counted in INSPIRE as of 29 Apr 2016
	
	%\cite{Bai:2016rmn}
	\bibitem{Bai:2016rmn}
	Y.~Bai and J.~Berger,
	{\it``Superbumps,''}
	arXiv:1603.07335 [hep-ph].
	%%CITATION = ARXIV:1603.07335;%%
	
	%\cite{Deppisch:2016qqd}
	\bibitem{Deppisch:2016qqd}
	F.~F.~Deppisch, S.~Kulkarni, H.~Päs and E.~Schumacher,
	{\it``Leptoquark patterns unifying neutrino masses, flavor anomalies and the diphoton excess,''}
	arXiv:1603.07672 [hep-ph].
	%%CITATION = ARXIV:1603.07672;%%
	%1 citations counted in INSPIRE as of 29 Apr 2016
	
	%\cite{Hewett:2016omf}
	\bibitem{Hewett:2016omf}
	J.~L.~Hewett and T.~G.~Rizzo,
	{\it``750 GeV Diphoton Resonance in Warped Geometries,''}
	arXiv:1603.08250 [hep-ph].
	%%CITATION = ARXIV:1603.08250;%%
	%6 citations counted in INSPIRE as of 29 Apr 2016
	
	%\cite{Ko:2016sht}
	\bibitem{Ko:2016sht}
	P.~Ko, C.~Yu and T.~C.~Yuan,
	{\it``750 GeV Diphoton Excess as a Composite (Pseudo)scalar Boson from New Strong Interaction,''}
	arXiv:1603.08802 [hep-ph].
	%%CITATION = ARXIV:1603.08802;%%
	%2 citations counted in INSPIRE as of 29 Apr 2016
	
	%\cite{Carmona:2016jhr}
	\bibitem{Carmona:2016jhr}
	A.~Carmona,
	{\it``A 750 GeV graviton from holographic composite dark sectors,''}
	arXiv:1603.08913 [hep-ph].
	%%CITATION = ARXIV:1603.08913;%%
	%5 citations counted in INSPIRE as of 29 Apr 2016
	
	%\cite{Howe:2016mfq}
	\bibitem{Howe:2016mfq}
	K.~Howe, S.~Knapen and D.~J.~Robinson,
	{\it``Diphotons from an Electroweak Triplet-Singlet,''}
	arXiv:1603.08932 [hep-ph].
	%%CITATION = ARXIV:1603.08932;%%
	%1 citations counted in INSPIRE as of 29 Apr 2016
	
	%\cite{Collins:2016pef}
	\bibitem{Collins:2016pef}
	J.~H.~Collins, C.~Csaki, J.~A.~Dror and S.~Lombardo,
	{\it``Novel kinematics from a custodially protected diphoton resonance,''}
	arXiv:1603.09350 [hep-ph].
	%%CITATION = ARXIV:1603.09350;%%
	%1 citations counted in INSPIRE as of 29 Apr 2016
	
	%\cite{Frandsen:2016bke}
	\bibitem{Frandsen:2016bke}
	M.~T.~Frandsen and I.~M.~Shoemaker,
	{\it``Asymmetric Dark Matter Models and the LHC Diphoton Excess,''}
	arXiv:1603.09354 [hep-ph].
	%%CITATION = ARXIV:1603.09354;%%
	%1 citations counted in INSPIRE as of 29 Apr 2016
	
	%\cite{Dillon:2016fgw}
	\bibitem{Dillon:2016fgw}
	B.~M.~Dillon and V.~Sanz,
	{\it``A Little KK Graviton at 750 GeV,''}
	arXiv:1603.09550 [hep-ph].
	%%CITATION = ARXIV:1603.09550;%%
	%3 citations counted in INSPIRE as of 29 Apr 2016
	
	%\cite{Ellis:2016yrj}
	\bibitem{Ellis:2016yrj}
	J.~Ellis,
	{\it``Prospects for Future Collider Physics,''}
	arXiv:1604.00333 [hep-ph].
	%%CITATION = ARXIV:1604.00333;%%
	
	%\cite{Liu:2016lkj}
	\bibitem{Liu:2016lkj}
	N.~Liu, W.~Wang, M.~Zhang and R.~Zheng,
	{\it``750 GeV Diphoton Resonance in a Vector-like Extension of Hill Model,''}
	arXiv:1604.00728 [hep-ph].
	%%CITATION = ARXIV:1604.00728;%%
	%2 citations counted in INSPIRE as of 29 Apr 2016
	
	%\cite{Chakrabarty:2016hxi}
	\bibitem{Chakrabarty:2016hxi}
	N.~Chakrabarty, B.~Mukhopadhyaya and S.~SenGupta,
	{\it``Diphoton excess via Chern-Simons interaction in a warped geometry scenario,''}
	arXiv:1604.00885 [hep-ph].
	%%CITATION = ARXIV:1604.00885;%%
	
	%\cite{Cynolter:2016jxv}
	\bibitem{Cynolter:2016jxv}
	G.~Cynolter, J.~.Kovács and E.~Lendvai,
	{\it``Diphoton excess and VV-scattering,''}
	arXiv:1604.01008 [hep-ph].
	%%CITATION = ARXIV:1604.01008;%%
	%2 citations counted in INSPIRE as of 29 Apr 2016
	
	%\cite{McDonald:2016cdh}
	\bibitem{McDonald:2016cdh}
	J.~McDonald,
	{\it``The 750 GeV Resonance as Non-Minimally Coupled Inflaton: Unitarity Violation and Why the Resonance is a Real Singlet Scalar,''}
	arXiv:1604.01711 [hep-ph].
	%%CITATION = ARXIV:1604.01711;%%
	
	%\cite{Lebiedowicz:2016lmn}
	\bibitem{Lebiedowicz:2016lmn}
	P.~Lebiedowicz, M.~Luszczak, R.~Pasechnik and A.~Szczurek,
	{\it``Can the diphoton enhancement at 750 GeV be due to a neutral technipion?,''}
	arXiv:1604.02037 [hep-ph].
	%%CITATION = ARXIV:1604.02037;%%
	%1 citations counted in INSPIRE as of 29 Apr 2016
	
	%\cite{Chala:2016mdz}
	\bibitem{Chala:2016mdz}
	M.~Chala, C.~Grojean, M.~Riembau and T.~Vantalon,
	{\it``Deciphering the CP nature of the 750 GeV resonance,''}
	arXiv:1604.02029 [hep-ph].
	%%CITATION = ARXIV:1604.02029;%%
	%1 citations counted in INSPIRE as of 29 Apr 2016
	
	%\cite{Kusenko:2016vcq}
	\bibitem{Kusenko:2016vcq}
	A.~Kusenko, L.~Pearce and L.~Yang,
	{\it``Leptogenesis via the 750 GeV pseudoscalar,''}
	arXiv:1604.02382 [hep-ph].
	%%CITATION = ARXIV:1604.02382;%%
	
	%\cite{Kamenshchik:2016tjz}
	\bibitem{Kamenshchik:2016tjz}
	A.~Y.~Kamenshchik, A.~A.~Starobinsky, A.~Tronconi, G.~P.~Vacca and G.~Venturi,
	{\it``Vacuum energy, Standard Model physics and the $750\; \rm{GeV}$ Diphoton Excess at the LHC,''}
	arXiv:1604.02371 [hep-ph].
	%%CITATION = ARXIV:1604.02371;%%
	
	%\cite{Barrie:2016ndh}
	\bibitem{Barrie:2016ndh}
	N.~D.~Barrie, A.~Kobakhidze, S.~Liang, M.~Talia and L.~Wu,
	{\it``Heavy Leptonium as the Origin of the 750 GeV Diphoton Excess,''}
	arXiv:1604.02803 [hep-ph].
	%%CITATION = ARXIV:1604.02803;%%
	%2 citations counted in INSPIRE as of 29 Apr 2016
	
	%\cite{Nilles:2016bjl}
	\bibitem{Nilles:2016bjl}
	H.~P.~Nilles and M.~W.~Winkler,
	{\it``750 GeV Diphotons and Supersymmetric Grand Unification,''}
	arXiv:1604.03598 [hep-ph].
	%%CITATION = ARXIV:1604.03598;%%
	%2 citations counted in INSPIRE as of 29 Apr 2016
	
	%\cite{Agarwal:2016gxe}
	\bibitem{Agarwal:2016gxe}
	B.~Agarwal, J.~Isaacson and K.~A.~Mohan,
	{\it``Minimal Dilaton Model and the Diphoton Excess,''}
	arXiv:1604.05328 [hep-ph].
	%%CITATION = ARXIV:1604.05328;%%
	
	%\cite{Duerr:2016eme}
	\bibitem{Duerr:2016eme}
	M.~Duerr, P.~Fileviez Pérez and J.~Smirnov,
	{\it``New Forces and the 750 GeV Resonance,''}
	arXiv:1604.05319 [hep-ph].
	%%CITATION = ARXIV:1604.05319;%%
	%2 citations counted in INSPIRE as of 29 Apr 2016
	
	%\cite{Gopalakrishna:2016tku}
	\bibitem{Gopalakrishna:2016tku}
	S.~Gopalakrishna and T.~S.~Mukherjee,
	{\it``The 750 GeV diphoton excess in a two Higgs doublet model and a singlet scalar model, with vector-like fermions, unitarity constraints, and dark matter implications,''}
	arXiv:1604.05774 [hep-ph].
	%%CITATION = ARXIV:1604.05774;%%
	%1 citations counted in INSPIRE as of 29 Apr 2016
	
	%\cite{Yamada:2016jgg}
	\bibitem{Yamada:2016jgg}
	M.~Yamada, T.~T.~Yanagida and K.~Yonekura,
	{\it``Diphoton Excess as a Hidden Monopole,''}
	arXiv:1604.07203 [hep-ph].
	%%CITATION = ARXIV:1604.07203;%%
	%1 citations counted in INSPIRE as of 29 Apr 2016
	
	%\cite{Takahashi:2016iph}
	\bibitem{Takahashi:2016iph}
	F.~Takahashi, M.~Yamada and N.~Yokozaki,
	{\it``Diphoton excess from hidden U(1) gauge symmetry with large kinetic mixing,''}
	arXiv:1604.07145 [hep-ph].
	%%CITATION = ARXIV:1604.07145;%%
	
	%\cite{Iwamoto:2016ral}
	\bibitem{Iwamoto:2016ral}
	S.~Iwamoto, G.~Lee, Y.~Shadmi and R.~Ziegler,
	{\it``Diphoton Signals from Colorless Hidden Quarkonia,''}
	arXiv:1604.07776 [hep-ph].
	%%CITATION = ARXIV:1604.07776;%%
	
	%\cite{Hamaguchi:2016umx}
	\bibitem{Hamaguchi:2016umx}
	K.~Hamaguchi and S.~P.~Liew,
	{\it``Models of 750 GeV quarkonium and the LHC excesses,''}
	arXiv:1604.07828 [hep-ph].
	%%CITATION = ARXIV:1604.07828;%%
	
	%\cite{Dutta:2016ach}
	\bibitem{Dutta:2016ach}
	B.~Dutta, Y.~Gao, T.~Ghosh, I.~Gogoladze, T.~Li and J.~W.~Walker,
	{\it``An SU(6) GUT Origin of the TeV-Scale Vector-like Particles Associated with the 750 GeV Diphoton Resonance,''}
	arXiv:1604.07838 [hep-ph].
	%%CITATION = ARXIV:1604.07838;%%
	
	%\cite{Bae:2016ewo}
	\bibitem{Bae:2016ewo}
	K.~J.~Bae, C.~R.~Chen, K.~Hamaguchi and I.~Low,
	{\it``From the 750 GeV Diphoton Resonance to Multilepton Excesses,''}
	arXiv:1604.07941 [hep-ph].
	%%CITATION = ARXIV:1604.07941;%%
	
	
	%\cite{Dev:2015vjd}
	\bibitem{Dev:2015vjd}
	P.~S.~B.~Dev, R.~N.~Mohapatra and Y.~Zhang,
	{\it``Quark Seesaw, Vectorlike Fermions and Diphoton Excess,''}
	JHEP {\bf 1602}, 186 (2016)
	%doi:10.1007/JHEP02(2016)186
	[arXiv:1512.08507 [hep-ph]].
	%%CITATION = doi:10.1007/JHEP02(2016)186;%%
	%73 citations counted in INSPIRE as of 11 May 2016
	
	%\cite{Bardhan:2016rsb}
	\bibitem{Bardhan:2016rsb}
	D.~Bardhan, P.~Byakti, D.~Ghosh and T.~Sharma,
	{\it``The 750 GeV diphoton resonance as an sgoldstino: a reappraisal,''}
	arXiv:1603.05251 [hep-ph].
	%%CITATION = ARXIV:1603.05251;%%
	%4 citations counted in INSPIRE as of 14 May 2016
	
	%\cite{Marzola:2015xbh}
	\bibitem{Marzola:2015xbh}
	L.~Marzola, A.~Racioppi, M.~Raidal, F.~R.~Urban and H.~Veermäe,
	{\it``Non-minimal CW inflation, electroweak symmetry breaking and the 750 GeV anomaly,''}
	JHEP {\bf 1603} (2016) 190
	%doi:10.1007/JHEP03(2016)190
	[arXiv:1512.09136 [hep-ph]].
	%%CITATION = doi:10.1007/JHEP03(2016)190;%%
	%45 citations counted in INSPIRE as of 14 May 2016
	
	%\cite{Ding:2015rxx}
	\bibitem{Ding:2015rxx}
	R.~Ding, L.~Huang, T.~Li and B.~Zhu,
	{\it``Interpreting $750$ GeV Diphoton Excess with R-parity Violation Supersymmetry,''}
	arXiv:1512.06560 [hep-ph].
	%%CITATION = ARXIV:1512.06560;%%
	%104 citations counted in INSPIRE as of 10 May 2016
	
	%\cite{Ding:2016ldt}
	\bibitem{Ding:2016ldt} 
	R.~Ding, Z.~L.~Han, Y.~Liao and X.~D.~Ma,
	{\it``Interpretation of 750 GeV Diphoton Excess at LHC in Singlet Extension of Color-octet Neutrino Mass Model,''}
	Eur.\ Phys.\ J.\ C {\bf 76}, no. 4, 204 (2016)
	%doi:10.1140/epjc/s10052-016-4052-6
	[arXiv:1601.02714 [hep-ph]].
	%%CITATION = doi:10.1140/epjc/s10052-016-4052-6;%%
	%37 citations counted in INSPIRE as of 10 May 2016
	
	%\cite{Dey:2015bur}
	\bibitem{Dey:2015bur}
	U.~K.~Dey, S.~Mohanty and G.~Tomar,
	{\it``750 GeV resonance in the dark left–right model,''}
	Phys.\ Lett.\ B {\bf 756} (2016) 384
	%doi:10.1016/j.physletb.2016.03.048
	[arXiv:1512.07212 [hep-ph]].
	%%CITATION = doi:10.1016/j.physletb.2016.03.048;%%
	%97 citations counted in INSPIRE as of 14 May 2016
	
	%\cite{Modak:2016ung}
	\bibitem{Modak:2016ung}
	T.~Modak, S.~Sadhukhan and R.~Srivastava,
	{\it``750 GeV diphoton excess from gauged $B - L$ symmetry,''}
	Phys.\ Lett.\ B {\bf 756}, 405 (2016)
	%doi:10.1016/j.physletb.2016.03.021
	[arXiv:1601.00836 [hep-ph]].
	%%CITATION = doi:10.1016/j.physletb.2016.03.021;%%
	%45 citations counted in INSPIRE as of 09 May 2016
	
	%\cite{Bonilla:2016sgx}
	\bibitem{Bonilla:2016sgx}
	C.~Bonilla, M.~Nebot, R.~Srivastava and J.~W.~F.~Valle,
	{\it``Flavor physics scenario for the 750 GeV diphoton anomaly,''}
	Phys.\ Rev.\ D {\bf 93}, no. 7, 073009 (2016)
	%doi:10.1103/PhysRevD.93.073009
	[arXiv:1602.08092 [hep-ph]].
	
	%\cite{Heckman:2015kqk}
	\bibitem{Heckman:2015kqk}
	J.~J.~Heckman,
	{\it``750 GeV Diphotons from a D3-brane,''}
	Nucl.\ Phys.\ B {\bf 906} (2016) 231
	%doi:10.1016/j.nuclphysb.2016.02.031
	[arXiv:1512.06773 [hep-ph]].
	%%CITATION = doi:10.1016/j.nuclphysb.2016.02.031;%%
	%99 citations counted in INSPIRE as of 14 May 2016
	
	%\cite{Moretti:2015pbj}
	\bibitem{Moretti:2015pbj} 
	S.~Moretti and K.~Yagyu,
	{\it``750 GeV diphoton excess and its explanation in two-Higgs-doublet models with a real inert scalar multiplet,''}
	Phys.\ Rev.\ D {\bf 93}, no. 5, 055043 (2016)
	[arXiv:1512.07462 [hep-ph]].
	%%CITATION = doi:10.1103/PhysRevD.93.055043;%%
	
	%\cite{Harigaya:2015ezk}
	\bibitem{Harigaya:2015ezk}
	K.~Harigaya and Y.~Nomura,
	{\it``Composite Models for the 750 GeV Diphoton Excess,''}
	Phys.\ Lett.\ B {\bf 754} (2016) 151
	%doi:10.1016/j.physletb.2016.01.026
	[arXiv:1512.04850 [hep-ph]].
	%%CITATION = doi:10.1016/j.physletb.2016.01.026;%%
	%196 citations counted in INSPIRE as of 14 May 2016
	
	%\cite{Hall:2015xds}
	\bibitem{Hall:2015xds}
	L.~J.~Hall, K.~Harigaya and Y.~Nomura,
	{\it``750 GeV Diphotons: Implications for Supersymmetric Unification,''}
	JHEP {\bf 1603} (2016) 017
	%doi:10.1007/JHEP03(2016)017
	[arXiv:1512.07904 [hep-ph]].
	%%CITATION = doi:10.1007/JHEP03(2016)017;%%
	%88 citations counted in INSPIRE as of 14 May 2016
	
	%\cite{Harigaya:2016pnu}
	\bibitem{Harigaya:2016pnu}
	K.~Harigaya and Y.~Nomura,
	{\it``A Composite Model for the 750 GeV Diphoton Excess,''}
	JHEP {\bf 1603} (2016) 091
	%doi:10.1007/JHEP03(2016)091
	[arXiv:1602.01092 [hep-ph]].
	%%CITATION = doi:10.1007/JHEP03(2016)091;%%
	%25 citations counted in INSPIRE as of 14 May 2016
	
	%\cite{Harigaya:2016eol}
	\bibitem{Harigaya:2016eol}
	K.~Harigaya and Y.~Nomura,
	{\it``Hidden Pion Varieties in Composite Models for Diphoton Resonances,''}
	arXiv:1603.05774 [hep-ph].
	%%CITATION = ARXIV:1603.05774;%%
	%7 citations counted in INSPIRE as of 14 May 2016
	
	
	%%%%%%%%%%%%% 3-body: gamgam+X %%%%%%%%%%%%%%
	
	%\cite{Kim:2015ron}
	\bibitem{Kim:2015ron} 
	J.~S.~Kim, J.~Reuter, K.~Rolbiecki and R.~R.~de Austri,
	{\it``A resonance without resonance: scrutinizing the diphoton excess at 750 GeV,''}
	[arXiv:1512.06083 [hep-ph]].
	%%CITATION = [arXiv:1512.06083;%%
	%22 citations counted in INSPIRE as of 25 Dec 2015
	
	%\cite{Alves:2015jgx}
	\bibitem{Alves:2015jgx} 
	A.~Alves, A.~G.~Dias and K.~Sinha,
	{\it``The 750 GeV $S$-cion: Where else should we look for it?,''}
	[arXiv:1512.06091 [hep-ph]].
	%%CITATION = [arXiv:1512.06091;%%
	%26 citations counted in INSPIRE as of 25 Dec 2015
	
	%\cite{Bernon:2015abk}
	\bibitem{Bernon:2015abk} 
	J.~Bernon and C.~Smith,
	{\it``Could the width of the diphoton anomaly signal a three-body decay ?,''}
	[arXiv:1512.06113 [hep-ph]].
	%%CITATION = [arXiv:1512.06113;%%
	%22 citations counted in INSPIRE as of 25 Dec 2015
	
	%\cite{Huang:2015evq}
	\bibitem{Huang:2015evq} 
	F.~P.~Huang, C.~S.~Li, Z.~L.~Liu and Y.~Wang,
	{\it``750 GeV Diphoton Excess from Cascade Decay,''}
	[arXiv:1512.06732 [hep-ph]].
	%%CITATION = [arXiv:1512.06732;%%
	%12 citations counted in INSPIRE as of 25 Dec 2015
	
	%\cite{Cho:2015nxy}
	\bibitem{Cho:2015nxy} 
	W.~S.~Cho, D.~Kim, K.~Kong, S.~H.~Lim, K.~T.~Matchev, J.~C.~Park and M.~Park,
	{\it``The 750 GeV Diphoton Excess May Not Imply a 750 GeV Resonance,''}
	[arXiv:1512.06824 [hep-ph]].
	%%CITATION = [arXiv:1512.06824;%%
	%14 citations counted in INSPIRE as of 25 Dec 2015
	
	%\cite{An:2015cgp}
	\bibitem{An:2015cgp} 
	H.~An, C.~Cheung and Y.~Zhang,
	{\it``Broad Diphotons from Narrow States,''}
	[arXiv:1512.08378 [hep-ph]].
	%%CITATION = [arXiv:1512.08378;%%
	
	
	%%%%%%%%%%%%% 4-photon jets %%%%%%%%%%%%%%
%\cite{Dobrescu:2000jt}
\bibitem{Dobrescu:2000jt}
B.~A.~Dobrescu, G.~L.~Landsberg and K.~T.~Matchev,
{\it``Higgs boson decays to CP odd scalars at the Tevatron and beyond,''}
Phys.\ Rev.\ D {\bf 63} (2001) 075003
%doi:10.1103/PhysRevD.63.075003
[hep-ph/0005308].
%%CITATION = doi:10.1103/PhysRevD.63.075003;%%
%145 citations counted in INSPIRE as of 05 Jun 2016

%\cite{Dobrescu:2000yn}
\bibitem{Dobrescu:2000yn}
B.~A.~Dobrescu and K.~T.~Matchev,
{\it``Light axion within the next-to-minimal supersymmetric standard model,''}
JHEP {\bf 0009} (2000) 031
%doi:10.1088/1126-6708/2000/09/031
[hep-ph/0008192].
%%CITATION = doi:10.1088/1126-6708/2000/09/031;%%
%137 citations counted in INSPIRE as of 05 Jun 2016


	%\cite{Chang:2015sdy}
	\bibitem{Chang:2015sdy} 
	J.~Chang, K.~Cheung and C.~T.~Lu,
	{\it``Interpreting the 750 GeV Di-photon Resonance using photon-jets in Hidden-Valley-like models,''}
	[arXiv:1512.06671 [hep-ph]].
	%%CITATION = [arXiv:1512.06671;%%
	%11 citations counted in INSPIRE as of 25 Dec 2015
	
	
	%\cite{Bi:2015lcf}
	\bibitem{Bi:2015lcf} 
	X.~J.~Bi {\it et al.},
	{\it``A Promising Interpretation of Diphoton Resonance at 750 GeV,''}
	[arXiv:1512.08497 [hep-ph]].
	%%CITATION = [arXiv:1512.08497;%%
	
	%\cite{Aparicio:2016iwr}
	\bibitem{Aparicio:2016iwr}
	L.~Aparicio, A.~Azatov, E.~Hardy and A.~Romanino,
	{\it``Diphotons from Diaxions,''}
	[arXiv:1602.00949 [hep-ph]].
	%%CITATION = [arXiv:1602.00949;%%
	%19 citations counted in INSPIRE as of 24 Mar 2016
	
	%\cite{Ellwanger:2016qax}
	\bibitem{Ellwanger:2016qax}
	U.~Ellwanger and C.~Hugonie,
	{\it``A 750 GeV Diphoton Signal from a Very Light Pseudoscalar in the NMSSM,''}
	[arXiv:1602.03344 [hep-ph]].
	%%CITATION = [arXiv:1602.03344;%%
	%17 citations counted in INSPIRE as of 24 Mar 2016
	
	%\cite{Domingo:2016unq}
	\bibitem{Domingo:2016unq}
	F.~Domingo, S.~Heinemeyer, J.~S.~Kim and K.~Rolbiecki,
	{\it``The NMSSM lives - with the 750 GeV diphoton excess,''}
	[arXiv:1602.07691 [hep-ph]].
	%%CITATION = [arXiv:1602.07691;%%
	
	%\cite{Badziak:2016cfd}
	\bibitem{Badziak:2016cfd}
	M.~Badziak, M.~Olechowski, S.~Pokorski and K.~Sakurai,
	{\it``Interpreting 750 GeV Diphoton Excess in Plain NMSSM,''}
	arXiv:1603.02203 [hep-ph].
	%%CITATION = ARXIV:1603.02203;%%
	%7 citations counted in INSPIRE as of 14 May 2016
	
	%\cite{Dasgupta:2016wxw}
	\bibitem{Dasgupta:2016wxw}
	B.~Dasgupta, J.~Kopp and P.~Schwaller,
	{\it``Photons, Photon Jets and Dark Photons at 750\,GeV and Beyond,''}
	arXiv:1602.04692 [hep-ph]].
	%%CITATION = [arXiv:1602.04692;%%
	%14 citations counted in INSPIRE as of 24 Mar 2016
	
	\bibitem{Agrawal:2015dbf} 
	P.~Agrawal, J.~Fan, B.~Heidenreich, M.~Reece and M.~Strassler,
	{\it``Experimental Considerations Motivated by the Diphoton Excess at the LHC,''}
	[arXiv:1512.05775 [hep-ph]].
	%%CITATION = [arXiv:1512.05775;%%
	%116 citations counted in INSPIRE as of 15 Mar 2016
	
	%\cite{Bernon:2014nxa}
	\bibitem{Bernon:2014nxa}
	J.~Bernon, J.~F.~Gunion, Y.~Jiang and S.~Kraml,
	{\it``Light Higgs bosons in Two-Higgs-Doublet Models"},
	Phys.\ Rev.\ D {\bf 91} (2015) 7,  075019
	%doi:10.1103/PhysRevD.91.075019
	[arXiv:1412.3385 [hep-ph]].
	
	
	%\cite{Ellis:2012sd}
	\bibitem{Ellis:2012sd}
	S.~D.~Ellis, T.~S.~Roy and J.~Scholtz,
	{\it``Jets and Photons,''}
	Phys.\ Rev.\ Lett.\  {\bf 110} (2013) no.12,  122003
	%doi:10.1103/PhysRevLett.110.122003
	[arXiv:1210.1855 [hep-ph]].
	%%CITATION = doi:10.1103/PhysRevLett.110.122003;%%
	%14 citations counted in INSPIRE as of 27 Apr 2016
	

	
	
	%%% Photon Jet: Collider %%%%
	
	
	\bibitem{Khachatryan:2015iwa} 
	V.~Khachatryan {\it et al.} [CMS Collaboration],
	{\it``Performance of Photon Reconstruction and Identification with the CMS Detector in Proton-Proton Collisions at sqrt(s) = 8 TeV,''}
	JINST {\bf 10} (2015) no. 08, P08010 
	%doi:10.1088/1748-0221/10/08/P08010
	[arXiv:1502.02702 [physics.ins-det]].
	%%CITATION = doi:10.1088/1748-0221/10/08/P08010;%%
	%46 citations counted in INSPIRE as of 15 Mar 2016
	
	\bibitem{Aad:2009wy} 
	G.~Aad {\it et al.} [ATLAS Collaboration],
	{\it``Expected Performance of the ATLAS Experiment - Detector, Trigger and Physics,''}
	[arXiv:0901.0512 [hep-ex]].
	%%CITATION = [arXiv:0901.0512;%%
	%1831 citations counted in INSPIRE as of 15 Mar 2016
	
	\bibitem{Aad:2010sp} 
	G.~Aad {\it et al.} [ATLAS Collaboration],
	{\it``Measurement of the inclusive isolated prompt photon cross section in $pp$ collisions at $\sqrt{s}=7$ TeV with the ATLAS detector,''}
	Phys.\ Rev.\ D {\bf 83} (2011) 052005 
	%doi:10.1103/PhysRevD.83.052005
	[arXiv:1012.4389 [hep-ex]].
	%%CITATION = doi:10.1103/PhysRevD.83.052005;%%
	%197 citations counted in INSPIRE as of 15 Mar 2016
	
	
	%%%% Introduction: Experiments %%%%%%%
	\bibitem{Aad:2013txa} 
	G.~Aad {\it et al.} [ATLAS Collaboration],
	{\it``Triggers for displaced decays of long-lived neutral particles in the ATLAS detector,''}
	JINST {\bf 8} (2013) P07015 
	%doi:10.1088/1748-0221/8/07/P07015
	[arXiv:1305.2284 [hep-ex]].
	
	\bibitem{Aad:2014yea} 
	G.~Aad {\it et al.} [ATLAS Collaboration],
	{\it``Search for long-lived neutral particles decaying into lepton jets in proton-proton collisions at $ \sqrt{s}=8 $ TeV with the ATLAS detector,''}
	JHEP {\bf 1411} (2014) 088 
	%doi:10.1007/JHEP11(2014)088
	[arXiv:1409.0746 [hep-ex]].
	
	\bibitem{Aad:2015asa} 
	G.~Aad {\it et al.} [ATLAS Collaboration],
	{\it``Search for pair-produced long-lived neutral particles decaying in the ATLAS hadronic calorimeter in $pp$ collisions at $\sqrt{s}$ = 8 TeV,''}
	Phys.\ Lett.\ B {\bf 743} (2015) 15 
	%doi:10.1016/j.physletb.2015.02.015
	[arXiv:1501.04020 [hep-ex]].
	
	
	%%% Dijet and tt 8TeV %%%%
	
	
	\bibitem{CMS:2015neg}
	CMS Collaboration [CMS Collaboration],
	{\it``Search for Resonances Decaying to Dijet Final States at $\sqrt{s} = 8$ TeV with Scouting Data,''}
	CMS-PAS-EXO-14-005.
	%%CITATION = CMS-PAS-EXO-14-005;%%
	%45 citations counted in INSPIRE as of 16 Mar 2016
	
	\bibitem{Aad:2015fna}
	G.~Aad {\it et al.} [ATLAS Collaboration],
	{\it``A search for $ t\overline{t} $ resonances using lepton-plus-jets events in proton-proton collisions at $ \sqrt{s}=8 $ TeV with the ATLAS detector,''}
	JHEP {\bf 1508} (2015) 148
	%doi:10.1007/JHEP08(2015)148
	[arXiv:1505.07018 [hep-ex]].
	%%CITATION = doi:10.1007/JHEP08(2015)148;%%
	%46 citations counted in INSPIRE as of 16 Mar 2016
	
	
	
	
	%%% Another 2HDM %%%%
	
	
	
	
	%\cite{Drozd:2014yla}
	\bibitem{Drozd:2014yla}
	A.~Drozd, B.~Grzadkowski, J.~F.~Gunion and Y.~Jiang,
	{\it Extending two-Higgs-doublet models by a singlet scalar field - the Case for Dark Matter},
	JHEP {\bf 1411} (2014) 105
	%doi:10.1007/JHEP11(2014)105
	[arXiv:1408.2106 [hep-ph]].
	
	
	\bibitem{ATLAS:2012soa} 
	[ATLAS Collaboration],
	{\it``Search for a Higgs boson decaying to four photons through light CP-odd scalar coupling using 4.9 fb$^{-1}$ of $7~\mathrm{TeV}$ $pp$ collision data taken with ATLAS detector at the LHC,''}
	ATLAS-CONF-2012-079.
	%%CITATION = ATLAS-CONF-2012-079;%%
	%27 citations counted in INSPIRE as of 15 Mar 2016
	%%% Higgs Handbook %%%%
	
	\bibitem{Dittmaier:2011ti} 
	S.~Dittmaier {\it et al.} [LHC Higgs Cross Section Working Group Collaboration],
	{\it``Handbook of LHC Higgs Cross Sections: 1. Inclusive Observables,''}
	%doi:10.5170/CERN-2011-002
	[arXiv:1101.0593 [hep-ph]].
	%%CITATION = doi:10.5170/CERN-2011-002;%%
	%1074 citations counted in INSPIRE as of 15 Mar 2016
	
	
	%% Sec6 Prospects %%%%%%%%%%%%%%%%
	
	%\cite{Dorsch:2014qja}
	\bibitem{Dorsch:2014qja}
	G.~C.~Dorsch, S.~J.~Huber, K.~Mimasu and J.~M.~No,
	{\it``Echoes of the Electroweak Phase Transition: Discovering a second Higgs doublet through $A_0 \rightarrow ZH_0$,''}
	Phys.\ Rev.\ Lett.\  {\bf 113} (2014) no.21,  211802
	%doi:10.1103/PhysRevLett.113.211802
	[arXiv:1405.5537 [hep-ph]].
	%%CITATION = doi:10.1103/PhysRevLett.113.211802;%%
	%21 citations counted in INSPIRE as of 22 Mar 2016
	
	%\cite{Basso:2015dka}
	\bibitem{Basso:2015dka}
	L.~Basso, P.~Osland and G.~M.~Pruna,
	{\it``Charged-Higgs production in the Two-Higgs-doublet model — the $\tau\nu$ channel,''}
	JHEP {\bf 1506} (2015) 083
	%doi:10.1007/JHEP06(2015)083
	[arXiv:1504.07552 [hep-ph]].
	%%CITATION = doi:10.1007/JHEP06(2015)083;%%
	%1 citations counted in INSPIRE as of 22 Mar 2016
	
	%\cite{Enberg:2014pua}
	\bibitem{Enberg:2014pua}
	R.~Enberg, W.~Klemm, S.~Moretti, S.~Munir and G.~Wouda,
	{\it``Charged Higgs boson in the $W^\pm$ Higgs channel at the Large Hadron Collider,''}
	Nucl.\ Phys.\ B {\bf 893} (2015) 420
	%doi:10.1016/j.nuclphysb.2015.02.001
	[arXiv:1412.5814 [hep-ph]].
	%%CITATION = doi:10.1016/j.nuclphysb.2015.02.001;%%
	%10 citations counted in INSPIRE as of 22 Mar 2016
	
	
\end{thebibliography}
%-----------------------------------------------------------------------------------------------------------------------------

%%%%%%%%%%%%%%%%%%%%%%%%%%%%%%%%%%%%%%%%%%%%%%%%%%%%%%
\end{document}